\documentclass[11pt]{article}
\usepackage{amsmath}
\usepackage{slashed}
\usepackage{amssymb}
\usepackage{axodraw}

\def\r2{\sqrt 2}
\def\beq{\begin{equation}}
\def\eeq{\end{equation}}
\def\beqn{\begin{eqnarray}}
\def\eeqn{\end{eqnarray}}

\def\PL{{1-\gamma_5\over 2}}
\def\PR{{1+\gamma_5\over 2}}
\def\sinW2{\sin^2\theta_W}

\def\mz2{M_{z}^2}
\def\c2b{\cos 2\beta}

\def\mz{M_z}

\def\Fq2{F_{2}(q^2)}
\begin{document}

\begin{titlepage}
\begin{center}
{\Large\bf A Healthy Electron/Neutron EDM in  D3/D7 $\mu$-Split-Like SUSY }
\vskip 0.1in Mansi Dhuria\footnote{email: mansidph@iitr.ac.in}
and
 Aalok Misra\footnote{e-mail: aalokfph@iitr.ac.in
}\\
Department of Physics, Indian Institute of Technology,
Roorkee - 247 667, Uttaranchal, India\\
 \vskip 0.5 true in
\date{\today}
\end{center}
\thispagestyle{empty}
\begin{abstract}

Within the framework of  ${\cal N}=1$ gauged supergravity, using a phenomenological model which can be obtained, locally, as a Swiss-Cheese  Calabi-Yau
string-theoretic compactification with a mobile $D3$-brane localized on a
nearly special Lagrangian three-cycle in the Calabi-Yau  and fluxed stacks of wrapped $D7$-branes, and which provides a natural realization of $\mu$-split SUSY, we show that  in addition to getting a significant value of (electron/neutron)  EDM at two-loop level, one can obtain a sizable contribution of (e/n) EDM even at one-loop level due to presence of heavy supersymmetric fermions nearly isospectral with heavy sfermions. Unlike traditional split SUSY models in which the one-loop diagrams do not give significant contribution to EDM of electron/neutron because of very heavy sfermions existing as propagators in the loop, we show that one obtains a `healthy' value of EDM in our model because of the presence of heavy higgsino, neutralino/chargino and gaugino as fermionic propagators in the loops. The independent CP-violating phases are generated from non-trivial distinct phase factors associated with four Wilson line moduli (identified with first generation leptons and quarks and their $SU(2)_L$-singlet cousins)  as well as the $D3$-brane position moduli (identified with two Higgses) and the same are sufficient to produce overall distinct phase factors corresponding to all possible effective Yukawas as well as effective gauge couplings that we have discussed in the context of ${\cal N}=1$ gauged supergravity action. However, the complex phases responsible to generate non-zero EDM at one-loop level mainly appear from off-diagonal contribution of sfermion as well as Higgs mass matrices at Electro-Weak scale. In our analysis, we obtain dominant contribution of electron/neutron EDM around $d_{e}/e \equiv {\cal O}(10^{-29})cm$ from two-loop diagrams involving heavy sfermions and a light Higgs, and $d_{e}/e \equiv {\cal O}(10^{-32})cm$ from one-loop diagram involving heavy chargino and a light Higgs as propagators in the loop.  The neutron EDM gets a dominant contribution of the order  $d_{n}/e \equiv {\cal O}(10^{-33})cm$ from one-loop diagram involving SM-like quarks and Higgs. To justify the possibility of obtaining a large EDM value in case of a Barr-Zee  diagram which involves $W^\pm$ and the Higgs (responsible to generate the non-trivial CP-violating phase) in the two-loop diagrams as discussed in \cite{Leigh et al}, we provide an analysis of the same in the context of our $D3/D7$ $\mu$-split SUSY model at the EW scale. By conjecturing that the CP-violating phase can appear from the diagonalization of the Higgs mass matrix obtained in the context of $\mu$-split SUSY, we also get an EDM of electron/neutron around ${\cal O}(10^{-27}) ecm$ in case of the two-loop diagram involving $W^\pm$ bosons.

\end{abstract}
\end{titlepage}

\section{Introduction}

 For the past few decades, string theoretic models have been considered to provide an excellent framework for possible unification of gravity with all other fundamental forces. To study the phenomenological implications of these models, the same must invoke a particular SUSY breaking mechanism (along with the SUSY breaking scale). The phenomenological models mainly rely on  ${\cal O}$(TeV) SUSY breaking scale because this helps to solve serious gauge-hierarchy problems which in fact have been considered as a primary motivation to introduce SUSY. However, low scale SUSY models give rise to many unwanted phenomenological problems, such as flavor changing neutral currents.  Motivated by obtaining an extremely small cosmological constant and the string landscape scenario, an alternative to these assumptions was proposed by Arkani-Hamed and Dimopoulos (dubbed as `Split SUSY') in  \cite{HamidSplitSUSY} according to which SUSY is broken at an energy scale way beyond the collider search and could be even near the scale of grand unification (GUT). The scenario is emerging to be quite interesting from the point of view of phenomenology because of the fact that heavy scalars mostly appearing as virtual particles in most of the particle decay studies, help to resolve many diverse issues of both particle physics and cosmology. The $\mu$- split SUSY model was proposed in \cite{mu split susy} to alleviate the famous $\mu$ problem by further splitting the split SUSY by raising $\mu$-paramter to a large value.  Though the exact signatures may not be foreseeable in the near future via precise measurements to be carried out at the Large Hadron Collider (LHC) however indirect methods can be made available to test some of the signatures of this scenario. In this context, the EDM of the electron/neutron serves as another testing ground for split SUSY scenario.  Recently, the ACME collaboration has reported a new experimental upper limit of $|d_e| < 8.7 \times 10^{-29}e$cm which is an order of magnitude improvement in sensitivity as compared to previous limits \cite{comminsetal,comminetal_1,regental,hudson_et_al}. The current experimental limit on neutron EDM \cite{bakeretal, beringer} is $|d_n/e| < 0.29 \times 10^{-25} cm$.

In Standard Model, the CP-odd phases generated through Cabibo-Kobayashi-Masakawa Matrix (CKM) give a theoretical bound on Electric Dipole Moment (EDM) which is far below the experimental limits. However, new CP violating phases can appear in supersymmetric theories model from complex soft SUSY breaking parameters. In addition to this, in string-inspired models, the CP-violating phases are associated with complex Yukawa couplings originating from string compactifications \cite{nath+ibrahim_CPV2,kane+lykken,nath+ibrahim_CPV3, kaneetal2}.  These CP violating phases associated with  complex soft SUSY breaking parameters as well as Yukawa couplings appearing in different supersymmetric models are typically large, i.e. ${\cal O}(1)$, and hence do not satisfy the current experimental bounds on the electron and on the neutron EDM. One hence has to put stringent constraints on the supersymmetric and in particular Supergravity (SUGRA)  models. More specifically,  the limits can be satisfied if  one  considers  (i) unnaturally small  CP violating phases of ${\cal O}({10}^{-2}-10^{-3})$, (ii) multi-TeV superpartners in the model or (iii)  internal cancellations between different supersymmetric contribution to EDM at loop-levels.  The constraints on the CP violating phases in the supersymmetric models have been discussed in \cite{kaneetal, Abeletal,pospelov} and the systematic analysis of EDM up to two loops in the context of MSSM is provided in \cite{Kizukurietal,changeetal,Pilaftsis_1,oliveetal}. In mSUGRA models discussed in literature \cite{nath+ibrahim_CPV4,nath+ibrahim_CPV1,ibrahim+nath, Huand+Wei, Accomondo+dutta,Accomondo+dutta1}, the EDMs bounds have been reconciled with the experimental limits by showing sufficient cancellations among different supersymmetric contributions without taking into account ${\cal O}(>TeV)$ superpartners and any fine-tuning in phase angles. The main difficulty in choosing multi-TeV scalars as an appropriate mechanism to generate EDM is because the same abandons naturalness and also requires severe fine tuning while satisfying radiative EW symmetry breaking.
However, the non-observation of sparticles at LHC may points toward some sort of fine-tuned natural SUSY \cite{Lhall_nsusy,Papucci_nsusy} or the high SUSY scale/split SUSY models \cite{giudice_hsusy,Lhall__hsusy,Arkanihamid_unnatural}. Therefore, it is interesting to probe high scale SUSY models, in particular $\mu$-split SUSY models to explain the EDM within the reach of experimental limits because the same also helps to satisfy radiative EW symmetry breaking condition by choosing natural value of $\mu$, hence alleviating $\mu$-problem.  The  region of parameter space satisfying EDM value of the order of experimental limits has been analysed in \cite{chattopadhyay} in the presence of  ${\cal O}(TeV)$ superpartners in the mSUGRA  model by considering moderate fine-tuning in $\tan{\beta}$.

 Our approach is quite different in that the SUGRA models discussed in literature even in the framework of string compactifications do not rely on high supersymmetry breaking scale. On the other hand, the typical split supersymmetry models used to study the EDM of electron/neutron include heavy sfermions but light gaugino and higgsino \cite{giudice_splitsusy, chang_splitsusy}. We analyse the EDM of electron and neutron in the supergravity limit of local large volume $D3/D7$ type IIB compactifications which provides, to our knowledge, the first realisation of $\mu$-split SUSY scenario (with large gaugino masses). In typical split SUSY models, all possible one-loop contributions to EDM are highly suppressed by the super heavy scalar masses in the loop and leading contributions to the EDM starts at the two-loop level due to presence of SM particles and EW charginos and neutralinos in the loops (for the analysis of two-loop Barr-Zee diagrams in different models, see \cite{giudice_splitsusy, chang_splitsusy, Pilaftsis_2loop,Fukuyama} and references therein).  \emph{However in our model, the gaugino and neutralino/chargino are almost as heavy as neutral scalars except one light Higgs. Based on that, one can not ignore the contribution of one-loop diagrams because of partial compensation of suppression factors appearing from heavy sfermion masses, by heavy fermions (neutralino, chargino and gaugino)' masses}.

 Therefore, in this paper, we perform a quantitative analysis of the neutron and electron EDMs for all possible one-loop as well as two-loop diagrams in the context of large volume $D3/D7$ $\mu$-split supersymmetry.  The non-zero imaginary phases that appear through  mixing between L-handed and R-handed sfermions (sfermions corresponding to left- and right-handed components of fermions) at Electro-Weak scale, play an important role.  In addition to discussing the one-loop diagrams that exhibit non-zero phases through mixing between sfermions, we also take into account the loop diagrams in which a unique phase appears through mixing between two Higgses at Electro-Weak scale. In the large volume $\mu$-split SUSY model of \cite{Dhuria+Misra_mu_Split SUSY,gravitino_DM}, we have already calculated the eigenvalues of the Higgs mass matrix at electroweak scale, which, with some  fine tuning, eventually leads to one light Higgs and one heavy Higgs. In this paper, we append the details of the complex phase associated with off-diagonal components of the Higgs mass matrix too. Because of the presence of a light and a heavy Higgs in our model, one can expect to get a reasonable order of magnitude of EDM of electron/neutron from one-loop diagrams involving Higgs and other SM/supersymmetric particles. The complete analysis has been carried out by including other interesting one-loop diagrams which involve sgoldstino's (identified, locally, with `big' divisor (bulk) volume modulus  in our set-up) as scalar particles in the loop. For two-loop diagrams, we mainly focus on the Barr-Zee diagrams which involve fermion, sfermions and $W^\pm$ as part of an internal loop, and are mediated through $h \gamma$ exchange except one R-parity violating diagram which involves fermion in the internal loop and is mediated through ${\nu}_{L}\gamma$ exchange. For the complete analysis, we also calculate the contribution of rainbow-type two-loop diagrams involving R-parity violating as well as R-parity conserving vertices. For all two-loop diagrams discussed in this paper, the complex effective Yukawa couplings (associated with the $e^{\frac{K}{2}}({\cal D}DW){\bar\chi}\chi$-term in the ${\cal N}=1$ gauged supergravity action of \cite{Wess_Bagger}) are sufficient to produce non-zero complex phases to generate non-zero EDM.

 The plan of the rest of the paper is as follows. In section {{\bf 2}}, we elaborate upon our large volume $D3/D7$ model discussed in \cite{gravitino_DM}. We discuss the details of our phenomenological model in {{\bf 2.1}} and show the same to be realizable, locally, as the large volume limit of  a type IIB Swiss-Cheese Calabi-Yau orientifold  involving a mobile space-time filling $D3$-brane localized at a nearly special Lagrangian three-cycle embedded in the `big' divisor (hence the local nature of the model's realization) and multiple fluxed stacks of  space-time filling $D7$-branes wrapping the same `big' divisor in {{\bf 2.2}}. After providing the geometrical framework of the model in {{\bf 2.2}}, we briefly mention the phenomenological results that describe the possible identification of Wilson line moduli with first generation leptons and quarks as well as their  $SU(2)_L$-singlet cousins, and $D3$-brane position moduli with two Higgses. Thereafter, we briefly summarize the calculation and results corresponding to values of soft SUSY breaking parameters as well as the supersymmetric fermionic masses. In section {{\bf 3}}, we explain the origin of non-zero complex phases obtained in the context of ${\cal N}=1$ gauged supergravity limit of our local $D3/D7$ model. We also argue that phases of effective Yukawa couplings do not change under a renormalization group flow from string scale down to the electroweak scale in our model. In section {{\bf 4}}, we turn towards order-of-magnitude estimates of EDM of electron/neutron for various possible one-loop diagrams. The effective vertices are calculated by considering the ${\cal N}=1$ gauged supergravity action of \cite{Wess_Bagger,Jockers_thesis}. The complex phases, as already explained, can be made to appear through the complex off-diagonal components of sfermion/Higgs mass matrix and complex effective Yukawa couplings appearing in all one-loop diagrams. We assume the phases of both off-diagonal components of scalar mass matrix as well as possible effective Yukawa's to lie in the range $(0, \frac{\pi}{2}]$ in all the calculations. The section has been divided into three subsections. In {{\bf 4.1}}, we give a detailed discussion of one-loop diagrams which involve sfermions as scalar propagators and gauginos, neutralinos and SM-like fermions as fermionic propagators respectively. In {{\bf 4.2}}, the one-loop diagrams are discussed which involve Higgs as scalar propagator and chargino and SM-like fermions as fermionic propagators respectively. Here, the non-zero imaginary phases appear through mixing between two Higgses  at electroweak scale in the Higgs mass matrix. In {\bf 4.3}, we evaluate the contribution of heavy gravitino and sgoldstino multiplet to EDM of electron/neutron. Though the loop diagrams involving the same are divergent, we pick out the finite contributions for the purpose of obtaining an estimate of EDM of electron/neutron in case of heavy gravitino. In {{\bf section 5}}, we consider two-loop Barr-Zee diagrams. The section has been divided into three subsections. In {{\bf 5.1}} - {{\bf 5.2}}, we compute the two-loop diagrams which involve an internal fermion loop and an internal sfermion loop. These diagrams are mediated by ${\gamma h}$ and ${\gamma {\nu_L}}$ exchange. In {{\bf 5.3}}, we carry out an analysis of two-loop diagrams involving W boson loop in our $\mu$-split SUSY model.  In {{\bf 5.4}}, we  discuss two-loop rainbow-type diagrams. Section {\bf 6} has the summary of our results and discussion.  There is one small {{\bf appendix A}}: in this, we evaluate the chargino mass matrix using ${\cal  N}=1$ gauged supergravity action in the context of large volume $D3/D7$ $\mu$-split SUSY set-up.

\section{The Setup}

In \cite{gravitino_DM}, within the context of type IIB string theory with a space-time filling $D3$-brane and fluxed stacks of $D7$-branes wrapping a divisor along with ED3/ED1-instanton generated super potential and world-sheet instanton-corrected K\"{a}hler potential, we worked locally, close to a nearly special Lagrangian three-cycle (\ref{sLag}) within a Swiss-Cheese type Calabi-Yau orientifold - various aspects of this setup will be summarised in {\bf 2.2} in this section. But before we do the same, we will first briefly describe in {\bf 2.1}, a model that could be locally realized as a large volume $D3/D7$ Swiss-Cheese setup of \cite{gravitino_DM}. In other words, {\bf 2.1} embeds the local model of \cite{gravitino_DM} into a phenomenological model, something which was not done in \cite{gravitino_DM}. Another way to put the same thing is that the phonomenological supergravity model discussed in {\bf 2.1} can be locally geometrically engineered via the construct of \cite{gravitino_DM}.

\subsection{The Model}

 For an ${\cal N}=1$ compactification, we will take the phenomenological K\"{a}hler  potential of our model to be:
\begin{eqnarray}
\label{K_pheno}
& & \hskip -0.5in K_{\rm Pheno} = - ln\left[-i(\tau-{\bar\tau})\right]
-ln\left(-i\int_{CY_3}\Omega\wedge{\bar\Omega}\right)\nonumber\\
& & \hskip -0.5in - 2\ ln\Biggl[a_B(\sigma_B + {\bar\sigma}_B - \gamma K_{\rm geom})^{\frac{3}{2}} - (\sum_{i}a_{S,i}(\sigma_{S,i} + {\bar\sigma}_{S,i} - \gamma K_{\rm geom}))^{\frac{3}{2}} + {\cal O}(1) {\cal V}\Biggr]
 \end{eqnarray}
where the divisor volumes $\sigma_\alpha$ are expressible in terms of ``K\"{a}hler" coordinates $T_\alpha, {\cal M}_{\cal I}$
\begin{eqnarray}
\label{sigma}
& & \sigma_\alpha\sim T_\alpha -\left[ i{\cal K}_{\alpha bc}c^b{\cal B}^c
+ i C^{{\cal M}_{\cal I}{\bar{\cal M}}_{\bar {\cal J}}}_\alpha({\cal V})Tr\left({\cal M}_{\cal I}{\cal M}^\dagger_{\bar {\cal J}}\right)\right],
\end{eqnarray}
$\alpha=\left(B,\{S,i\}\right)$ and ${\cal M}_{\cal I}$ being
  $SU(3_c)\times SU(2)_L$  bifundamental  matter field $a_{{\cal I}=2}$, $SU(3_c)\times U(1)_R$ bifundamental matter field
$a_{{\cal I}=4}$, $SU(2)_L\times U(1)_L$ bifundamental matter field $a_{{\cal I}=1}$,
$U(1)_L\times U(1)_R$ bifundamental matter field $a_{{\cal I}=3}$ along with
$SU(2)_L\times U(1)_L$ bifundamental $\tilde{z}_{1,2}$
 with the intersection matrix:
$ C^{a_I{\bar a}_{\bar J}}_\alpha\sim \delta^B_\alpha C^{I{\bar J}}_\alpha, C^{a_I\bar {\tilde{z}}_{\bar j}}_\alpha=0$, $\rho_{S,B}, {\cal G}^a = c^a - \tau b^a$ being complex axionic
fields ($\alpha,a$ running over the real dimensionality of a sub-space of the internal manifold's cohomology complex),  and the phenomenological  superpotential is given as under:
\begin{equation}
\label{W_pheno}
 W_{\rm Pheno}\sim\left(z_1^{18} + z_2^{18}\right)^{n^s}
e^{-n^s vol(\Sigma_S) - (\alpha_S z_1^2 + \beta_S z_2^2 + \gamma_S z_1z_2)},
\end{equation}
\noindent where the bi-fundamental $\tilde{z}_i$ in $K$ will be equivalent to the $z_{1,2}\in\mathbb C$ in $W$. It is
expected that ${\cal M}_{\cal I}, T_{S,B}, {\cal G}^a$ will constitute the ${\cal N}=1$ chiral coordinates. The
 intersection matrix elements $\kappa_{S/B ab}$ and the volume-dependent $C^{{\cal M}_{\cal I}{\bar{\cal M}}_{\bar {\cal  J}}}_\alpha({\cal V})$, are chosen in such a way that at a local (meta-stable) minimum:
 \begin{eqnarray}
 \label{extremum_i}
& &  \langle \sigma_S\rangle \sim \langle (T_S + {\bar T}_S)\rangle  - i C^{\tilde{z}_i\bar {\tilde{z}}_{\bar j}}({\cal V}) Tr\left(\langle\tilde{z}_i\rangle \langle \bar {\tilde{z}}_{\bar j}\rangle\right)\sim {\cal O}(1)
\nonumber\\
& &  \langle \sigma_B\rangle \sim \langle (T_B + {\bar T}_B)\rangle  - i C^{\tilde{z}_i\bar {\tilde{z}}_{\bar j}}({\cal V}) Tr\left(\langle\tilde{z}_i\rangle \langle\bar {\tilde{z}}_{\bar j}\rangle\right)
- i C^{a_I{\bar a}_{\bar J}}({\cal V}) Tr\left(\langle a_I\rangle \langle{\bar a}_{\bar J}\rangle\right)\sim e^{f \langle \sigma_S\rangle},\nonumber\\
& &
 \end{eqnarray}
 where $f$ is a fraction not too small as compared to 1, and the stabilized values of $T_\alpha$ around the meta-stable local minimum:
 \begin{equation}
 \label{extremum_ii}
 \langle\Re e T_S\rangle,\langle\Re e T_B\rangle\sim {\cal O}(1).
 \end{equation}
 In the context of ${\cal N}=1$ type IIB orientifolds, in (\ref{extremum_i}), $\alpha,a$ index respectively involutively even, odd sectors of $h^{1,1}(CY_3)$ under a holomorphic, isometric involution.
If the volume ${\cal V}$ of the internal manifold is large in string length units, one sees that one obtains a hierarchy between the stabilized values $\langle\Re e\tau_{S,B}\rangle$ but not $\langle\Re e T_{S,B}\rangle$.

\subsection{Local Realisation of the Model of {\bf 2.1} }

We {\it review}  the local $D3$-$D7$ brane framework presented in \cite{gravitino_DM} which  realizes  the  aforementioned phenomenological supergravity model [(\ref{K_pheno}) - (\ref{extremum_ii})], locally, in string theory. In this, we consider type IIB compactified on the orientifold of a Swiss-Cheese Calabi-Yau in the L(arge) V(olume) S(cenarios) limit that includes non-(perturbative) $\alpha^{\prime}$ corrections and non-perturbative instanton-corrections in superpotential \cite{dSetal} in addition to a space-time filling $D3$-brane and multiple fluxed stacks of $D7$-branes wrapping the `big' divisor.  We elaborate a little more than what was done in \cite{gravitino_DM} on some algebraic geometric aspects.

The `bottom-up' approach to  phenomenological models in the context of $D$-brane models to realize SM spectrum was initiated in \cite{Aldazabal}  by considering $D3$-branes on the top of orbifold singularities of $\mathbb C^3/ {\mathbb Z_{3}}$ with additional intersecting $D7$-branes (with their world volumes transverse to respective complex planes). In this model, quarks and one of the Higgs doublets are obtained from strings stretching between different $D3$-branes while the other Higgs doublet, leptons and  right handed quark ($d_R$) are obtained from strings stretching between $D3$ and $D7$-branes; the adjoint gauge fields correspond to open strings starting and ending on the same $D7$-brane. Motivated by this approach,  different models were constructed in the context of compact Calabi-Yau compactifications by following configurations of intersecting $D7$-branes wrapping different four-cycles (see \cite{ibanez_ints, blumenhagen_ints, Marchesano+shiu,watari1,watari2,shiu1} and references therein).  With the progress of  large volume moduli stabilisation \cite{lvs}, realistic constructions reproducing SM spectrum via $D$-branes were obtained by wrapping of $D7$-branes around blown-up cycle(s) \cite{conlonetal} (small divisor $\Sigma_s$ in the geometry of Swiss-cheese Calabi-Yau orientifold), similar to the techniques used in models of branes at singularities.

 The configuration of $D3$-$D7$ branes as described in \cite{gravitino_DM} was also obtained, locally, in the context of large volume scenarios. However the setup of \cite{gravitino_DM}, is  different from the aforementioned large volume scenarios constructs because: (i)  it considers four stacks  of multiple (magnetized) $D7$-branes in groups of 3(corresponding to $U(1)\times SU(2)_c$), 2 (corresponding to $U(1)\times SU(2)_L$),1 (corresponding to a $U(1)$) and 1(corresponding to another $U(1)$) with the hypercharge corresponding to a linear combination of the four $U(1)$s, wrapping around the `big' divisor  in the rigid limit of the same  (given that it was possible to locally stabilize the moduli corresponding to the fluctuations normal to the `big divisor' $\Sigma_B$ around which $D7$-branes are wrapped, at null values)  but with different choices of
two-form fluxes turned on the different two-cycles homologously non-trivial from the point of view of this four-cycle's
Homology and not the ambient Swiss-cheese Calabi-Yau; (ii) it takes into account the non-perturbative corrections in the K{\" a}hler potential \cite{dSetal} in  type IIB Swiss-cheese Calabi-Yau orientifold  compactification, not considered in the `Large Volume Scenario' proposed in \cite{lvs}.

Further, similar in spirit to \cite{softsusy2} - \cite{intbranes-luest}, by turning on different but small two-form fluxes on the different two-cycles homologously non-trivial from the point of view of the `big' divisor's geometry as a result of which initially adjoint-valued matter fields decompose into bi-fundamental matter fields corresponding to the SM gauge groups, we provided explicit matrix-valued representations in \cite{gravitino_DM} for $SU(3)_c \times SU(2)_L$ bifundamental first-generation quarks, their right-handed EW singlet cousins, $SU(2)_L \times U(1)_L$ bifundamental first-generation leptons and Higgs, as well as the the right-handed EW-singlet leptonic cousins in \cite{gravitino_DM}. All aforementioned matter fields arise from strings stretched between $D7$-branes stacks with different two-form fluxes turned on.  The leptons and quarks get identified with the sermonic super-partners of Wilson line moduli ${\cal A}^I$ and the Higgs  with the $D3$-brane's position moduli $z_i$; $\tau$ is the axion-dilaton modulus and ${\cal G}^a$ are NS-NS and RR two-form axions complexified by the axion-dilaton modulus. In the orientifold-limit of F-theory, one considers an orientifold of the Calabi-Yau involving a holomorphic isometric involution. Though the contribution to the K\"{a}hler potential from the matter fields  `$C_{37}$' coming from open strings stretched between the $D3$- and $D7$-branes wrapping $\Sigma_B$ for Calabi-Yau orientifolds is not known, but based on results for orientifolds of $(T^2)^3$ - see \cite{d3-d7} - we guess the following expression: $\frac{|C_{37}|^2}{T_B}\sim {\cal V}^{-\frac{1}{18}} |C_{37}|^2$ [using (\ref{moduli-stabilized-locally})]. Assuming $C_{37}$ to be stabilized at ${\cal V}^{-c_{37}}, c_{37} > 0$) this contribution would be sub-dominant relative to other contributions to the K\"{a}hler potential. We will henceforth ignore $D3-D7$-matter fields.

We will assume that in the coordinate patch
(but not globally):
$
|z_1|\sim {\cal V}^{\frac{1}{36}},\ |z_2|\sim{\cal V}^{\frac{1}{36}},\ |z_3|\sim{\cal V}^{\frac{1}{6}},$
 the Calabi-Yau looks like the Swiss-Cheese $\mathbb {WCP}^4_{1,1,1,6,9}[18]$. The
defining hypersurface for the same is: $u_1^{18}+u_2^{18}+u_3^{18}+u_4^3+u_5^2
-18\psi\prod_{i=1}^5u_i-3\phi (u_1u_2u_3)^6=0$. This can be thought of as the following
hypersurface in an ambient complex four-fold: $P(x_1,...,x_5;\xi)=0$ after resolution
of the $\mathbb Z_3$-singularity \cite{Rummel_et_al} [the $x_4$ and $x_5$ have been switched
relative to \cite{DDF}; $n=6$ $\mathbb {CP}^1$-fibration over $\mathbb{CP}^2$ with
projective coordinates $x_{1,2,3},x_4,x_5$ of \cite{DDF} is equivalent to $n=-6$ with projective coordinates
$x_{1,2,3,},x_5,x_4$ - see \cite{Denef_Les_Houches}] with the toric data for the same given by:
$$\begin{array}{c|cccccc}
& x_1 & x_2 & x_3 & x_4 & x_5 & \xi \\ \hline
Q^1 & 1 & 1 & 1 & 6 & 0 & 9 \\
Q^2 & 0 & 0 & 0 & 1 & 1 & 2 \\
\end{array}.$$
In the coordinate patch: $x_2\neq0$(implying one is away
from the $\mathbb Z_3$-singular $(0,0,0,x_4,x_5)$ in
$\mathbb{WCP}^{4}_{1,1,1,6,9}[18]$), $\xi\neq0$, one sees that the following are
the gauge-invariant coordinates: $z_1=\frac{x_1}{x_2}, z_2=\frac{x_3}{x_2},
z_3=\frac{x_4^2}{x_2^3\xi}, z_4=\frac{x_5^2x_2^9}{\xi}.$
We henceforth assume
the Calabi-Yau hypersurface to be written in this coordinate patch as:
$z_1^{18} + z_2^{18} + {\cal P}(z_{1,2,3,4};\psi,\phi)=0$.
The divisor $\left\{x_5=0\right\}\cap\left\{P(x_{1,2,3,4,5};\xi)=0\right\}$
is rigid with $h^{0,0}=1$  (See \cite{Rummel_et_al})  satisfying the Witten's
unit-arithmetic genus condition, and that the Calabi-Yau volume can
be written as
${\rm vol}(CY_3)=\frac{\tau_4^{\frac{3}{2}}}{18} -
\frac{\sqrt{2}\tau_5^{\frac{3}{2}}}{9},$  implying that the `small divisor'
$\Sigma_s$ is $\left\{x_5=0\right\}\cap\left\{z_1^{18} + z_2^{18} + {\cal P}(z_{1,2,3},z_4=0;\psi,\phi)=0\right\}$
and the `big' divisor $\Sigma_B$ is  $\left\{x_4=0\right\}\cap\left\{z_1^{18} + z_2^{18} + {\cal P}(z_{1,2,4},z_3=0;\psi,\phi)=0\right\}.$ Alternatively, using
the toric data of \cite{candelas_et_al_11169}:
$$\begin{array}{c|cccccc}
& x_1 & x_2 & x_3 & x_4 & x_5 & \xi \\ \hline
Q^1 & 1 & 1 & 1 & 0 & 0 & -3 \\
Q^2 & 0 & 0 & 0 & -2 & -3 & -1 \\
\end{array},$$ one can verify that
$\left\{\xi=0\right\}\cap\left\{P^\prime(x_{1,2,3,4,5};\xi)=0
\right\}$ is the rigid blow-up mode with $h^{0,0}=1$ (which can be easily
verified using cohomCalg\footnote{We thank P.Shukla for verifying the same.})
and one can define gauge-invariant coordinates in the $x_2\neq0,x_4\neq0$
coordinate-patch: $z_1=\frac{x_1}{x_2}, z_2=\frac{x_3}{x_2},
z_3=\frac{(x_5x_1)^2}{x_4^3}, z_4=\frac{(x_6x_1^3)^2}{x_4}.$
Interestingly, we found in \cite{gravitino_DM} that the  three-cycle
\begin{equation}
\label{sLag}
C_3:|z_1|\equiv {\cal V}^{\frac{1}{36}},\ |z_2|\equiv{\cal V}^{\frac{1}{36}},\ |z_3|\equiv{\cal V}^{\frac{1}{6}}
\end{equation}
 (the Calabi-Yau can be thought of,
locally, as a complex three-fold ${\cal M}_3$ which is a $T^3$(swept out by ($arg z_1$,$arg z_2$, $arg z_3$)-fibration over a large base $(|z_1|,|z_2|,|z_3|)$; precisely apt for application of mirror symmetry as three T-dualities a la S(trominger) Y(au) Z(aslow)), $C_3$ is almost a s(pecial) Lag(rangian) sub-manifold because it satisfies the requirement that  $$f^*J\approx0,\ \Re e\left(f^*e^{i\theta}\Omega\right)\biggr|_{\theta=\frac{\pi}{2}}\approx {\rm vol}(C_3),\ \Im m\left(f^* e^{i\theta}\Omega\right)\biggr|_{\theta=\frac{\pi}{2}}\approx0$$ where $f:C_3\rightarrow CY_3$). As the defining hypersurface of the Swiss-Cheese Calabi-Yau in $x_2\neq0$-coordinate patch
will be $z_1^{18} + z_2^{18} + ...$ which near $C_3$ (implying that the other two coordinates will scale like
${\cal V}^{\frac{1}{6}}, {\cal V}^{\frac{1}{6}} - {\cal V}^{\frac{1}{4}}$) receives the most dominant contributions from the monomials $z_1^{18}$ and
$z_2^{18}$ it is sufficient to consider ${\cal P}_{\Sigma_S}\Bigr|_{D3|_{{\rm near}\ C_3\hookrightarrow\Sigma_B}},{\cal P}_{\Sigma_B}|_{{\rm near}\ C_3\hookrightarrow\Sigma_B}\sim z_1^{18} + z_2^{18}$ with the understanding $|{\cal P}(z_{1,2,3},z_4=0;\phi,\psi)|_{C_3},|{\cal P}(z_{1,2,4},z_3;\phi,\psi)|_{C_3}<|z_1^{18}+z_2^{18}|$.

The set of  ${\cal N}=1$ chiral co-ordinates (in particular the `divisor volume') gets modified in the presence of  $D3$- and $D7$-branes \cite{Jockers_thesis}. To evaluate the Wilson line moduli contribution in one of the ${\cal N}=1$ chiral coordinates $T_B$, due to inclusion of four Wilson line moduli on the world volume of space-time filling $D7$-branes wrapped around the `big' divisor restricted to (nearly) a special Lagrangian sub-manifold, we constructed distribution harmonic one-forms localized along the mobile space-time filling $D3$-brane (restricted to the 3-cycle). Here,  we review the construction of involutively odd harmonic distribution one-forms in the large volume limit, as given in \cite{gravitino_DM}. [The most non-trivial example of involutions which are meaningful only at large volumes is mirror symmetry implemented as three T-dualities in \cite{SYZ} to a Calabi-Yau which locally can be thought of
as a $T^3$-fibration over a (large) base; all Calabi-Yau's with mirrors (in the conventional
sense) are expected to have such a local fibration.] Harmonic distribution one-forms can be constructed by integrating:
$dA_I=\left(P_{\Sigma_B}(z_{1,2})\right)^Idz_1\wedge dz_2 {\rm  with}~ (I=1,2,3,4),$
near $C_3\hookrightarrow\Sigma_B$; $A_I$ is harmonic only within $\Sigma_B$ and not at any other generic locus outside $\Sigma_B$ in the Calabi-Yau manifold. Four such
distribution one-forms on $\Sigma_B$ localized along $C_3$ corresponding to the
location of the $D3$-brane can be written as:
$A_I\sim\delta\left(|z_1|-{\cal V}^{\frac{1}{36}}\right)  \delta\left(|z_2|-{\cal V}^{\frac{1}{36}}\right)\left[\omega_I(z_1,z_2)dz_1 + \tilde{\omega}_I(z_1,z_2)dz_2\right].$
Writing $A_I(z_1,z_2)=\omega_I(z_1,z_2)dz_1+\tilde{\omega}_I(z_1,z_2)dz_2$
\footnote{Intuitively, these distribution one-forms could be thought of as
the holomorphic square-root of a Poincare dual of a four-cycle.} where $\omega(-z_1,z_2)=\omega(z_1,z_2), \tilde{\omega}(-z_1,z_2)=-\tilde{\omega}(z_1,z_2)$ and $\partial_1\tilde{\omega}=-\partial_2\omega$, one obtains (See
\cite{gravitino_DM}):
\begin{eqnarray}
\label{eq:A_1234}
& & A_1|_{C_3}\sim - z_1^{18}z_2^{19}dz_1 + z_1^{19}z_2^{18}dz_2,  A_2|_{C_3}\sim - z_1^{18}z_2dz_1 + z_2^{18}z_1 dz_2,\nonumber\\
& & A_3|_{C_3}\sim -z_1^{18}z_2^{37}dz_1 -z_2^{18}z_1^{37}dz_1 ,  A_4|_{C_3}\sim -z_1^{36}z_2^{37}dz_1 + z_2^{36}z_1^{37}dz_2.
\end{eqnarray}
\vskip 0.3in
\noindent{\bf Yang-Mills coupling constant:}  We now summarise the discussion on obtaining an ${\cal O}(1)$ gauge coupling constant. The Yang-Mills gauge coupling constant squared for the $i$-th gauge group ($i:SU(3), SU(2), U(1))$ will be given as:
\begin{equation}
\frac{1}{g^{2}_{j=SU(3) {\rm or} SU(2)}}=\Re e\left(T_{S/B}\right)+ ln\left(\left.P(\Sigma_S)\right|_{D3|_{\Sigma_B}}\right) + ln\left(\left.{\bar P}(\Sigma_S)\right|_{D3_{\Sigma_B}}\right) + {\cal O} (F^{2}_{j})\tau
\end{equation}
where $Re(T_{S/B})$ corresponds to size of divisor volume around which $D7$-branes are wrapped and $F^{2}_{j}=F^{\alpha}_jF^{\beta}_{j} {\kappa}_{\alpha\beta}+{\tilde F}^{\alpha}_j{\tilde F}^{\beta}_{j} {\kappa}_{\alpha\beta}$ are the components of the two-form fluxes for the ${\rm j^{th}}$-stack expanded out in the basis of $i^{*}w_{\alpha}$, $w_{\alpha} \in H^{1,1}_{-}(CY_{3})$, and ${\tilde F}^{a}_{j}$ are the components of  two-form fluxes for the jth-stack expanded out in the basis ${\tilde w}_{a}\in coker\left(H^{(1,2)}_{-}(CY_3)\rightarrow H^{(1,1)}_{-}(\sigma_{B})\right)$. In dilute flux approximation, $g_{YM}$ is mainly governed by size of divisor volume around which $D7$-branes are wrapped. Using the distribution one-forms of (\ref{eq:A_1234}), the ${\cal N}=1$ chiral co-ordinates  with the inclusion of mobile $D3$-brane position moduli $z_{1,2}$ (which we identify with the $\Sigma_B$ coordinates) and mulitple matrix-valued $D7$-branes Wilson line moduli
${a_I}$ were guessed in  \cite{gravitino_DM}. The quadratic contribution arising in $T_B$ (the `big' divisor) due to Wilson line moduli contribution is of the form: $i\kappa_4^2\mu_7 C_{I\bar{J}}^{B}a^I{\bar a}^{\bar J}$ with $C_{I\bar{J}}^{B}=\int_{\Sigma_B}i^*\omega\wedge A^I\wedge{\bar A}^{\bar J}$, where $\omega\in H^{(1,1)}_+(\Sigma_B)$. In
\cite{gravitino_DM}, we estimated the intersection matrices $C_{I\bar{J}}^{B}$ by
 constructing harmonic one forms using equation (\ref{eq:A_1234}). Also, coefficient
of quadratic term  $ \left(\omega_\alpha\right)_{i{\bar j}}z^i\left({\bar z}^{\bar j}
- \frac{i}{2}\left({\cal P}_{\tilde{a}}\right)^{\bar j}_{\ l}{\bar z}^{\tilde{a}}z^l\right)$
arising in $T_B$ due to inclusion of position moduli $z_i$ was shown in \cite{gravitino_DM} to be
${\cal O}(1)$ by calculating $(\omega_B)_{i{\bar j}}\sim(\omega_S)_{i{\bar j}}\sim{\cal O}(1)$ near $z_{1,2}\sim\frac{{\cal V}^{\frac{1}{36}}}{\sqrt{2}}$ (See \cite{gravitino_DM}).  Using the same, it was argued that, in the dilute flux approximation, gauge couplings corresponding to the gauge theories living on stacks of $D7$ branes wrapping the `big' divisor $\Sigma_B$ in the large volume limit,  will given by:
$$g_{YM}^{-2}\sim\Re e(T_B)\sim Vol(\Sigma_B)+C_{I{\bar J}}a_I{\bar a}_{\bar J} + h.c.\sim {\cal V}^{\frac{1}{18}}\sim O(1).$$(justified by the partial cancelation between between $\Sigma_B$ and $C_{I{\bar J}}a_I{\bar a}_{\bar J}$ with some fine-tuning).

\vskip 0.3in
\noindent {\bf Stabilized potential of ${\cal N}=1$ local Large Volume D3-D7 set-up:} As we do not have a global picture, we content ourselves with a local bulk and open-string moduli stabilization near (\ref{sLag}). We showed in \cite{gravitino_DM} that near (\ref{sLag}), the moduli can be stabilised as under:
\begin{eqnarray}
\label{moduli-stabilized-locally}
& & {\rm vol}(\Sigma_S)\sim {\cal V}^{\frac{1}{18}},\ {\rm vol}(\Sigma_B)\sim {\cal V}^{\frac{2}{3}};\ {\cal G}^a\sim\frac{\pi}{{\cal O}(1)k^a(\sim{\cal O}(10))}M_P;\nonumber\\
& & |z_{1,2}|\equiv{\cal V}^{\frac{1}{36}}M_P,|z_3|\equiv{\cal V}^{\frac{1}{6}}M_P;\  |a_1|\equiv{\cal V}^{-\frac{2}{9}}M_P,
|a_2|\equiv{\cal V}^{-\frac{1}{3}}M_P, |a_3|\equiv{\cal V}^{-\frac{13}{18}}M_P,|a_4|\equiv{\cal V}^{-\frac{11}{9}}M_P;\nonumber\\
& & \zeta^{A=1,...,h^{0,2}_-(\Sigma_B|_{C_3})}\equiv0 ({\rm implying~rigidity~ of~the~non-rigid~\Sigma_B})
\end{eqnarray}
such that $\Re e T_S\sim\Re e T_B\sim{\cal V}^{\frac{1}{18}}$ and implying the possibility of obtaining a local meta-stable dS-like minimum corresponding to the positive minimum of the potential $e^KG^{T_S{\bar T}_S}|D_{T_S}W|^2$ near (\ref{moduli-stabilized-locally}), and realising (\ref{extremum_ii}) and thereby the supergravity model of {\bf 2.1} for ${\cal V}\sim 10^5$ in $l_s=1$ units.

The K\"{a}hler potential relevant to all the calculations (using modified ${\cal N}=1$ chiral co-ordinates) in this paper (without being careful about ${\cal O}(1)$ constant factors) is given as under \cite{gravitino_DM}:
 \begin{eqnarray}
\label{eq:K}
& &  K\sim-2 ln\left(a_B\Biggl[\frac{T_B+{\bar T}_B}{M_P} - \mu_3(2\pi\alpha^\prime)^2
\frac{\left\{|z_1|^2 + |z_2|^2 + z_1{\bar z}_2 + z_2{\bar z}_1\right\}}{M_P^2}
 +{\cal V}^{\frac{10}{9}}\frac{|a_1|^2}{M_P^2}+{\cal V}^{\frac{11}{18}}
 \frac{\left(a_1{\bar a}_2+h.c.\right)}{M_P^2}\right.\nonumber\\
 & & +{\cal V}^{\frac{1}{9}}\frac{|a_2|^2}{M_P^2} + {\cal V}^{\frac{29}{18}}
 \frac{\left(a_1{\bar a}_3+h.c.\right)}{M_P^2}+ {\cal V}^{\frac{10}{9}}
 \frac{\left(a_2{\bar a}_3+h.c.\right)}{M_P^2} + {\cal V}^{\frac{19}{9}}\frac{|a_3|^2}{M_P^2} + {\cal V}^{\frac{19}{9}}\frac{\left(a_1{\bar a}_4 + a_4{\bar a}_1\right)}{M_P^2} + \nonumber \\
 & &   {\cal V}^{\frac{29}{18}}\frac{\left(a_2{\bar a}_4 + a_4{\bar a}_2\right)}{M_P^2}+ {\cal V}^{\frac{47}{18}}\frac{\left(a_3{\bar a}_4 + a_4{\bar a}_3\right)}{M_P^2} + {\cal V}^{\frac{28}{9}}\frac{|a_4|^2)}{M_P^2}
\Biggr]^{3/2}- \nonumber\\
& & \left. a_S\left(\frac{{T_S+{\bar T}_S}}{M_P}-\mu_3(2\pi\alpha^\prime)^2 \frac{\left\{|z_1|^2 + |z_2|^2 + z_1{\bar z}_2 + z_2{\bar z}_1\right\}}{M_P^2}\right)^{3/2} +\sum n^0_\beta(...)\right)
\end{eqnarray}
\noindent and ED3/ED1 generated non-perturbative superpotential used in \cite{gravitino_DM} is given by:
\vskip -0.3in
\begin{eqnarray}
\label{eq:W}
\begin{array}{r}
  W\sim\left({\cal P}_{\Sigma_S}\Bigr|_{D3|_{{\rm near}\ C_3\hookrightarrow\Sigma_B}}\sim z_1^{18} + z_2^{18}\right)^{n^s}\sum_{m_a}e^{i\tau\frac{m^2}{2}}e^{in^s G^am_a} e^{i n^s T_s},
\end{array}
\end{eqnarray}
\vskip -0.09in
\noindent which is like (\ref{W_pheno}) assuming $G^a, \tau$ has been stabilized. The  genus-zero Gopakumar-Vafa invariants (which for projective varieties are very large) prefix the $h^{1,1}_-$-valued real axions $b^a,c^a$. In general there are no known globally defined involutions, valid for all Calabi-Yau volumes, for which $h^{1,1}_-(CY_3)\neq0, h^{0,1}_-(\Sigma_B
)\neq0$.
However, as mentioned earlier, in the spirit of the involutive mirror symmetry
implemented a la SYZ prescription in terms of a triple of T dualities along a
local $T^3$ in the large volume limit, we argued in \cite{Misra+Shukla_invlargeV},
e.g., $z_1\rightarrow-z_1$ would, restricted to $C_3$, generate non-zero
$h^{1,1}_-\left(\begin{array}{c} T^3(arg z_{1,2,3})\rightarrow{\cal M}_3(z_{1,2,3})\\
\hskip1in\downarrow \\
\hskip 0.3in M_3(|z_1|,|z_2|,|z_3|) \end{array}  \right)$. An example of holomorphic involutions near $C_3$  not requiring a
large Calabi-Yau volume has been discussed in \cite{gravitino_DM}. However, even if $h^{1,1}_-=0$, one can self-consistently stabilize $c^a,b^a$ to zero and $\sigma_s,\sigma_b$ to ${\cal V}^{\frac{1}{18}}, {\cal V}^{\frac{2}{3}}$ such that the K\"{a}hler potential continues to be stabilized at $- 2 ln {\cal V}$.

 The evaluation of ``physical"/normalized Yukawa couplings, soft SUSY breaking parameters and various 3-point vertices needs the matrix generated from the mixed double derivative of the K\"{a}hler potential to be a diagonalized matrix. After diagonalization  the corresponding eigenvectors of the same were found in \cite{gravitino_DM} to be given by:
\begin{eqnarray*}
\label{eq:eigenvectors}
\begin{array}{l}
  {\cal A}_4\sim a_4 + {\cal V}^{-\frac{3}{5}}a_3 + {\cal V}^{-\frac{6}{5}} a_1+{\cal V}^{-\frac{9}{5}} a_2 + {\cal V}^{-2}\left(z_1+z_2\right) ;\\
{\cal A}_3\sim -a_3 + {\cal V}^{-\frac{3}{5}}a_4 - {\cal V}^{-\frac{3}{5}}a_1 - {\cal V}^{-\frac{7}{5}}a_2 + {\cal V}^{-\frac{8}{5}}\left(z_1+z_2\right);\\
  {\cal A}_1\sim a_1 - {\cal V}^{-\frac{3}{5}}a_3 + {\cal V}^{-1}a_2 - {\cal V}^{-\frac{6}{5}}a_4+ {\cal V}^{-\frac{6}{5}}\left(z_1+z_2\right);\\
{\cal A}_2\sim - a_2 - {\cal V}^{-1}a_1 + {\cal V}^{-\frac{7}{5}}a_3 - {\cal V}^{-\frac{3}{5}}\left(z_1+z_2\right);\\
{\cal Z}_2\sim - \frac{\left(z_1+z_2\right)}{\sqrt 2}  - {\cal V}^{-\frac{6}{5}}a_1 + {\cal V}^{-\frac{3}{5}}a_2 + {\cal V}^{-\frac{8}{5}}a_3+ {\cal V}^{-2}a_4;\\
{\cal Z}_1\sim \frac{\left(z_1-z_2\right)}{\sqrt 2}  - {\cal V}^{-\frac{6}{5}}a_1 + {\cal V}^{-\frac{3}{5}}a_2 + {\cal V}^{-\frac{8}{5}}a_3+ {\cal V}^{-2}a_4.
\end{array}
\end{eqnarray*}
For  ${\cal V}=10^5$, the numerical eigenvalues are estimated to be:
\begin{eqnarray}
\label{eq:eigenvals}
 && K_{{\cal Z}_1{\cal Z}_1} \sim10^{-5},K_{{\cal Z}_2{\cal Z}_2}\sim 10^{-3}, K_{{\cal A}_1{\cal A}_1}\sim 10^4,  K_{{\cal A}_2{\cal A}_2}\sim10^{-2}, K_{{\cal A}_3{\cal A}_3}\sim 10^7, K_{{\cal A}_4{\cal A}_4}\sim 10^{12}.
\end{eqnarray}
\vskip 0.3in
\noindent{\bf Mass scales of SM-like particles:}
The effective Yukawa couplings can be calculated using $\hat{Y}^{\rm eff}_{C_i C_j C_k}\equiv\frac{e^{\frac{K}{2}}Y^{\rm eff}_{C_i C_j C_k}}{\sqrt{K_{C_i{\bar C}_i}K_{C_j{\bar C}_j}K_{C_k{\bar C}_k}}}, C_i$ being an open string modulus which for us is $\delta{\cal Z}_{1,2},\delta{\cal A}_{1,2,3,4}$. where $Y^{\rm eff}_{{\cal Z}_i{\cal A}_I{\cal A}_J}$ is given by  ${\cal O}({\cal Z}_i)$-coefficient  in the mass term $e^{\frac{K}{2}}{\cal D}_{{\bar{\cal A}}_I}D_{{\bar{\cal A}}_J}{\bar W}{\bar\chi}^{{\cal A}_I}\chi^{{\cal A}_J}$ in the ${\cal N}=1$ SUGRA action of \cite{Wess_Bagger}. By  estimating in the large volume limit, all possible Yukawa couplings corresponding to four Wilson line moduli and showing that the RG-flow of the effective physical Yukawa's change almost by ${\cal O}(1)$ under an RG flow from the string scale down to the EW scale \cite{gravitino_DM}, we see that for ${\cal V}\sim 10^5,\langle {\cal Z}_i\rangle\sim 246 GeV$:
$$\frac{{\cal O}({\cal Z}_i)\ {\rm term\ in}\
e^{\frac{K}{2}}{\cal D}_{{\cal A}_1}D_{{\cal A}_3}W}{\sqrt{K_{{\cal Z}_i\bar{\cal Z}_i}K_{{\cal A}_1\bar{\cal A}_1}K_{{\cal A}_3\bar{\cal A}_3}}}\equiv
\hat{Y}^{\rm eff}_{{\cal Z}_i{\cal A}_1{\cal A}_3}\sim 10^{-3}\times {\cal V}^{-\frac{4}{9}},
$$
giving $\langle {\cal Z}_i\rangle \hat{Y}_{{\cal Z}_1{\cal A}_1{\cal A}_3}\sim MeV$ - about the mass of the electron;
$$\frac{{\cal O}({\cal Z}_i)\ {\rm term\ in}\
e^{\frac{K}{2}}{\cal D}_{{\cal A}_2}D_{{\cal A}_4}W}{\sqrt{K_{{\cal Z}_i\bar{\cal Z}_i}K_{{\cal A}_2\bar{\cal A}_2}K_{{\cal A}_4\bar{\cal A}_4}}}\equiv
\hat{Y}^{\rm eff}_{{\cal Z}_i{\cal A}_2{\cal A}_4}\sim 10^{-\frac{5}{2}}\times {\cal V}^{-\frac{4}{9}},$$
giving $\langle z_i\rangle \hat{Y}_{{\cal Z}_i{\cal A}_2{\cal A}_4}\sim10MeV$ - close to the mass of the up quark!
The above shows that fermionic superpartners of ${\cal A}_1$ and ${\cal A}_3$ correspond respectively to first generation of left- handed $SU(2)$ and right-handed $U(1)$ leptons  while fermionic superpartners of ${\cal A}_2$ and ${\cal A}_4$ correspond respectively to left-handed $SU(2)$ and right-handed U(1) quarks. The diagonalised basis (\ref{eq:eigenvectors}) was shown to also work out for appropriately chosen matrix-valued $a_I$ and $z_i$ for multiple fluxed $D7$-brane stacks.
\vskip 0.3in
\noindent {\bf Computation of Soft Terms:}
By using the appropriate ${\cal N} = 1$ coordinates as obtained in \cite{Jockers_thesis} due to the presence of a single $D3$-brane and a single $D7$-brane wrapping the four-cycle (`big' divisor $\Sigma_B$ in a swiss-Cheese Calabi-Yau) along with $D7$-brane fluxes, the soft SUSY breaking parameters were calculated in \cite{gravitino_DM}. The value of scalar masses identified with the masses of squarks and leptons, so obtained, turns out to be quite high but at the same time one gets one light Higgs, thus indicating possibility of ``split SUSY-like scenario" in a local large volume  $D3/D7$ model.

We briefly review the evaluation of various soft supersymmetric as well as supersymmetry breaking parameters in the model involving four-Wilson line moduli as described in \cite{gravitino_DM}. The various soft terms are calculated by power series expansion of  superpotential as well as K\"{a}hler potential,
\begin{eqnarray}
\label{eq:KWexp}
&& W= {\hat W}(\Phi) + \mu(\Phi) {\cal Z}_I {\cal Z}_{J}+ \frac{1}{6} Y_{IJK}(\Phi){\cal M}^{I}{\cal M}^{J} {\cal M}^{K}+...,\nonumber\\
 && K = {\hat K}(\Phi,\bar\Phi) + K_{I {\bar J}}(\Phi,\bar\Phi){\cal M}^{I}{\cal M}^{\bar J} + Z(\Phi,\bar\Phi){\cal M}^{I}{\cal M}^{\bar J}+....,
 \end{eqnarray}
where ${\cal M}^{I}=({\cal Z}^{I},{\cal A}^{I}).$
The soft SUSY breaking parameters are calculated by expanding ${\cal N}=1$ supergravity potential, $V= e^{K}\left(K^{I \bar J}D_{I}W D_{\bar J}{\bar W}- 3 \left|W\right|^2\right)$, in the powers of matter fields ${\cal M}^{I}$, after expanding superpotential as well as K\"{a}hler potential as according to equation (\ref{eq:KWexp}).
In gravity-mediated supersymmetry breaking, SUSY gets spontaneously broken in bulk sector by giving a vacuum expectation value to auxiliary F-terms. Hence, to begin with, one needs to evaluate the bulk $F$-terms which in turn entails evaluating the bulk metric. Writing the K\"{a}hler sector of the K\"{a}hler potential in terms of the bulk moduli as:
 \begin{eqnarray}
&& K\sim-2 ln\left[\left(\sigma_B + \bar{\sigma}_B - \gamma K_{\rm geom}\right)^{\frac{3}{2}} - \left(\sigma_S + \bar{\sigma}_S - \gamma K_{rm geom}\right)^{\frac{3}{2}}\right.\nonumber\\
  && \left. + \sum_{\beta\in H_2^-(CY_3)}n^0_\beta\sum_{(n,m)} cos\left( i n k\cdot(G - \bar{G})g_s + m k\cdot (G + \bar{G}) \right)\right],
  \end{eqnarray}
 disregarding $K_{\rm geom}$ (introduced due to the presence of a mobile space-time filling $D3$-brane) in the large volume limit (See \cite{D3_D7_Misra_Shukla}, \cite{Donaldson_Metric})  and working near $sin \left( i n k\cdot(G - \bar{G})g_s + m k\cdot (G + \bar{G}) \right)=0$ - corresponding to a local minimum -  using the stabilized VEV of $\sigma_{S/B}$ and ${\cal G}^{S,B}$ as given above equation (\ref{eq:K})- generated the following components of the bulk metric's inverse in \cite{gravitino_DM}:
\begin{equation}
G^{m\bar{n}}\sim\left(\begin{array}{cccc} {\cal V}^{\frac{37}{36}} & {\cal V}^{\frac{13}{18}} & 0 & 0 \\
{\cal V}^{\frac{13}{18}} & {\cal V}^{\frac{4}{3}} & 0 & 0 \\
0 & 0 & {\cal O}(1) & {\cal O}(1) \\
0 & 0 & {\cal O}(1) & {\cal O}(1)
\end{array}\right).
\end{equation}
Given that bulk $F$-terms are defined as \cite{softsusy2}: $F^m=e^{\frac{K}{2}}G^{m\bar{n}}D_{\bar{n}}\bar{W}$, one obtained in \cite{gravitino_DM}:
\begin{equation}
\label{eq:bulk_F}
F^{\sigma_S}\sim{\cal V}^{-\frac{n^s}{2} + \frac{1}{36}}M_P^2,\ F^{\sigma_B}\sim{\cal V}^{-\frac{n^s}{2} - \frac{5}{18}}M_P^2,\ F^{G^a}\sim{\cal V}^{-\frac{n^s}{2}-1}M_P^2.
\end{equation}
Hence after spontaneous supersymmetry breaking in the bulk, the gravitino mass is given by:
\begin{equation}
m_{3/2}= e^{K}\left|W\right|^2\sim {\cal V}^{-\frac{n^s}{2}-1}M_P.
\end{equation}

The gaugino mass is given as:
\begin{equation}
\label{eq:gaugino_mass}
m_{\tilde{g}}=\frac{F^m\partial_m T_B}{Re T_B}\lesssim{\cal V}^{\frac{2}{3}}m_{3/2}.
\end{equation}
The analytic form of scalar masses obtained via spontaneous symmetry breaking is given as \cite{softsusy2}:
$m^{2}_{I}= (m^{2}_{\frac{3}{2}} +V_{0}) - F^{\bar m} F^{n} {\partial_{\bar m}}{\partial_n}\log K_{I {\bar I}}$.
These were calculated in \cite{gravitino_DM} to yield
\begin{equation}
\label{eq:mass_Zi}
m_{{\cal Z}_i}\sim {\cal V}^{\frac{59}{72}}m_{3/2}, m_{{\cal A}_1}\sim\sqrt{{\cal V}}m_{3/2},
\end{equation}
implying a non-universality in the open-string moduli masses.
 Further in \cite{gravitino_DM}, we showed  the universality in the trilinear $A$-couplings \cite{softsusy2},
\begin{eqnarray}
\label{ew:A}
& & A_{{\cal I}{\cal J}{\cal K}}=F^m\left(\partial_mK + \partial_m ln Y_{{\cal I}{\cal J}{\cal K}}
+ \partial_m ln\left(K_{{\cal I}\bar{\cal I}}K_{{\cal J}\bar{\cal J}}K_{{\cal K}\bar{\cal K}}\right)\right) \sim{\cal V}^{\frac{37}{36}}m_{3/2}\sim\hat{\mu}_{{\cal Z}_1{\cal Z}_2}.
\end{eqnarray}
The physical higgsino mass parameter $\hat{\mu}_{{\cal Z}_1{\cal Z}_2}$ turned out to be given by:
\begin{equation}
\label{eq:muhat_Z1Z2}
\hat{\mu}_{{\cal Z}_1{\cal Z}_2}=\frac{e^{\frac{K}{2}}\mu_{{\cal Z}_1{\cal Z}_2}}{\sqrt{K_{{\cal Z}_1\bar{\cal Z}_1}K_{{\cal Z}_2\bar{\cal Z}_2}}}\sim{\cal V}^{\frac{19}{18}}m_{3/2}.
\end{equation}
 Further,
\begin{eqnarray}
\label{eq:muhatB_Z1Z2}
& & \left(\hat{\mu}B\right)_{{\cal Z}_1{\cal Z}_2}=\frac{e^{-i arg(W)+\frac{K}{2}}}{\sqrt{K_{{\cal Z}_1\bar{\cal Z}_1}K_{{\cal Z}_2\bar{\cal Z}_2}}}F^m\left(\partial_mK \mu_{{\cal Z}_1{\cal Z}_2} + \partial_m\mu_{{\cal Z}_1{\cal Z}_2} - \mu_{{\cal Z}_1{\cal Z}_2}\partial_m ln\left(K_{{\cal Z}_1\bar{\cal Z}_1}K_{{\cal Z}_2\bar{\cal Z}_2}\right)\right)\nonumber\\
& &  \sim\hat{\mu}_{{\cal Z}_1{\cal Z}_2}\left(F^m\partial_mK+F^{\sigma_S} - F^m\partial_m ln\left(K_{{\cal Z}_1\bar{\cal Z}_1}K_{{\cal Z}_2\bar{\cal Z}_2}\right)\right) \sim{\cal V}^{\frac{19}{18}+\frac{37}{36}}m^2_{3/2}\sim\hat{\mu}^2_{{\cal Z}_1{\cal Z}_2},
\end{eqnarray}
an observation which will be very useful in obtaining a light Higgs of mass $125 \ GeV$.
\paragraph{Realizing  light SM-like Higgs:}
 We calculated in \cite{gravitino_DM} the mass of light Higgs formed by linear combination of two Higgs doublets (using the prescription as given in \cite{HamidSplitSUSY}, to realize split SUSY)  by first calculating the masses of the latter which after soft supersymmetry breaking are given by $M_{H_{u,d}}=(m_{{\cal Z}_{1,2}}^{2}+\hat{\mu}_{{\cal Z}_1{\cal Z}_2}^{2})^{1/2}$ and thereafter using RG solution to Higgs mass discussed in \cite{Dhuria+Misra_mu_Split SUSY}, we obtained the contribution of Higgs doublets as well as the higgsino mass parameter  $\hat{\mu}_{{\cal Z}_1{\cal Z}_2}$ at the EW scale.  The Higgs mass eigenstates are defined as:
\beqn
\label{massmat}
&& H_{1}=D_{h_{11}} H_{u} +D_{h_{12}}H_{d}, H_{2}=D_{h_{21}}H_{u} +D_{h_{22}} H_{d}.
\eeqn
where $D_h=\left(\begin{array}{ccccccc}
\cos \frac{\theta_h}{2} & -\sin \frac{\theta_h}{2} e^{-i\phi_{h}}\\
\sin \frac{\theta_h}{2} e^{i\phi_{h}} & \cos \frac{\theta_h}{2}
 \end{array} \right),
D_{h}^\dagger M_{h}^2 D_h={\rm diag}(M^{2}_{H_1},M^{2}_{H_2})$,
and $\tan \theta_h=
\frac{2|M_{h_{21}}^2|}{M_{h_{11}}^2-M_{h_{22}}^2}$ for a particular range of ${{-\pi}\over {2}} \leq  \theta_h \leq {{\pi}\over {2}}$.

 The RG solution to Higgs mass formed after soft supersymmetry breaking in large $tan\beta$ (but less than 50)-limit are given  \cite{Dhuria+Misra_mu_Split SUSY, gravitino_DM}  (assuming that $m^2_{{\cal Z}_2}(M_s)\equiv  m^2_0 \equiv  {\cal V}^{\frac{59}{72}}m_{\frac{3}{2}}$ (implying $\delta_2=0$ but $ \delta_{1,3,4}\neq0$) and non-universality  w.r.t. to both $D3$-brane position moduli masses ($m_{Z_{1,2}}$) given by $\delta_1$) as:
\begin{eqnarray}
\label{eq:muhat_II}
&& {\hskip -0.4in}\hat{\mu}^2({EW})\equiv -\biggl[-m_0^2-{(0.01)}(n^s)^2\hat{\mu}_{{\cal Z}_1{\cal Z}_2}^2+{(0.32)}{\cal V}^{\frac{4}{3}}m_{3/2}^2-1/2 M_{EW}^2+{(0.03)}{\cal V}^{\frac{2}{3}}n^s\hat{\mu}_{{\cal Z}_1{\cal Z}_2}m_{3/2}+\frac{19\pi}{2200}S_0\biggr],\nonumber\\
& & {\hskip -0.4in} m^2_{H_u}({EW})  \equiv   m^2_{0}(1+ \delta_1)+\frac{1}{2}M_{EW}^2 + m_0^2 -{(0.03)}{\cal V}^{\frac{2}{3}}n^s\hat{\mu}_{{\cal Z}_1{\cal Z}_2}m_{3/2}  + {(0.01)}(n^s)^2\hat{\mu}_{{\cal Z}_1{\cal Z}_2}^2, \nonumber\\
&& {\hskip -0.4in}m^2_{H_d}({EW})\equiv   2m^2_{0}-{(0.06)}{\cal V}^{\frac{2}{3}}n^s\hat{\mu}_{{\cal Z}_1{\cal Z}_2}m_{3/2}+\frac{1}{2}M_{EW}^2-\frac{19\pi}{1100}S_0.\nonumber
\end{eqnarray}
where $S_0$ is hypercharge weighted sum of squared soft scalar mass having value  around $m_0^2$. Assuming $\hat{\mu}B \equiv  \xi\hat{\mu}_{{Z_i}{Z_j}}(\xi\equiv  O(1))$, the Higgs mass matrix is given as:
\begin{eqnarray}
\label{eq:Higss_mass_matrix}
& & \left(\begin{array}{cc}
m^2_{H_u} & \hat{\mu}B\\
\hat{\mu}B & m^2_{H_d}\end{array}\right) \sim\left(\begin{array}{cc}
m^2_{H_u} & \xi\hat{\mu}^2\\
\xi\hat{\mu}^2 & m^2_{H_d}
\end{array}\right),\nonumber
\end{eqnarray}
 and the eigenvalues are given by:
$ \frac{1}{2}\biggl(m^2_{H_u}+m^2_{H_d}\pm\sqrt{\left(m^2_{H_u}-m^2_{H_d}\right)^2+4\xi^2\hat{\mu}^4}\biggr)$.
Using equation (\ref{eq:muhat_II}), for $\delta_1 ={\cal O}(0.1)$ and ${\cal O}(1)\ n^s$, we have
\begin{eqnarray}
\label{eq:evs_1}
& & m^2_{H_u}+m^2_{H_d}\sim   m_0^2-0.06S_0+... \nonumber\\
& & m^2_{H_u}-m^2_{H_d}\sim  m_0^2+0.06S_0+..., \hat{\mu}_{{\cal H}_u{\cal H}_d}^2\sim   m_0^2 -0.03S_0+...\nonumber
\end{eqnarray}
Utilizing above, one sees that the eigenvalues are:
\begin{eqnarray}
\label{eq:evs_2}
&   m^2_{H_{1,2}}= m_0^2 -0.06S_0+.... &\pm\sqrt{\left(  m_0^2 +0.06S_0+...\right)^2+4\xi^2\left(   m_0^2- 0.03S_0\right)^2}. \nonumber
\end{eqnarray}
Considering   $S_0\sim -4.2 m_0^2$ and $ \xi^2 \sim  {\frac{1}{5}} + \frac{1}{16}\frac{m^2_{EW}}{m_0^2}$, we obtain one light Higgs  (corresponding to the negative sign of the square root) of order 125 GeV and one heavy Higgs (corresponding to the positive sign of the square root) whereas  the squared higgsino mass parameter $\hat{\mu}_{{\cal Z}_1{\cal Z}_2}$ then turns out to be heavy with a value, at the EW scale of around ${\cal V}m_{3/2}$.
\vskip 0.3in
\noindent {\bf Realization of $\mu$-split like SUSY:} We summarized above the  different mass scales corresponding to different supersymmetric particles as mentioned in the above paragraphs, and actually calculated in \cite{gravitino_DM}  by considering Calabi-Yau volume ${\cal V}=10^5$ (The justification behind constraining  a value of Calabi-Yau ${\cal V}$ to be ${\cal O}(10^5)$ was based on the right identification of Wilson line moduli and position moduli with SM particle spectrum).  The gravitino appears to be the lightest supersymmetric particle with mass around $10^8$ GeV. The sfermion masses corresponding to first generation of quarks and leptons (identifiable as Wilson line moduli mass in our framework as mentioned  above) are very heavy of the order $10^{10}$ GeV at string scale.
Similarly, the gaugino masses also turns out to be heavy of the order ${10}^{11}$ GeV. However, the higgsino masses turns out to be heavier of the order $10^{13}$ GeV. One of the Higgs doublets  was shown to have mass of the order $125$ GeV,  thus showing the possibility of realizing  $\mu$-split like SUSY scenario (though there is a  `split' between mass of higgsino, and gaugino and sfermions at very high energy scale; the SM fermions are light) in the context of our local LVS $D3$-$D7$.  The  fine-tuning involved in the hyper charge weighted sum of soft scalar masses  ($S_0$) as well as ${\cal O}(1)$ proportionality constant between the higgsino mass parameter squared $ \mu^2$ and soft SUSY parameter $\mu B$ to obtain Higgs of the order $125$ GeV seems acceptable at such high energy scales.  The results of mass scales of all SM as well as superpartners are summarized in Table \ref{table:mass scales} also.
 \begin{table}[htbp]
\centering
\begin{tabular}{|l|l|} \hline
Quark mass & $ M_{q}\sim O(10) MeV $\\
Lepton mass & $ M_{l}\sim {\cal O}(1) MeV $\\  \hline
Gravitino mass &  $ m_{\frac{3}{2}}\sim{\cal V}^{-\frac{n^s}{2} - 1} M_{P}; n_s=2$ \\
Gaugino mass & $ M_{\tilde g}\sim  {\cal V}^{\frac{2}{3}}m_{\frac{3}{2}}$\\
(Lightest) Neutralino/Chargino & $ M_{{\chi^0_3}/{\chi^{\pm}_3}}\sim {\cal V}^{\frac{2}{3}}m_{\frac{3}{2}}$\\
mass & \\
\hline
$D3$-brane position moduli  & $ m_{{\cal Z}_i}\sim {\cal V}^{\frac{59}{72}}m_{\frac{3}{2}}$ \\
(Higgs) mass & \\
Wilson line moduli & $ m_{\tilde{\cal A}_I}\sim {\cal V}^{\frac{1}{2}}m_{\frac{3}{2}}$\\
(sfermion mass ) & ${I =1,2,3,4}$\\ \hline
A-terms & $A_{pqr}\sim n^s{\cal V}^{\frac{37}{36}}m_{\frac{3}{2}}$\\
& $\{p,q,r\} \in \{{{\tilde{\cal A}_I}},{{\cal Z}_i}\}$\\ \hline
Physical $\mu$-terms & $\hat{\mu}_{{\cal Z}_i{\cal Z}_j}$ \\
 (Higgsino mass) & $\sim{\cal V}^{\frac{37}{36}}m_{\frac{3}{2}}$ \\ \hline
Physical $\hat{\mu}B$-terms & $\left(\hat{\mu}B\right)_{{\cal Z}_1{\cal Z}_2}\sim{\cal V}^{\frac{37}{18}}m_{\frac{3}{2}}^2$\\ \hline
\end{tabular}

\caption{Mass scales of first generation of SM as well supersymmetric, and soft SUSY breaking parameters.}
\label{table:mass scales}
\end{table}

\paragraph{Modified ${\cal N}=1$ gauged supergravity action in case of multiple D7-brane:} We will be using the following terms (written out in four-component notation or their two-component analogs and utilizing/generalizing results of \cite{Jockers_thesis}) in the ${\cal N}=1$ gauged supergravity action of Wess and Bagger \cite{Wess_Bagger} with the understanding that $m_{{\rm moduli/modulini}}<<m_{\rm KK}\left(\sim\frac{M_s}{{\cal V}^{\frac{1}{6}}}\Biggr|_{{\cal V}\sim10^{5/6}}\sim10^{14}GeV\right)$, $M_s=\frac{M_P}{\sqrt{{\cal V}}}\Biggr|_{{\cal V}\sim10^{5/6}}\sim10^{15}GeV$, and that for multiple $D7$-branes, the non-abelian gauged isometry group\footnote{As explained in  \cite{Jockers_thesis}, one of the two Pecci-Quinn/shift symmetries along the RR two-form axions $c^a$ and the four-form axion $\rho_B$ gets gauged due to the dualization of the Green-Schwarz term $\int_{{\bf R}^{1,3}}dD^{(2)}_B\wedge A$ coming from the KK reduction of the Chern-Simons term on $\Sigma_B\cup\sigma(\Sigma_B)$ - $D^{(2)}_B$ being an RR two-form axion. In the presence of fluxes for multiple $D7$-brane fluxes, the aforementioned Green-Schwarz is expected to be modified to $Tr\left(Q_B\int_{{\bf R}^{1,3}}dD^{(2)}_B\wedge A\right)$, which after dualization in turn modifies the covariant derivative of $T_B$ and hence the killing isometry.}, corresponding to the killing vector $6i\kappa_4^2\mu_7\left(2\pi\alpha^\prime\right)Q_B\partial_{T_B}, Q_B=\left(2\pi\alpha^\prime\right)\int_{\Sigma_B}i^*\omega_B\wedge P_-\tilde{f}$ arising due to the elimination of of the two-form axions $D_B^{(2)}$ in favor of the zero-form axions
$\rho_B$ under the KK-reduction of the ten-dimensional four-form axion \cite{Jockers_thesis} (which results in a modification of the covariant derivative of $T_B$ by an additive shift given by $6i\kappa_4^2\mu_7\left(2\pi\alpha^\prime\right)Tr(Q_B A_\mu)$)  can be identified with the SM group (i.e. $A_\mu$ is the SM-like adjoint-valued gauge field \cite{Wess_Bagger}):
\begin{eqnarray}
\label{eq:WB_gSUGRA N=1}
& & {\hskip -0.3in}{\cal L} = g_{YM}g_{T_B {\bar{\cal J}}}Tr\left(X^{T_B}{\bar\chi}^{\bar{\cal J}}_L\lambda_{\tilde{g},\ R}\right) +ig_{{\cal I}\bar{\cal J}}Tr\left({\bar\chi}^{\bar{\cal I}}_L\left[\slashed{\partial}\chi^{\cal I}_L+\Gamma^i_{Mj}\slashed{\partial} a^M\chi^{\cal J}_L
+\frac{1}{4}\left(\partial_{a_M}K\slashed{\partial} a_M - {\rm c.c.}\right)\chi^{\cal I}_L\right] \right) \nonumber\\
& & {\hskip -0.3in}+\frac{e^{\frac{K}{2}}}{2}\left({\cal D}_{\bar{\cal I}}D_{\cal J}\bar{W}\right)Tr\left(\chi^{\cal I}_L\chi^{\cal J}_R\right) + g_{T_B\bar{T}_B}Tr\left[\left(\partial_\mu T_B - A_\mu X^{T_B}\right)\left(\partial^\mu T_B - A^\mu X^{T_B}\right)^\dagger\right] \nonumber\\
& & + g_{T_B{\cal J}}Tr\left(X^{T_B}A_\mu\bar{\chi}^{\cal J}_L\gamma^\nu\gamma^\mu\psi_{\nu,\ R}\right) + \bar{\psi}_{L,\ \mu}\sigma^{\rho\lambda}\gamma^\mu\lambda_{\tilde{g},\ L}F_{\rho\lambda} + \bar{\psi}_{L,\ \mu}\sigma^{\rho\lambda}\gamma^\mu\lambda_{\tilde{g},\ L}W^+_\rho W^-_\lambda\nonumber\\
 & & + Tr\left[\bar{\lambda}_{\tilde{g},\ L}\slashed{A}\left(6\kappa_4^2\mu_7(2\pi\alpha^\prime)Q_BK + \frac{12\kappa_4^2\mu_7(2\pi\alpha^\prime)Q_Bv^B}{\cal V}\right) \lambda_{\tilde{g},\ L}\right]\nonumber\\
 & & + \frac{e^K G^{T_B\bar{T}_B}}{\kappa_4^2}6i\kappa_4^2(2\pi\alpha^\prime)Tr\left[Q_BA^\mu\partial_\mu
 \left(\kappa_4^2\mu_7(2\pi\alpha^\prime)^2C^{I\bar{J}}a_I\bar{a}_{\bar J}\right)\right] + {\rm h.c.}\nonumber\\
 & & -\frac{f_{ab}}{4}F^a_{\mu\nu}F^{b\ mu\nu} + \frac{1}{8}f_{ab}\epsilon^{\mu\nu\rho\lambda}F^a_{\mu\nu}F^b_{\rho\lambda} - \frac{i\sqrt{2}}{4} g \partial_{i/I}f_{ab} Tr\left(\frac{12\kappa_4^2\mu_7(2\pi\alpha^\prime)Q_B^av^B}{\cal V}{\bar\lambda}_{\tilde{g},L}^b\chi^{i/I}_R\right) + {\rm h.c.}\nonumber\\
 & & - \frac{\sqrt{2}}{4}\partial_{i/I}f_{ab}Tr\left({\bar\lambda}_{\tilde{g},R}^a\sigma^{\mu\nu}\chi^{i/I}_L\right)F_{\mu\nu}^b + {\rm h.c.},
\end{eqnarray}

\section{CP Violating Phases}

In this section we explain the possible origin of CP-violating phases in the ${\cal N}=1$ gauged supergravity limit of large volume $D3/D7$-$\mu$ split SUSY model. The electric dipole moment of a spin-$\frac{1}{2}$ particle is defined by the effective CP violating dimension-5 operator given as: $
{\cal L}_I=-\frac{i}{2} d_f \bar{\psi} \sigma_{\mu\nu} \gamma_5 \psi F^{\mu\nu}.$
Given that the effective operator is non-renormalizable, the same can be realized at the loop level provided the theory contains a source of CP violation. In Standard Model, CP violating phases in general appear from the complex Kobayashi Maskawa (CKM) phases in the quark mass matrix but the same get a non-zero contribution only at three-loop level in Standard Model. However, in supersymmetric theories, instead of CKM phase generated in Standard Model, one can consider the new phases appearing from complex parameters of soft-SUSY breaking terms, complex effective Yukawa couplings as well as supersymmetric mass terms.

We consider the existence of non-zero phases appearing from complex effective Yukawa couplings present in ${\cal N}=1$ gauged supergravity action. As discussed in \cite{gravitino_DM}, the position as well as Wilson line moduli identification with SM-like particles  generate  effective Yukawa couplings including R-parity conserving  as well as R-parity violating ones in the context of ${\cal N}=1$  gauged supergravity action \cite{gravitino_DM}, and the solution of RG evolution of effective Yukawa couplings at one-loop level yields:
\beqn
\label{eq:Yhat_sol-2}
\hat{Y}_{\Lambda\Sigma\Delta}(t)\sim\hat{Y}_{\Lambda\Sigma\Delta}(M_s)
\prod_{(a)=1}^3\left(1 + \beta_{(a)}t\right)^{\frac{-2\left(C_{(a)}(\Lambda)
    +C_{(a)}(\Sigma) + C_{(a)}(\Delta)\right) }{b_{(a)}}}. \eeqn
Using the fact that quadratic Casimir invariants as well as beta functions are real, we see that magnitudes of Yukawa couplings $\hat{Y}_{\Lambda\Sigma\Delta}$s change only by ${\cal O}(1)$ while phases of all Yukawas do not change at all as one RG-flows down from the string to the EW scale. Also, given that all four-Wilson line moduli ${\cal A}_I$ as well as position moduli ${\cal Z}_I$ are stabilized at different values; we make an assumption that there will be a distinct phase factor associated with all position as well as Wilson line moduli superfields which produces an overall distinct phase factor for each possible effective Yukawa coupling corresponding to four Wilson line moduli as well as position moduli.

The other important origin of generation of non-zero phases are given by complex soft SUSY breaking parameters (${m}^{2}_i, {\cal A}_{IJK}, \mu B$)  as well supersymmetric mass term $\mu$. The soft SUSY scalar mass terms can be made real by phase redefinition. However, in addition to the diagonal entries of sfermions corresponding to fermions with L-handed as well as R-handed chirality in the sfermion mass matrix, one gets an off-diagonal contribution because of mixing between L-R sfermion masses  after EW-symmetry breaking. The contribution of the same is governed by complex trilinear couplings as well as supersymmetric mass parameter $\mu$ at EW scale. Therefore, the scalar (sfermion) fields  $\tilde{f}_L$ and $\tilde{f}_R$ have been considered as linear combinations of the mass eigenstates which are obtained by diagonalizing sfermion $(mass)^2$  matrices \cite{ibrahim+nath} i.e.
\beqn
&& \tilde{f}_L=D_{f_{11}} \tilde{f}_1 +D_{f_{12}} \tilde{f}_2, \tilde{f}_R=D_{f_{21}} \tilde{f}_1 +D_{f_{22}} \tilde{f}_2.
\eeqn
where $f$ corresponds to first generation leptons and quarks and
\begin{equation}
D_f=\left(\begin{array}{ccccccc}
\cos \frac{\theta_f}{2} &-\sin \frac{\theta_f}{2} e^{-i\phi_{f}}\\
 \sin \frac{\theta_f}{2} e^{i\phi_{f}} & \cos \frac{\theta_f}{2}
 \end{array} \right),
\end{equation}
and  the mass matrix is given as follows:
\begin{equation}
\label{eq:massmatrix}
M_{\tilde{f}}^2=\left(\begin{array}{ccccccc}
 M_{\tilde{f_L}}^2 & m_u({\cal A}_{f}^{*}-\mu \cot\beta)\\
 m_u({\cal A}_{f}-\mu^{*} \cot\beta) & M_{\tilde{f_R}}^2
 \end{array} \right)_{EW},
\end{equation}
where  ${{\cal A}_{IJK}}$ corresponds to complex trilinear coupling parameter. Diagonalizing the above matrix by performing unitary transformation: $D_{f}^\dagger M_{\tilde{f}}^2 d_f={\rm diag}(M_{\tilde{f}1}^2, M_{\tilde{f}2}^2)$, where $\tan \theta_f=
\frac{2|M_{\tilde{f}21}^2|}{M_{\tilde{f}11}^2-M_{\tilde{f}22}^2}$.
The eigenvalues $M_{\tilde{f}1}^2$ and $M_{\tilde{f}2}^2$ are as follows:
\beq
\label{eq:Me1e2}
M_{\tilde{f}(1)(2)}^2=\frac{1}{2} (M_{\tilde{f}11}^2+M_{\tilde{f}22}^2)
	(+)(-)\frac{1}{2}[(M_{\tilde{f}11}^2-M_{\tilde{f}22}^2)^2 +
		4|M_{\tilde{f}21}^2|^2]^{\frac{1}{2}}.
\eeq
 For $f=e, {\cal A}_{e}^{*}= {{\cal A}_{{\cal Z}_I {\cal A}_1 {\cal A}_3}} $; for  $f=({u,d}), {\cal A}_{u/d}^{*}= {{\cal A}_{{\cal Z}_I {\cal A}_2 {\cal A}_4}} $. In our model as discussed in section {{\bf 2}}, we have universality in trilinear couplings w.r.t position as well as Wilson line moduli. Assuming the same to be true at EW scale, the values of trilinear coupling parameters are ${\cal A}_{{\cal Z}_I {\cal A}_1 {\cal A}_3}= {\cal A}_{{\cal Z}_I {\cal A}_2 {\cal A}_4}= {\cal V}^{\frac{37}{36}}m_{\frac{3}{2}}$.
As given in section {{\bf 2}}, the value of  supersymmetric mass parameter  $\mu$ at EW scale is ${\cal V}^{\frac{59}{72}}m_{\frac{3}{2}}$. Also we have universality in slepton (squark) masses of first two generations. Therefore, $M_{\tilde{e}11}^2= M_{\tilde{e}22}^2= M_{\tilde{u}11}^2 =M_{\tilde{u}22}^2 \sim  {\cal V} m^{2}_{\frac{3}{2}}$, and
\beqn
\label{me21}
&& |M_{\tilde{e}21}^2|^2= m_e|{\cal A}_{e}^{*}-{\mu}\cot{\beta}| \equiv  ({\cal V}^{\frac{37}{36}})m_{e}m_{\frac{3}{2}} << M_{\tilde{e}11}^2, \nonumber\\
&& |M_{\tilde{u}21}^2|^2= m_u|{\cal A}_{u}^{*}-{\mu}\cot{\beta}| \equiv  ({\cal V}^{\frac{37}{36}})m_{e}m_{\frac{3}{2}} << M_{\tilde{u}11}^2.
\eeqn
Using the above, one can show that eigenvalues of sfermion mass matrix $M_{\tilde{f}(1)(2)}^2 \sim  M_{{\tilde f}^{2}_{L,R}}={\cal V}m^{2}_{\frac{3}{2}}$. The aforementioned mass eigenstates can be utilized to produce non-zero phase responsible to generate finite  EDM of electron as well as neutron in the one-loop diagrams involving sfermions as scalar propagators, and  gaugino and neutralino as fermionic propagators.

\section{One-Loop Contribution to Electric Dipole Moment}
At one-loop level, for a theory of fermion $\psi_f$ interacting with other heavy fermions $\psi_i$'s and heavy scalars $\phi_k$'s
with masses $m_i$, $m_k$ and charges $Q_i $, $Q_k$, the interaction that contains CP violation in
general is given by \cite{ibrahim+nath}:
\beqn
\label{eq:eff1loop}
-{\cal L}_{int}=\sum_{i,k} \bar{\psi}_f
                \left(K_{ik} \PL +L_{ik} \PR\right) \psi_i\phi_k+ h.c
\eeqn
Here  ${\cal L}$ violates CP invariance iff
${\rm Im}(K_{ik} L_{ik}^*)\neq 0$.
and one-loop EDM of the fermion f in this  case is given by
\beq
\sum_{i,k} \frac{m_i }{(4\pi)^2 m_k^2}{\rm Im}(K_{ik} L_{ik}^*)
        \left(Q_i A\left(\frac{m_i^2}{m_k^2}\right)+Q_k    B\left(\frac{m_i^2}{m_k^2}\right)\right),
\eeq
where $A(r)$ and $B(r)$ are defined by
\beq
\label{eq:A}
 A(r)=\frac{1}{2(1-r)^2}\Bigl(3-r+\frac{2lnr}{1-r}\Bigr),
B(r)=\frac{1}{2(r-1)^2}\Bigl(1+r+\frac{2rlnr}{1-r}\Bigr),
\eeq
where, $Q_k=Q_f-Q_i$.

We use the above-mentioned results to get an order-of-magnitude estimate of EDM of electron/quark in the context of ${\cal N}=1$ gauged supergravity by including all SM as well as supersymmetric particles in the loop diagram. The EDMs of the neutron can be estimated by calculating the contribution of $u$ and $d$ quarks  by using relation $d_n = (4d_{d}-d_{u})/3$. Since in our model, we have identified both up- as well as down-quark with single Wilson line modulus, we will have same contribution of EDM for both up and down-quarks. Hence, the neutron EDM is same as up-quark EDM. Therefore, in the calculations below, we will estimate EDM of electron and up quark only.

\subsection{One-Loop Diagrams involving Neutral Sfermions in the Loop} {\bf{Gaugino contribution:}} In this subsection, we estimate the contribution of  electron/neutron EDM at one-loop level due to presence of heavy gaugino nearly isospectral with heavy sfermions (for the Calabi-Yau volume ${\cal V}=10^5$ in string-length units). In traditional split SUSY models discussed in literature, the masses of sfermions are very heavy while  gaugino as well as higgsino's are kept light because of the gauge coupling unification. Therefore, one-loop diagrams involving sfermion-gaugino exchange do not give any significant contribution to EDM of fermion . However in the large volume $D3-D7$ set-up that we have discussed, the gaugino as well as higgsino also turn out to be very heavy. As it is clear from the equation (\ref{eq:eff1loop}), the order of magnitude of EDM at one-loop level is directly proportional to fermion mass and inversely proportional to sfermion masses circulating in the loop whereas the one-loop function can almost be of  $ {\cal O}(0.1-1)$ provided either the difference between fermion and sfermion mass is of ${\cal O}(1)$  or the fermion mass is very light as compared to sfermion mass.  Therefore, naively one would expect an enhancement in the order of magnitude of one-loop EDM due to presence of heavy fermions circulating in a loop. In view of this, we estimate the contribution of one-loop EDM of electron as well as neutron in $N=1$ gauged supergravity limit of large volume $D3/D7$ $\mu$-split SUSY model discussed in {{\bf section 2}}.  However, the CP violation (imaginary phases) can be induced in a loop diagram by considering diagonalized eigenstates of sfermion mass matrix as propagators in the loop. The loop diagram is given in Figure 1.

\begin{figure}
\begin{center}
\begin{picture}(145,127) (130,40)
   \Line(100,50)(330,50)
   \DashCArc(215,50)(80,0,180){4}
   \Photon(222,130)(230,190){4}{4}
   \Text(150,110)[]{${\tilde f}_{i}$}
   \Text(290,100)[]{${\tilde f}_{i}$}
   \Text(90,50)[]{{$f_{L}$}}
   \Text(234,150)[]{{$\gamma$}}
   \Text(340,50)[]{{$f_{R}$}}
   \Text(235,40)[]{{{$\tilde{\lambda}^0$}}}
   \end{picture}
\end{center}
\caption{One-loop diagram involving gaugino.}
 \end{figure}
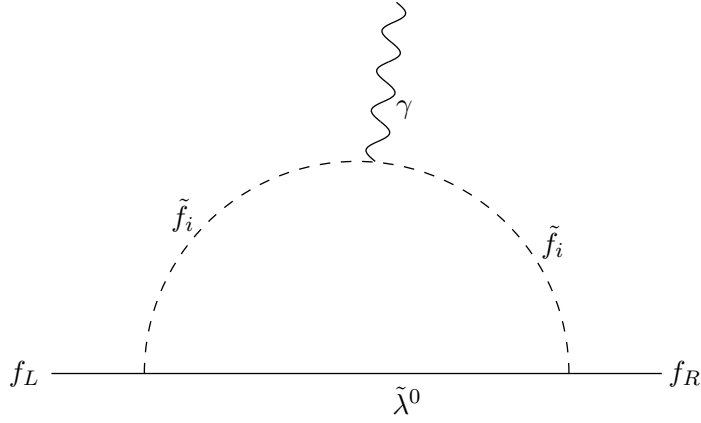
 The effective one-loop operator given in equation (\ref{eq:eff1loop}) can be recasted in the following form:
\beq
\label{eq:eff1loop1}
{\cal L}_{int}=\sum_{i=e,u,d} \bar{\psi}_{f_i}
                \left(K_{i} \PL +L_{i} \PR\right) {\phi_{{\tilde f_i}}}{\tilde \lambda^0}+ h.c
\eeq
For $i=1,2$, above equation can be expanded as:
 \beq
 \label{eq:eff1loop2}
-{\cal L}_{int}= \bar{\psi}_{f}
                \left(K_{1} \PL +L_{1} \PR\right) {\phi_{\tilde f_1}}{\tilde \lambda^0}+\bar{\psi}_{f}
                \left(K_{2} \PL +L_{2} \PR\right) {\phi_{\tilde f_2}}{\tilde \lambda^0}+ h.c.,
\eeq
and one-loop EDM of the fermion $f$ in this  case will be given as:
\beq
\label{eq:EDM}
\frac{m_{\tilde{\lambda^{0}_i}}}{(4\pi)^2 } \left[\frac{1}{m_{{\tilde f_1}^2}}   {\rm Im}(K_{1} L_{1}^*)\left(Q'_{\tilde {f_1}} B\left(\frac{m_{\tilde {\lambda^0}}^2}{m_{{\tilde f_1}^2}}\right)\right)+  \frac{1}{m_{{\tilde f_2}^2}}  {\rm Im}(K_{2} L_{2}^*)
       \left(Q_{\tilde {f_2}}B\left(\frac{m_{\tilde {\lambda^0}}^2}{m_{{\tilde f_2}^2}}\right)\right)\right].
\eeq
where $m_{\tilde {\lambda^0}}$ corresponds to gaugino mass, $m_{\tilde f_1}$ and $m_{\tilde f_2}$ correspond to masses of eigenstates of diagonalized sfermion mass matrix. and $Q'_{\tilde {f_i}}$ corresponds to effective charge defined as: $Q'_{\tilde {f_i}} \sim  Q_{\tilde {f_i}} C_{{\tilde {f_i}}{\tilde {f_i}}{\gamma}}$, where
$C_{{\tilde {f_i}}{\tilde {f_i}}{\gamma}}$ will be volume suppression factor coming from sfermion-photon-sfermion vertex.

To determine  the value of one-loop EDM in this case, we first calculate the contribution of required vertices involved in Figure 1.
In ${\cal N}=1$ gauged supergravity, lepton(quark)-slepton(squark)-gaugino interaction are governed by \cite{Wess_Bagger} following term:
\beqn
{\cal L}_{f-\tilde{f}-\tilde{\lambda^{0}}}= g_{YM}g_{{ J}\bar{T}_B}X^{{*}B}{\bar\chi}^{\bar J}\tilde{\lambda^{0}}+ \partial_{J}T_B D^{B}{\bar\chi}^{\bar J}\lambda^{0} .\nonumber
\eeqn
where ${\bar\chi}^{\bar J}$ corresponds to spin $\frac{1}{2}$ fermion,  $X^{{*}B}$ is killing isometry vector and $\tilde{\lambda^{0}}$ corresponds to $SU(2)$-singlet component of neutral gaugino. Though the gauge coupling $ g_{YM}$ is real, the non-zero phase factor is produced from moduli space metric component $g_{{ J}\bar{T}_B}$ and is associated with the volume suppression factor arising from the same. Hence, effective gauge coupling interaction vertex generate a particular phase factor which we consider to be of the order ${\cal O}(1)$.

 We repeatedly mention that to get the numerical estimate of contribution of aforementioned vertices, we use the identification described in \cite{gravitino_DM} according to which  fermionic superpartners of ${\cal A}_1$ and ${\cal A}_3$ can be identified, respectively with $e_L$ and $e_R$, and the fermionic superpartners of ${\cal A}_2$ and ${\cal A}_4$ can be identified, respectively with the first generation quarks: $u_L/d_L$ and $u_R/d_R$. In principle, incoming left-handed electron(quark) can couple with  scalar superpartners of both left-handed as well as right-handed leptons(quarks). Therefore, for a  left handed electron $e_L$ interacting with slepton as well as gaugino, interaction vertex will be given as
\beqn
{\cal L}_{e_{L}-\tilde{e}-\tilde{\Lambda^{0}}}= g_{YM}g_{{\cal A}_1\bar{T}_B}X^{{*}B}{\bar\chi}^{\bar {\cal A}_1}\tilde{\lambda^{0}}+ \partial_{a_1}T_B D^{B}{\bar\chi}^{\bar a_1}\lambda^{0}. \eeqn
To calculate the contribution of $e_{L}-\tilde{e_L}-\tilde{\lambda^{0}}$ vertex, we expand $g_{{\cal A}_1\bar{T}_B}$ in the fluctuations linear in ${\cal A}_1$ around its stabilized VEV. In terms of undiagonalized basis, we have: $ g_{{T_B} {\bar a}_1}\rightarrow-{\cal V}^{-\frac{1}{4}} (a_1-{\cal V}^{-\frac{2}{9}})$. Using $T_{B}=Vol(\sigma_B)- C_{I{\bar J}}a_I{\bar a}_{\bar J} + h.c.$, where the values of intersection matrices $C_{I{\bar J}}$ are given in the appendix of \cite{gravitino_DM}. Utilizing those values,  we get $\partial_{a_1} T_B\rightarrow {\cal V}^{\frac{10}{9}} (a_1-{\cal V}^{-\frac{2}{9}})$. Using the argument that $g_{YM}g_{{T_B} {\bar a}_1} \sim {\cal O}(1)g_{YM}g_{{T_B} {\bar {\cal A}}_1}$ as shown in \cite{gravitino_DM}; incorporating values of  $X^{B}=-6i\kappa_4^2\mu_7Q_{T_B}, \kappa_4^2\mu_7\sim \frac{1}{\cal V}, D^B=\frac{4\pi\alpha^\prime\kappa_4^2\mu_7Q_Bv^B}{\cal V}$ and  $Q_{T_B}\sim{\cal V}^{\frac{1}{3}}(2\pi\alpha^\prime)^2\tilde{f}$, we get the contribution of physical gaugino$(\tilde{\lambda^{0}})$- lepton$(e_{L})$-slepton$(\tilde{e_L})$ interaction vertex as follows :
\begin{equation}
\label{eq:eLeLlambda}
|C_{e_{L}\tilde{e_L}\tilde{\lambda^{0}}}|\equiv \frac{ {\cal V}^{-\frac{2}{9}} {\tilde f}}{{ \sqrt{\hat{K}_{{\cal A}_1{\bar {\cal A}}_1}} }{ \sqrt{\hat{K}_{{\cal A}_1{\bar {\cal A}}_1}} }}{\tilde {\cal A}_1}{\bar\chi}^{\bar {\cal A}_1}\tilde{\lambda^{0}}
 \equiv\tilde{f}\left({\cal V}^{-1}\right)
 \end{equation},
where ${\tilde f}$ is dilute flux and upper limit of the same as calculated in \cite{Dhuria+Misra_mu_Split SUSY}  is  ${\cal V}^{-\frac{23}{30}}\sim{\cal O}(10^{-4})$ for Calabi-Yau volume ${\cal V}\sim 10^5$.

Similarly, contribution of physical gaugino$(\tilde{\lambda^{0}})$- quark$(u_{L})$-squark$(\tilde{u_L})$ interaction vertex will be given by expanding $g_{{\cal A}_2\bar{T}_B}$ in the fluctuations linear in ${\cal A}_2$ around its stabilized VEV. Doing so, one will get: $ g_{{T_B} {\bar a}_2}\rightarrow-{\cal V}^{-\frac{5}{4}} (a_2-{\cal V}^{-\frac{1}{3}}), \partial_{a_2}T_B \rightarrow {\cal V}^{\frac{1}{9}} (a_2-{\cal V}^{-\frac{1}{3}})$,
 and
\begin{equation}
\label{eq:uLuLlambda}
 |C_{u_{L}\tilde{u_L}\tilde{\lambda^{0}}}| \equiv \frac{  {\cal V}^{-\frac{11}{9}} {\tilde f}}{{ \sqrt{\hat{K}_{{\cal A}_2{\bar {\cal A}}_2}} }{ \sqrt{\hat{K}_{{\cal A}_2{\bar {\cal A}}_2}} }}{\tilde {\cal A}_2}{\bar\chi}^{\bar {\cal A}_2}\tilde{\lambda^{0}}
 \equiv\tilde{f}\left({\cal V}^{-\frac{4}{5}}\right).
 \end{equation}

The gaugino$(\tilde{\lambda^{0}})$ - fermion($f_L$)-sfermion(${\tilde f}_R$) vertex does not possess $SU(2)$  electroweak symmetry. However, terms in supergravity Lagrangian preserve $SU(2)$ EW symmetry. Therefore, we first generate a term of the type $f_{L} \tilde{f_R} \tilde{\lambda^{0}}H_L$ wherein  $H_L$ is $SU(2)_L$  Higgs doublet.  After spontaneous breaking of the EW symmetry when $H_L$ acquires a non-zero vev $\langle H^0\rangle$, this term generates: $\langle H^0\rangle f_{L} \tilde{f_R} \tilde{\lambda^{0}}$.
For $f_{L,R}= e_{L,R}$, by expanding $g_{{\cal A}_1\bar{T}_B}$ in the fluctuations linear in ${\cal Z}_i$ and then linear in ${\cal A}_3$ around their stabilized value, we have: $ g_{{T_B} {\bar {\cal A}}_1}\rightarrow {\cal V}^{-\frac{13}{36}} \langle {\cal Z}_1 \rangle ({\cal A}_3- {\cal V}^{-\frac{13}{18}})$. The contribution of physical gaugino$(\tilde{\lambda^{0}})$- lepton$(e_{L})$-slepton$(\tilde{e_R})$ interaction vertex will be as follows :
\beq
\label{eq:eLeRlambda}
| C_{e_{L}\tilde{e_{R}}\tilde{\lambda^{0}}}| \equiv \frac{g_{YM}g_{{T_B} {\bar {\cal A}_1}}X^{T_B}\sim {\cal V}^{-\frac{19}{18}} {\tilde f}}{{ \sqrt{\hat{K}_{{\cal Z}_1{\bar {\cal Z}}_1}\hat{K}_{{\cal A}_1{\bar {\cal A}}_1}\hat{K}_{{\cal A}_3{\bar {\cal A}}_3}}}}{\tilde {\cal A}_3 }{\bar\chi}^{\bar {\cal A}_1}\tilde{\lambda^{0}}
 \equiv\tilde{f}\left({\cal V}^{-\frac{15}{9}}\right).
 \eeq
For $f_{L,R}= u_{L,R}$, by expanding $g_{{\cal A}_2\bar{T}_B}$ in the fluctuations linear in ${\cal Z}_i$ and then linear in ${\cal A}_4$ around their stabilized value, we have: $ g_{{T_B} {\bar {\cal A}}_2}\rightarrow {\cal V}^{-\frac{13}{36}} \langle{\cal Z}_1\rangle ({\cal A}_4- {\cal V}^{-\frac{11}{9}})$,
and
\beq
\label{eq:uLeRlambda}
|C_{u_{L}\tilde{u_{R}}\tilde{\lambda^{0}}}|\equiv \frac{g_{YM}g_{{T_B} {\bar {\cal A}_2}}X^{T_B}\sim {\cal V}^{-\frac{19}{18}} {\tilde f}}{{ \sqrt{\hat{K}_{{\cal Z}_1{\bar {\cal Z}}_1}\hat{K}_{{\cal A}_2{\bar {\cal A}}_2}\hat{K}_{{\cal A}_4{\bar {\cal A}}_4}}}}{\tilde {\cal A}_4}{\bar\chi}^{\bar {\cal A}_2}\tilde{\lambda^{0}}
 \equiv\tilde{f}\left({\cal V}^{-\frac{14}{9}}\right).
\eeq
Similarly, outgoing right-handed electron (quark) can couple with both left handed as well as right handed sleptons(squarks) and include the gaugino$(\tilde{\lambda^{0}})$-fermion($f_R$)-sfermion(${\tilde f}_L$) vertex in a loop diagram. The same does not possess $SU(2)$ EW symmetry. For $f_{L,R}= e_{L,R}$, by expanding $g_{{\cal A}_3\bar{T}_B}$ first in the fluctuations linear in ${\cal Z}_1$ and then linear in ${\cal A}_1$ around their stabilized VEV's, we have: $ g_{{T_B} {\bar {\cal A}}_3}\rightarrow-{\cal V}^{-\frac{13}{36}} \langle {\cal Z}_1\rangle ({\cal A}_1- {\cal V}^{-\frac{2}{9}})$. The physical gaugino$(\tilde{\lambda^{0}})$- lepton$(e_{R})$-slepton$(\tilde{e_L})$ interaction vertex will be given as :
\beq
\label{eq:eReLlambda}
|C_{e_{R}\tilde{e_{L}}\tilde{\lambda^{0}}}|\equiv \frac{g_{YM}g_{{T_B} {\bar {\cal A}_3}}X^{T_B}\sim {\cal V}^{-\frac{19}{18}} {\tilde f}}{{ \sqrt{\hat{K}_{{\cal Z}_1{\bar {\cal Z}}_1}\hat{K}_{{\cal A}_1{\bar {\cal A}}_1}\hat{K}_{{\cal A}_3{\bar {\cal A}}_3}}}}{\tilde {\cal A}_1}{\bar\chi}^{\bar {\cal A}_3}\tilde{\lambda^{0}}
 \equiv\tilde{f}\left({\cal V}^{-\frac{15}{9}}\right).
\eeq
For $f_{L,R}= u_{L,R}$, one gets: $ g_{{T_B} {\bar {\cal A}}_4}\rightarrow-{\cal V}^{-\frac{13}{36}} \langle {\cal Z}_1\rangle ({\cal A}_2- {\cal V}^{-\frac{1}{3}})$, and
\beq
\label{eq:uRuLlambda}
|C_{u_{R}\tilde{u_{L}}\tilde{\lambda^{0}}}|\equiv \frac{g_{YM}g_{{T_B} {\bar {\cal A}_2}}X^{T_B}\sim {\cal V}^{-\frac{19}{18}} {\tilde f}}{{ \sqrt{\hat{K}_{{\cal Z}_1{\bar {\cal Z}}_1}\hat{K}_{{\cal A}_2{\bar {\cal A}}_2}\hat{K}_{{\cal A}_4{\bar {\cal A}}_4}}}}{\tilde {\cal A}_2}{\bar\chi}^{\bar {\cal A}_4}\tilde{\lambda^{0}}
 \equiv\tilde{f}\left({\cal V}^{-\frac{14}{9}}\right).
\eeq
 To calculate the contribution of $e_{R}-\tilde{e_R}-\tilde{\lambda^{0}}$ vertex, we expand  $g_{{\cal A}_3\bar{T}_B}$ in the fluctuations linear in ${\cal A}_3$, and obtain: $ g_{{T_B} {\bar {\cal A}}_3}\rightarrow-{\cal V}^{\frac{7}{9}} ({\cal A}_3-{\cal V}^{\frac{13}{18}}),  \partial_{{\cal A}_3}T_B \rightarrow {\cal V}^{\frac{19}{9}} ({\cal A}_3-{\cal V}^{-\frac{13}{18}})$. Utilizing this, the physical gaugino$(\tilde{\lambda^{0}})$- lepton$(e_{R})$-slepton$(\tilde{e_R})$ interaction vertex will be given as :
\beq
\label{eq:eReRlambda}
|C_{e_{R}\tilde{e_{R}}\tilde{\lambda^{0}}}|\equiv \frac{{\cal V}^{ \frac{7}{9}} {\tilde f}}{{\sqrt{\hat{K}_{{\cal A}_3{\bar {\cal A}}_3} \hat{K}_{{\cal A}_3{\bar {\cal A}}_3}}} }{\tilde {\cal A}_3}{\bar\chi}^{\bar {\cal A}_3}\tilde{\lambda^{0}}
 \equiv\tilde{f}\left({\cal V}^{-\frac{3}{5}}\right).
\eeq
Similarly, by expanding $g_{{\cal A}_4 \bar{T}_B}$ in the fluctuations linear in ${\cal A}_4$, we will have
$ g_{{T_B} {\bar {\cal A}}_4}\rightarrow-{\cal V}^{\frac{16}{9}} ({\cal A}_4-{\cal V}^{\frac{11}{9}}), \partial_{{\cal A}_4}T_B \rightarrow {\cal V}^{\frac{28}{9}} ({\cal A}_4-{\cal V}^{-\frac{11}{9}})$, and
\beq
\label{eq:uRuRlambda}
|C_{u_{R}\tilde{u_{R}}\tilde{\lambda^{0}}}|\equiv \frac{{\cal V}^{ \frac{16}{9}} {\tilde f}}{{\sqrt{\hat{K}_{{\cal A}_4{\bar {\cal A}}_4}{\hat{K}_{{\cal A}_4{\bar {\cal A}}_4}} }} }{\tilde {\cal A}_4}{\bar\chi}^{\bar {\cal A}_4}\tilde{\lambda^{0}}
 \equiv\tilde{f}\left({\cal V}^{-\frac{3}{5}}\right).
\eeq
 To determine the contribution of effective charge $Q'_i$, we need to evaluate the contribution of sfermion(${\tilde f_i}$)-photon($\gamma$)-sfermion({$\tilde f_i$}) vertices which are expressed  in terms of ${\tilde f_{L/R}}$ basis as below:
\beqn
\label{eq:Ce1e1gamma}
&& C_{{\tilde {f_1}}{\tilde {f_1}}{\gamma}}\sim D_{f_{11}} D_{f_{11}}^*C_{{\tilde {f_L}}{\tilde {f_L}^*}{\gamma}}+ (D_{f_{11}} D_{f_{12}}^*+ D_{f_{12}} D_{f_{12}}^*) C_{{\tilde {f_L}}{\tilde {f_R}^*}{\gamma}}+ D_{f_{12}} D_{f_{12}}^*C_{{\tilde {f_R}}{\tilde {f_R}^*}{\gamma}}, \nonumber\\
&& C_{{\tilde {f_2}}{\tilde {f_2}}{\gamma}}\sim D_{f_{21}} D_{f_{21}}^*C_{{\tilde {f_L}}{\tilde {f_L}^*}{\gamma}}+ ( D_{f_{21}} D_{f_{22}}^*+ D_{f_{22}} D_{f_{21}}^*)C_{{\tilde {f_L}}{\tilde {f_R}^*}{\gamma}}+ D_{f_{22}} D_{f_{22}}^*C_{{\tilde {f_R}}{\tilde {f_R}^*}{\gamma}}.
\eeqn
The sfermion-sfermion-photon vertex can be evaluated from the bulk kinetic term in the ${\cal N}=1$ gauged supergravity action as given below:
 \beqn
 {\cal L}= {\frac{1}{\kappa_4^2{\cal V}^2}}G^{T_B{\bar T}_B}\tilde{\bigtriangledown}_\mu T_B\tilde{\bigtriangledown}^\mu {\bar T}_{\bar B},
  \eeqn
  {where}
\begin{eqnarray}
\label{eq:sigmas_TB}
&& \tilde{\bigtriangledown}_\mu T_B = \partial_\mu T_B + 6i \kappa_4^2\mu_7 lQ_{T_B}A_{\mu}; \nonumber\\
& & T_B \sim  \sigma_B + \left( i\kappa_{B bc}c^b{\cal B}^c + \kappa_B + \frac{i}{(\tau - {\bar\tau})}\kappa_{B bc}{\cal G}^b({\cal G}^c
- {\bar {\cal G}}^c) i\delta^B_B\kappa_4^2\mu_7l^2C_B^{I{\bar J}}a_I{\bar a_{\bar J}} + \frac{3i}{4}\delta^B_B\tau Q_{\tilde{f}}\right.\nonumber\\
 & & \left.  + i\mu_3l^2(\omega_B)_{i{\bar j}} z^i\left({\bar z}^{\bar j}-\frac{i}{2}{\bar z}^{\tilde{a}}({\bar{\cal P}}_{\tilde{a}})^{\bar j}_lz^l\right)\right). \eeqn
The form of expression that eventually leads to give the contribution of required sfermion-sfermion-photon vertex is as given below:
\begin{eqnarray}
\label{eq:sq sq gl}
& &  C_{f_{L/R} f^{*}_{L/R} \gamma} \sim {\frac{6i\kappa_4^2\mu_72\pi\alpha^\prime Q_BG^{T_B{\bar T}_B}}{\kappa_4^2{\cal V}^2}}A^\mu\partial_\mu\left(\kappa_4^2\mu_7(2\pi\alpha^\prime)^2C_{i{\bar j}}{\cal A}_i{\bar {\cal A}}_{\bar j}\right)
\end{eqnarray}
Using $G^{T_B{\bar T}_B}(EW)\sim {\cal V}^{\frac{7}{3}}$(the large value is justified by obtaining ${\cal O}(1)$ SM fermion-fermion-photon coupling vertex in ${\cal N}=1$ gauged supergravity action; see details therein), $Q_B\sim {\cal V}^{\frac{1}{3}}{\tilde f}, \kappa_4^2\mu_7\sim \frac{1}{\cal V}$, the above expression reduces to $ |C_{f_{L/R} f^{*}_{L/R} \gamma}|\equiv  {\cal V}^{\frac{1}{3}}A^\mu\partial_\mu\left(\kappa_4^2\mu_7(2\pi\alpha^\prime)^2C_{i{\bar j}}{\tilde {\cal A}_i} {\tilde {\cal A}}_{\bar j}\right)$.
For $i=j=1$,  $\kappa_4^2\mu_7(2\pi\alpha^\prime)^2C_{1{\bar 1}}\sim {\cal V}^{\frac{10}{9}} $ as given in appendix of \cite{gravitino_DM}. Using the same
\beq
\label{eq:Celelgamma}
|C_{{\tilde {e_L}}{\tilde {e_L}^*}{\gamma}}|\equiv \frac{{\cal V}^{\frac{16}{9}} {\tilde f}} { \sqrt{\hat{K}_{{\cal A}_1{\bar {\cal A}}_1}\hat{K}_{{\cal A}_1{\bar {\cal A}}_1}} }\equiv ({\cal V}^{\frac{44}{45}}{\tilde f}){\tilde {\cal A}_1}  A^\mu\partial_\mu {\tilde {\cal A}_{1}}.
\eeq
For $i=1,j=3$;  $\kappa_4^2\mu_7(2\pi\alpha^\prime)^2C_{1{\bar 3}}\sim {\cal V}^{\frac{29}{18}}$, we have
\beq
\label{eq:CeleRgamma}
|C_{{\tilde {e_L}}{\tilde {e_R}^*}{\gamma}}|\equiv \frac{{\cal V}^{\frac{41}{18}}{\tilde f}} { \sqrt{\hat{K}_{{\cal A}_1{\bar {\cal A}}_1}\hat{K}_{{\cal A}_3{\bar {\cal A}}_3}} }\equiv ({\cal V}^{\frac{53}{45}}{\tilde f}){\tilde {\cal A}_1} A^\mu\partial_\mu {\tilde {\cal A}}_{3}.
\eeq
For $i=j=3$; $\kappa_4^2\mu_7(2\pi\alpha^\prime)^2C_{3{\bar 3}}\sim {\cal V}^{\frac{19}{9}}$ and
\beq
\label{eq:CeReRgamma}
|C_{{\tilde {e_R}}{\tilde {e_R}^*}{\gamma}}|\equiv \frac{{\cal V}^{\frac{25}{9}}{\tilde f}} { \sqrt{\hat{K}_{{\cal A}_3{\bar {\cal A}}_3}\hat{K}_{{\cal A}_3{\bar {\cal A}}_3}}}\equiv({\tilde f}{\cal V}^{\frac{62}{45}}){\tilde {\cal A}_3} A^\mu\partial_\mu {\tilde {\cal A}}_{3}.
\eeq
For i=j=2, $\kappa_4^2\mu_7(2\pi\alpha^\prime)^2C_{2{\bar 2}}\sim {\cal V}^{\frac{1}{9}}$ and
\beq
\label{eq:Cululgamma}
|C_{{\tilde {u_L}}{\tilde {u_L}^*}{\gamma}}|\equiv \frac{{\cal V}^{\frac{7}{9}}{\tilde f}}  { \sqrt{\hat{K}_{{\cal A}_2{\bar {\cal A}}_2}\hat{K}_{{\cal A}_2{\bar {\cal A}}_2}} }\equiv ({\tilde f}{\cal V}^{\frac{53}{45}}){\tilde {\cal A}_2} A^\mu\partial_\mu {\tilde {\cal A}}_{2}.
\eeq
For $i=2,j=4$,  $\kappa_4^2\mu_7(2\pi\alpha^\prime)^2C_{2{\bar 4}}\sim {\cal V}^{\frac{29}{18}}$ and
\beq
\label{eq:CuluRgamma}
|C_{{\tilde {u_L}}{\tilde {u_R}^*}{\gamma}}|\equiv \frac{{\cal V}^{\frac{41}{18}}{\tilde f}} { \sqrt{\hat{K}_{{\cal A}_2{\bar {\cal A}}_2}\hat{K}_{{\cal A}_4{\bar {\cal A}}_4}} }\equiv ({\tilde f}{\cal V}^{\frac{23}{18}}){\tilde {\cal A}_2} A^\mu\partial_\mu {\tilde {\cal A}}_{4}.
\eeq
For $i=j=4$, $\kappa_4^2\mu_7(2\pi\alpha^\prime)^2C_{4{\bar 4}}\sim {\cal V}^{\frac{28}{9}}$ and
\beq
\label{eq:CuRuRgamma}
|C_{{\tilde {u_R}}{\tilde {u_R}^*}{\gamma}}|\equiv \frac{ {\cal V}^{\frac{34}{9}}{\tilde f}} { \sqrt{\hat{K}_{{\cal A}_4{\bar {\cal A}}_4}\hat{K}_{{\cal A}_4{\bar {\cal A}}_4}}}\equiv({\tilde f}{\cal V}^{\frac{62}{45}}){\tilde {\cal A}_4} A^\mu\partial_\mu {\tilde {\cal A}}_{4}.
\eeq
Substituting the results given in eqs.~(\ref{eq:Celelgamma})-(\ref{eq:CuRuRgamma}) in equation~(\ref{eq:Ce1e1gamma}), the volume suppression factors corresponding to scalar-scalar-photon vertices are given as follows:
\beqn
\label{eq:Ce1e1gamma1}
&& C_{{\tilde {e_1}}{\tilde {e_1}}{\gamma}}\equiv {\tilde f}  \left( {\cal V}^{\frac{44}{45}}\cos^2  {\theta_e}  -  {\cal V}^{\frac{53}{45}} \cos {\theta_e} \sin {\theta_e}  (e^{i\phi_{e}} + e^{-i\phi_{e}})e^{i\phi_{g_e}} +  {\cal V}^{\frac{62}{45}}\sin^2  {\theta_e} \right) ,\nonumber\\
&&C_{{\tilde {e_2}}{\tilde {e_2}}{\gamma}}\equiv {\tilde f}  \left(  {\cal V}^{\frac{44}{45}} \sin^2  {\theta_e}  + {\cal V}^{\frac{53}{45}} \cos  {\theta_e} \sin {\theta_e}  (e^{i\phi_{e}} + e^{-i\phi_{e}}))e^{-i\phi_{g_e}}+ {\cal V}^{\frac{62}{45}} \cos^2 {\theta_e}\right), \nonumber\\
&& C_{{\tilde {u_1}}{\tilde {u_1}}{\gamma}}\equiv {\tilde f}  \left( {\cal V}^{\frac{53}{45}}\cos^2  {\theta_u}  -  {\cal V}^{\frac{23}{18}} \cos {\theta_u} \sin {\theta_e}  (e^{i\phi_{u}} + e^{-i\phi_{u}}))e^{i\phi_{g_u}}+  {\cal V}^{\frac{62}{45}}\sin^2  {\theta_u} \right),\nonumber\\
&&C_{{\tilde {u_2}}{\tilde {u_2}}{\gamma}}\equiv {\tilde f}  \left(  {\cal V}^{\frac{53}{45}} \sin^2  {\theta_u}  + {\cal V}^{\frac{23}{18}} \cos  {\theta_u} \sin {\theta_u}  (e^{i\phi_{u}} + e^{-i\phi_{u}}))e^{-i\phi_{g_u}} + {\cal V}^{\frac{62}{45}} \cos^2 {\theta_u}\right),
\eeqn
where $\phi_{g_e}$ and $\phi_{g_u}$ are phase factors associated with $C_{{\tilde {e_L}}{\tilde {e^{*}_R}}{\gamma}}$ and $C_{{\tilde {u_L}}{\tilde {u^{*}_R}}{\gamma}}$- we consider the same to be ${\cal O}(1).$
Now,  the Lagrangian relevant to couplings involved in one-loop diagram shown in Figure~1 is given as:
\begin{eqnarray}
\label{eq:LagLR}
{\cal L}= C_{f_L {\tilde f}^{*}_L {\tilde \lambda}^{0}_i} f_{L}\tilde{f_{L}}\tilde{\lambda^{0}}+C_{f_L {\tilde f}^{*}_R {\tilde \lambda}^{0}_i} f_{L}\tilde{f_{R}}\tilde{\lambda^{0}}+ C_{f^{*}_R {\tilde f}_L {\tilde \lambda}^{0}_i} f_{R}\tilde{f_{L}}\tilde{\lambda^{0}}+  C_{f^{*}_R {\tilde f}_R {\tilde \lambda}^{0}_i} f_{R}\tilde{f_{R}}\tilde{\lambda^{0}}.
\end{eqnarray}
where, from  eqs.~(\ref{eq:eLeLlambda})-(\ref{eq:uRuRlambda}), we have:
\beqn
\label{verticesgaugino}
&& | C_{e_L {\tilde e}^{*}_L {\tilde \lambda}^{0}_i}|\equiv{\tilde f} {\cal V}^{-1}, |C_{e_R {\tilde e}^{*}_R {\tilde \lambda}^{0}_i}|\equiv{\tilde f}{\cal V}^{-\frac{3}{5}}, |C_{e^{*}_R {\tilde e}_L {\tilde \lambda}^{0}_i}| \equiv |C_{e^{*}_L {\tilde e}_R {\tilde \lambda}^{0}_i}|\equiv{\tilde f}{\cal V}^{-\frac{15}{9}} \nonumber\\
&& |C_{u_L {\tilde u}^{*}_L {\tilde \lambda}^{0}_i}|\equiv{\tilde f} {\cal V}^{-\frac{4}{5}}, |C_{u_R {\tilde u}^{*}_R {\tilde \lambda}^{0}_i}|\equiv{\tilde f}{\cal V}^{-\frac{3}{5}}, |C_{u^{*}_R {\tilde u}_L {\tilde \lambda}^{0}_i}| \equiv |C_{u^{*}_L {\tilde u}_R {\tilde \lambda}^{0}_i}|\equiv{\tilde f}{\cal V}^{-\frac{14}{9}}.
\eeqn
Writing $\tilde{f}_L$ as well as $\tilde{f}_R$ given in equation (\ref{eq:LagLR}) in terms of diagonalized basis $\tilde{f}_1$ and $\tilde{f}_2$, the equation takes the form as of equation (\ref{eq:eff1loop2}):
\beqn
\label{eq:eff1loopg}
&& {\cal L}_{int}= \bar{\chi}_{f}
                \Bigl(( C_{{\lambda}^0_i {f_L}\tilde {f_L}} D_{f_{11}}+ C_{{\lambda}^0_i {f_L}\tilde {f_R}}  D_{f_{21}}) \PR +(C_{{\lambda}^0_i {f_R}\tilde {f_L}}  D_{f_{11}}+ C_{{\lambda}^0_i {f_R}\tilde {f_R}} D_{f_{21}}) \PL\Bigr) {\phi_{f_1}}{\tilde \lambda^0}\nonumber\\
                &&{\hskip -0.4in} +  \bar{\chi}_{f}
                \Bigl((C_{{\lambda}^0_i {f_L}\tilde {f_L}}  D_{f_{12}}+ C^{\chi^0_i {f_L}\tilde {f_R}} D_{f_{22}}) \PR +( C_{{\lambda}^0_i {f_R}\tilde {f_L}} D_{f_{12}}+ C_{{\lambda}^0_i {f_R}\tilde {f_R}}D_{f_{22}}) \PL \Bigr)  {\phi_{f_2}}{\tilde \lambda^0}+ h.c..
\eeqn
Using equation (\ref{eq:EDM}), the EDM expression will take the form:
\beqn
&& \frac{d_f}{e}|_{\lambda^{0}_i}=  \frac{m_{\tilde{\lambda}_i^0}}{{(4\pi)^2 }} \Bigl[ \frac{1}{m^{2}_{\tilde {f_1}}} {\rm Im}\left(C_{{\lambda}^0_i {f_L}\tilde {f_L}}C_{{\lambda}^0_i {f_R}\tilde {f_R}} D_{f_{11}} D_{f_{21}}^*+ C_{{\lambda}^0_i {f_L}\tilde {f_R}}C_{\chi^0_i {f_R}\tilde {f_L}} D_{f_{21}}D_{f_{11}}^*\right) Q'_{\tilde {f_1}} B\Bigl(\frac{m_{\tilde{\lambda}_i^0}^2}{m_{{\tilde f_1}^2}}\Bigr)\nonumber\\
&&+  {\frac{1}{m^{2}_{\tilde {f_2}}}} {\rm Im}\left(C_{{\lambda}^0_i {f_L}\tilde {f_L}}C_{\chi^0_i {f_R}\tilde {f_R}}  D_{f_{12}}D_{f_{22}}^*+ C_{{\lambda}^0_i {f_L}\tilde {f_R}}C_{{\lambda}^0_i {f_R}\tilde {f_L}} D_{f_{22}}D_{f_{12}}^*\right)  Q'_{\tilde {f_2}} B\Bigl(\frac{m_{\tilde{\lambda}_i^0}^2}{m_{{\tilde f_2}^2}}\Bigr)\Bigr].
\eeqn
Considering $f_{L,R}=e_{L,R}$, incorporating results of interaction vertices as given in equation (\ref{verticesgaugino}) and using the assumption that phase factors associated with effective gauge couplings are ${\cal O}(1)$, the dominant contribution of electron EDM is given as:
\beqn
&& {\hskip -0.2in} \frac{d_e}{e}|_{\lambda^{0}_i}\equiv \frac{m_{\tilde \lambda^0} ({\tilde f}^2 {\cal V}^{-\frac{8}{5}}sin {\theta_e} \sin  {\phi_e})}{(4\pi)^2} \left[\frac{C_{{\tilde {e_2}}{\tilde {e_2}^*}{\gamma}}}{m_{\tilde {e_2}}^2}B\left(\frac{m_{\tilde {\lambda^0}}^2}{m_{{\tilde e_2}^2}}\right)-\frac{C_{{\tilde {e_1}}{\tilde {e_1}^*}{\gamma}}}{m_{\tilde {e_1}}^2}B\left(\frac{m_{\tilde {\lambda^0}}^2}{m_{{\tilde e_1}^2}}\right) \right].
\eeqn
For $f_{L,R}=u_{L,R}$, quark EDM will be given as:
\beqn
&& {\hskip -0.2in}  \frac{d_u}{e}|_{\lambda^{0}_i}\equiv \frac{m_{\tilde \lambda^0} ({\tilde f}^2 {\cal V}^{-\frac{7}{5}}sin {\theta_u} \sin  {\phi_u})}{(4\pi)^2} \left[\frac{C_{{\tilde {u_2}}{\tilde {u_2}^*}{\gamma}}}{m_{\tilde {u_2}}^2}B\left(\frac{m_{\tilde {\lambda^0}}^2}{m_{{\tilde u_2}^2}}\right)-\frac{C_{{\tilde {u_1}}{\tilde {u_1}^*}{\gamma}}}{m_{\tilde {u_1}}^2}B\left(\frac{m_{\tilde {\lambda^0}}^2}{m_{{\tilde u_1}^2}}\right) \right].
\eeqn
Putting the values \footnote{We only incorporate the volume suppression coming from $ C_{{\tilde {e_i}}{\tilde {e_i}^*}{\gamma}}$ and $C_{{\tilde {u_i}}{\tilde {u_i}^*}{\gamma}}$. The momentum dependence of both vertices have already been included in the one-loop functions $A(r)$ and $B(r)$.}  of $C_{{\tilde {e_i}}{\tilde {e_i}^*}{\gamma}}$  and $C_{{\tilde {u_i}}{\tilde {u_i}^*}{\gamma}}$ as given in equation no (\ref{eq:Ce1e1gamma1}), we get
\beqn
\label{eq:delambda}
&&  {\hskip -0.25in}\frac{d_e}{e}|_{\lambda^0}\equiv \frac{m_{\tilde \lambda^0}({\tilde f}^2 {\cal V}^{-\frac{8}{5}}sin {\theta_e} \sin  {\phi_e}) }{(4\pi)^2 }{\cal V}^{\frac{62}{45}}{\tilde f} \left[\frac{\cos^2 {\theta_e}}{m_{\tilde {e_2}}^2}B\left(\frac{m_{\tilde {\lambda^0}}^2}{m_{{\tilde e_2}^2}}\right)-\frac{\sin^2 {\theta_e}}{m_{\tilde {e_1}}^2}B\left(\frac{m_{\tilde {\lambda^0}}^2}{m_{{\tilde e_1}^2}}\right) \right], \nonumber\\
&&  {\hskip -0.25in}{\rm and}\nonumber\\
&&  {\hskip -0.25in}\frac{d_u}{e}|_{\lambda^0}\equiv \frac{m_{\tilde \lambda^0}({\tilde f}^2 {\cal V}^{-\frac{7}{5}} sin {\theta_u} \sin  {\phi_u}) }{(4\pi)^2 }{\cal V}^{\frac{62}{45}}{\tilde f} \left[\frac{\cos^2 {\theta_u}}{m_{\tilde {u_2}}^2}B\left(\frac{m_{\tilde {\lambda^0}}^2}{m_{{\tilde u_2}^2}}\right)-\frac{\sin^2 {\theta_u}}{m_{\tilde {u_1}}^2}B\left(\frac{m_{\tilde {\lambda^0}}^2}{m_{{\tilde u_1}^2}}\right) \right].
\eeqn
 Here, $\sin \theta_f=
\frac{2|M_{\tilde{f_{21}^2}}|}{\sqrt{\left(M_{\tilde{f_L}}^2-M_{\tilde{f_R}}^2\right)^2+4 M_{\tilde{f}21}^4}}$.  As discussed in section {\bf 2}, in our model, we have $M_{\tilde{e_L}}^2= M_{\tilde{e_R}}^2= M_{\tilde{u_L}}^2= M_{\tilde{u_R}}^2 \sim {\cal V}m^{2}_{\frac{3}{2}}$. Using the same, we get $\sin \theta_e=\sin \theta_u=1$.  Also, we assume $\phi_{e,u}= (0,\frac{\pi}{2}]$. As explained in section {{\bf 3}},  ${m^{2}_{\tilde {f_1}}}={m^{2}_{\tilde {f_2}}}= m_{\tilde{f_L}/\tilde{f_R}}^2 ={\cal V} m^{2}_{\frac{3}{2}}$. Utilizing the same and the value of gaugino mass $m_{\tilde {\lambda^0}}^2= {\cal V}^{\frac{4}{3}}m^{2}_{\frac{2}{3}}$, we get:
\beqn
 {\hskip -0.2in}B\Bigl(\frac{m_{\tilde {\lambda^0}}^2}{m_{{\tilde f_i}^2}}\Bigr)=\frac{1}{2\Bigl(\frac{m_{\tilde {\lambda^0}}^2}{m_{{\tilde f_i}^2}}-1\Bigr)^2}\Bigl(1+\frac{m_{\tilde {\lambda^0}}^2}{m_{{\tilde f_i}^2}}+\frac{2\frac{m_{\tilde {\lambda^0}}^2}{m_{{\tilde f_i}^2}}ln\Bigl(\frac{m_{\tilde {\lambda^0}}^2}{m_{{\tilde f_i}^2}}\Bigr)}{1-\frac{m_{\tilde {\lambda^0}}^2}{m_{{\tilde f_i}^2}}}\Bigr) \sim \frac{m_{{\tilde f_i}^2}}{m_{\tilde {\lambda^0}}^2}\sim~{\cal V}^{-\frac{1}{3}},
\eeqn
where for $i=1,2$, $f_i=({e_1},{e_2}),({u_1},{u_2})$. Incorporating the value of masses in equation (\ref{eq:delambda}), using ${\tilde f}\sim {\cal V}^{-\frac{23}{30}}$ as obtained in \cite{Dhuria+Misra_mu_Split SUSY}, and Calabi-Yau volume ${\cal V}\sim 10^5$, the dominant contribution of EDM of electron will be given as:
\beqn
{\hskip -0.2in} \frac{d_e}{e}|_{\lambda^0}\equiv \frac{{\cal V}^{\frac{2}{3}}m_{\frac{3}{2}}\Bigl({\tilde f}^2 {\cal V}^{-\frac{8}{5}} \Bigr)}{(4\pi)^2 } \times{\tilde f} {\cal V}^{\frac{62}{45}}\Bigl( \frac{ {\cal V}^{-\frac{1}{3}}}{{\cal V} m^{2}_{\frac{3}{2}}} \Bigr) \equiv \frac{{\tilde f}^3 {\cal V}^{\frac{2}{3}+ \frac{62}{45}-\frac{8}{5}-\frac{1}{3}-1}}{(4\pi)^2 m_{\frac{3}{2}}} \equiv 10^{-39} cm,
\eeqn
and the dominant contribution of EDM of neutron/quark will be given as:
\beqn
{\hskip -0.2in} \frac{d_n}{e}|_{\lambda^0}\equiv \frac{{\cal V}^{\frac{2}{3}}m_{\frac{3}{2}}\left({\tilde f}^2 {\cal V}^{-\frac{7}{5}} \right)}{(4\pi)^2 } \times{\tilde f} {\cal V}^{\frac{62}{45}}\left( \frac{ {\cal V}^{-\frac{1}{3}}}{{\cal V} m^{2}_{\frac{3}{2}}} \right) \equiv \frac{{\tilde f}^3 {\cal V}^{\frac{2}{3}+ \frac{62}{45}-\frac{7}{5}-\frac{1}{3}-1}}{(4\pi)^2 m_{\frac{3}{2}}} \equiv 10^{-38} cm.
\eeqn
\\
{\bf{Neutralino contribution:}}   The physical eigenstates of neutralino mass matrix in the context of ${\cal N}$=1 gauged supergravity action are given as \cite{gravitino_DM}:
  \begin{eqnarray}
\label{eq:neutralinos_I}
& & \tilde{\chi}_1^0\sim\frac{-\tilde{H}^{0}_u+\tilde{H}^{0}_d}{\sqrt{2}};\  m_{{\chi}_1^0} \sim{\cal V}^{\frac{59}{72}} m_{\frac{3}{2}},\nonumber\\
& & \tilde{\chi}_2^0\sim \left(\frac{v}{M_P}{\tilde f}{\cal V}^{\frac{5}{6}}\right)\lambda^0+\frac{\tilde{H}^{0}_u+\tilde{H}^{0}_d}{\sqrt{2}};\ m_{{\chi}_2^0}\sim{\cal V}^{\frac{59}{72}} m_{\frac{3}{2}}, \nonumber\\
&&\tilde{\chi}_3^0\sim - \lambda^0+\left(\frac{v}{M_P}{\tilde f}{\cal V}^{\frac{5}{6}}\right) \left(\tilde{H}^{0}_u+\tilde{H}^{0}_d\right); \ m_{{\chi}_3^0} \sim
{\cal V}^{\frac{2}{3}} m_{\frac{3}{2}}.
\end{eqnarray}
where v is value of Higgs VEV at electroweak scale. $\tilde{H}^{0}_u$ and $\tilde{H}^{0}_d$ correspond to SU(2)-doublet higgsino. $\tilde{\chi}_1^0$ is purely a higgsino and $\tilde{\chi}_2^0$ ($\tilde{\chi}_3^0$) are formed by linear combination of gaugino (higgsino) with a very small admixture of higgsino (gaugino). Since  neutralinos are also very heavy, we evaluate the contribution of the same to one-loop electron/neutron EDM involving heavy sfermions. Though the neutralino ($\chi^{0}_{1,2}$)-fermion-sfermion couplings are complex in this case, the phase disappears due to presence of both complex coupling as well as its conjugate in the EDM expression. Therefore, the non-zero EDM arises due to CP violating phases appearing from mass eignstates of sfermion mass matrix only.  The one-loop diagram is given in Figure~2.

We have already calculated the contribution of gaugino-lepton(quark)-slepton(quark) vertices in the {{\bf subsection 4.1}}. Now  we estimate of coefficients of vertices corresponding to  higgsino-lepton(quark)-slepton(squark) interaction vertices.
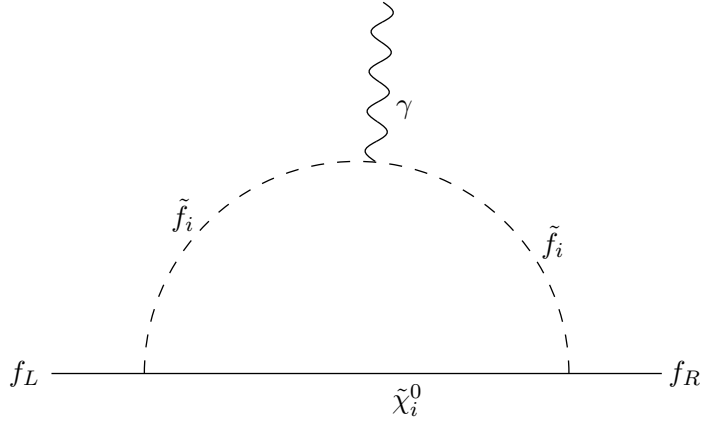
\begin{figure}
\begin{center}
\begin{picture}(145,127) (130,40)
   \Line(100,50)(330,50)
   \DashCArc(215,50)(80,0,180){5}
   \Photon(222,130)(225,190){4}{4}
    \Text(234,150)[]{{$\gamma$}}
   \Text(150,110)[]{${\tilde f}_{i}$}
   \Text(290,100)[]{${\tilde f}_{i}$}
   \Text(90,50)[]{{$f_{L}$}}
   \Text(340,50)[]{{$f_{R}$}}
   \Text(235,40)[]{{{$\tilde{\chi}_i^0$}}}
   \end{picture}
\end{center}
\caption{One-loop diagram involving neutralino.}
 \end{figure}
In ${\cal N}=1$ gauged supergravity, higgsino-fermion-sfermion interaction is governed by \cite{Wess_Bagger}:
\beqn
&& {\hskip -0.3in} {\cal L}_{f-\tilde{f}-\tilde{H^{0}_i}}= \frac{e^{\frac{K}{2}}}{2}\left({\cal D}_iD_{\bar J}W\right)\chi^i_{L}{\chi^{j}}^{c}_{L}+  ig_{i{\bar J}}{\bar\chi}^{\bar J}\left[{\bar\sigma}\cdot\partial\chi^i+\Gamma^i_{Lk}{\bar\sigma}\cdot\partial a^L\chi^k
+\frac{1}{4}\left(\partial_{a_L}K{\bar\sigma}\cdot a_L - {\rm c.c.}\right)\chi^i\right].
\eeqn
Using this, we evaluate the coefficients of higgsino-lepton(quark)-slepton(squark) interaction vertices. For an incoming electron ($e^{-}_L$) interacting with slepton as well as neutralino, the contribution of higgsino-lepton ($e^{-}_L$)-slepton (${\tilde e_L}$) vertex in gauged supergravity action of Wess and Bagger \cite{Wess_Bagger}, is given by: $\frac{e^{\frac{K}{2}}}{2}\left({\cal D}_{{\cal Z}_1}D_{\bar {\cal A}_1}W\right)\chi^{{\cal A}_i}{\chi^{c^{{\cal Z}_I}}} +ig_{{\bar I}{ {{\cal A}_1}}}{\bar\chi}^{Z_i}\left[{\bar\sigma}\cdot\partial{{\chi^{{\cal A}_1}}}+\Gamma^{{\cal A}_1}_{{\cal A}_1{\bar A_1}}{\bar\sigma}\cdot\partial {\cal A}_1{\chi^{{\cal A}_1}}
+\frac{1}{4}\left(\partial_{{\cal A}_3}K{\bar\sigma}\cdot {\cal A}_1 - {\rm c.c.}\right){\chi^{{\cal {A}}_1}}\right]$, $\chi^{\cal {Z}}$ and $\chi^{c^{{\cal Z}_1}}$ correspond to $SU(2)_L$ higgsino and its charge conjugate, $\chi^{{\cal A}_1}$ corresponds to $SU(2)_L$ electron and ${\tilde {\cal A}_1}$ corresponds to left-handed slepton. In diagonalized set of basis, $g_{ I{\bar {\cal A}_1}}=0$. Since  $SU(2)$ EW symmetry is not conserved for higgsino-lepton-slepton vertex, therefore to calculate the contribution of same, we  generate a term of the type $e_L \tilde{e}_L{\tilde H^{c}}_L H_L$ wherein $e_L$ and $H_L$ are respectively the $SU(2)_L$ electron and Higgs doublets, $\tilde{l}_L$ is also an $SU(2)_L$ doublet and ${\tilde H^{c}}_L$ is $SU(2)_L$  higgsino doublet. After giving VEV to one of the Higgs doublet $H_L$, one gets required vertex. By considering $a_1\rightarrow a_1+{\cal V}^{-\frac{2}{9}}{M_P}$, and further picking up the component of $ {\cal D}_i D_{{\bar a}_1}W$ linear in $z_i$ as well as linear in fluctuation $ (a_1-{\cal V}^{-\frac{2}{9}}{M_P})$, we see that $e^{\frac{K}{2}}{\cal D}_i D_{{\bar a}_1}W \sim  {\cal V}^{-\frac{31}{18}} z_i (a_1-{\cal V}^{-\frac{2}{9}}{M_P})$. As was shown in \cite{gravitino_DM}, $ e^{\frac{K}{2}}{\cal D}_I D_{\bar {{\cal A}_1}}W \sim {\cal O}(1) e^{\frac{K}{2}}{\cal D}_i D_{{\bar a}_1}W$. Utilizing the same, magnitude of the physical  higgsino($\tilde {H^{c}_L}$)-lepton (${e_L}$)-slepton(${\tilde {e_L}}$) vertex after giving  VEV to ${\cal Z}_I$  will be given as :
\begin{eqnarray}
\label{eq:elelZi}
& &  |C_{{\tilde H^{c}}_L {e_L} \tilde {e_L}}|\equiv\frac{{\cal V}^{-\frac{31}{18}}\langle{\cal Z}_I\rangle}{{\sqrt{\hat{K}^{2}_{{\cal Z}_1{\bar{\cal Z}}_1}{\hat{K}_{{\cal A}_1{\bar {\cal A}}_1}}{\hat{K}_{{\cal A}_1{\bar {\cal A}}_1}}}}}{\tilde {\cal A}_1}\chi^{c^{{\cal Z}_I}}{\chi^{{\cal A}_1}} \equiv {\cal V}^{-\frac{3}{2}}.
\end{eqnarray}
The coefficient of  higgsino($\tilde {H^{c}_L}$)-lepton(${u_L}$)-slepton(${\tilde {u_L}}$) vertex can be determined by expanding ${\cal D}_I D_{\bar {{\cal A}_2}}W$ linear in ${\cal Z}_I$ as well as linear in fluctuation $ ({\cal A}_2-{\cal V}^{-\frac{1}{3}}{M_P})$. The magnitude of the same has already been calculated in \cite{gravitino_DM} and given as:
\begin{eqnarray}
\label{eq:ululZi}
& &  |C_{{\tilde H^{c}}_L {u_L} \tilde {u_L}}|\equiv\frac{{\cal V}^{-\frac{20}{9}}\langle{\cal Z}_i\rangle}{{\sqrt{\hat{K}^{2}_{{\cal Z}_1{\bar{\cal Z}}_1}{\hat{K}_{{\cal A}_2{\bar {\cal A}}_2}}{\hat{K}_{{\cal A}_2{\bar {\cal A}}_2}}}}}{\tilde {\cal A}_2}\chi^{c^{{\cal Z}_I}}{\chi^{{\cal A}_2}} \equiv {\cal V}^{-\frac{4}{5}}.
\end{eqnarray}
To determine the contribution of higgsino-lepton($e^{-}_L$)-slepton(${\tilde e_R}$) vertex, one needs to expand\\  $\frac{e^{\frac{K}{2}}}{2}\left({\cal D}_{{\cal Z}_I}D_{\bar {\cal A}_1}W\right) $ in the fluctuations linear in ${\cal A}_3$ about its stabilized value. Considering  $a_3\rightarrow a_3+{\cal V}^{-\frac{13}{18}}{M_P}$ and picking up the component of ${\cal D}_I D_{\bar {a_1}}W$ linear in $a_3$, we have: $e^{\frac{K}{2}}{\cal D}_i D_{{\cal A}_1}W\equiv e^{\frac{K}{2}}{\cal D}_I D_{\bar {a_1}}W\sim {\cal V}^{-\frac{43}{36}} (a_3-{\cal V}^{-\frac{13}{18}}{M_P})$ the contribution of physical higgsino($\tilde H^{c}_L$)-lepton(${e_L}$)-slepton(${\tilde {e_R}}$) vertex  will be given as :
\begin{eqnarray}
\label{eq:eleRZi}
& &  |C_{{\tilde H^{c}}_L {e_L} \tilde {e_R}}|\equiv\frac{{\cal V}^{-\frac{43}{36}} }{{\sqrt{{\hat{K}_{{\cal Z}_1{\bar {\cal Z}}_1}}{\hat{K}_{{\cal A}_1{\bar {\cal A}}_1}}{\hat{K}_{{\cal A}_3{\bar {\cal A}}_3}}}}}{\tilde {\cal A}_3}\chi^{Z_i}{\chi^{c^{{\cal A}_1}}}\equiv{\cal V}^{-\frac{9}{5}}.
\end{eqnarray}
Similarly, one can calculate higgsino($\tilde H^{c}_L$)-lepton (${u_L}$)-slepton(${\tilde {u_R}}$) vertex by expanding $\frac{e^{\frac{K}{2}}}{2}\left({\cal D}_{{\cal Z}_I}D_{\bar {\cal A}_2}W\right)$ in the fluctuations linear in ${\cal A}_4$ about its stabilized value. Considering  $a_4\rightarrow a_4+{\cal V}^{-\frac{11}{9}}{M_P}$ and picking up the component of ${\cal D}_i D_{\bar {a_1}}W$ linear in $a_4$, we have:  $\frac{e^{\frac{K}{2}}}{2}\left({\cal D}_{{\cal Z}_I}D_{\bar {\cal A}_2}W\right)\equiv e^{\frac{K}{2}}{\cal D}_i D_{\bar {a_1}}W\sim {\cal V}^{-\frac{43}{36}} (a_4-{\cal V}^{-\frac{11}{9}}{M_P})$, the coefficient of higgsino($\tilde H^{c}_L$)-lepton(${u_L}$)-slepton(${\tilde {u_R}}$) vertex will be given as:
\begin{eqnarray}
\label{eq:uluRZi}
& &  |C_{{\tilde H^{c}}_L {u_L} \tilde {u_R}}|\equiv\frac{{\cal V}^{-\frac{43}{36}} }{{\sqrt{{\hat{K}_{{\cal Z}_1{\bar {\cal Z}}_1}}{\hat{K}_{{\cal A}_2{\bar {\cal A}}_2}}{\hat{K}_{{\cal A}_4{\bar {\cal A}}_4}}}}}{\tilde {\cal A}_4}\chi^{Z_i}{\chi^{c^{{\cal A}_2}}} \equiv{\cal V}^{-\frac{5}{3}}.\end{eqnarray}
For an outgoing electron $e^{-}_R$ interacting with slepton as well as neutralino, the  contribution of  higgsino-lepton ($e^{-}_R$)-slepton(${\tilde e_L}$) vertex is given by expanding  $\frac{e^{\frac{K}{2}}}{2}\left({\cal D}_{{\cal Z}_1}D_{{\cal A}_3}W\right) $ linear in ${\cal A}_1$. Considering  $a_1\rightarrow a_1+{\cal V}^{-\frac{2}{9}}{M_P}$ and picking up the component of ${\cal D}_i D_{a_3}W$ linear in $a_1$, we have: $e^{\frac{K}{2}}{\cal D}_I D_{{\cal A}_3}W\equiv e^{\frac{K}{2}}{\cal D}_i D_{a_3}W \sim {\cal V}^{-\frac{43}{36}} (a_1-{\cal V}^{-\frac{2}{9}}{M_P})$. The contribution of physical  higgsino($\tilde H^{c}_L$)-lepton (${e_R}$)-slepton(${\tilde {e_L}}$) vertex  will be given as :
\begin{eqnarray}
\label{eq:eRelZi}
& &  |C_{{\tilde H^{c}}_L {e_R} \tilde {e_L}}|\equiv\frac{{\cal V}^{-\frac{37}{72}}}{{\sqrt{\hat{K}_{{\cal Z}_1{\bar{\cal Z}}_1}{\hat{K}_{{\cal A}_1{\bar {\cal A}}_1}}{\hat{K}_{{\cal A}_3{\bar {\cal A}}_3}}}}}{\tilde {\cal A}_1}\chi^{Z_i}{\chi^{c^{{\cal A}_3}}} \equiv {\cal V}^{-\frac{9}{5}}.\end{eqnarray}
Similarly, considering  $a_4\rightarrow a_4+{\cal V}^{-\frac{11}{9}}{M_P}$ and picking up the component of above term linear in $a_1$, we have, $e^{\frac{K}{2}}{\cal D}_I D_{{\cal A}_4}W\equiv e^{\frac{K}{2}}{\cal D}_i D_{a_4}W\sim {\cal V}^{-\frac{43}{36}} (a_2-{\cal V}^{-\frac{1}{3}}{M_P})$, the contribution of physical  higgsino($\tilde H_L$)-quark (${u_R}$)-squark(${\tilde {u_L}}$) vertex is given as :
\begin{eqnarray}
\label{eq:uRulZi}
& &  |C_{{\tilde H}_L {u_R} \tilde {u_L}}|\equiv\frac{{\cal V}^{-\frac{43}{36}}}{{\sqrt{\hat{K}_{{\cal Z}_1{\bar{\cal Z}}_1}{\hat{K}_{{\cal A}_2{\bar {\cal A}}_2}}{\hat{K}_{{\cal A}_4{\bar {\cal A}}_4}}}}}{\tilde A_2}\chi^{Z_i}{\chi^{c^{{\cal A}_4}}} \equiv {\cal V}^{-\frac{5}{3}}.
\end{eqnarray}
The higgsino-lepton($e^{-}_R$)-slepton(${\tilde e_R}$) vertex also does not possess
  $SU(2)$ EW symmetry. Therefore, to calculate the contribution of same, we  generate a term of the type $e_R \tilde{e}_R{\tilde H}_L H_L$, where $H_L$ is one of the $SU(2)_L$ Higgs doublets. Thereafter, we expand  $\frac{e^{\frac{K}{2}}}{2}\left({\cal D}_{{\cal Z}_1}D_{\bar {\cal A}_3}W\right)\chi^{{\cal Z}_i}{\chi^{ {{\cal A}}_3}}$ linear in ${\cal Z}_1$ and then linear in ${\cal A}_3$ about their stabilized VEV's. Considering  $a_3\rightarrow a_3+{\cal V}^{-\frac{13}{18}}{M_P}$ and further picking up the component linear in $z_i$ as well as linear in fluctuation $(a_3-{\cal V}^{-\frac{2}{9}}{M_P})$, we get: $ e^{\frac{K}{2}}{\cal D}_i D_{{\cal A}_3}W\equiv e^{\frac{K}{2}}{\cal D}_i D_{{\bar a}_3}W\sim {\cal V}^{-\frac{13}{18}} \langle z_i \rangle (a_3-{\cal V}^{-\frac{13}{18}}{M_P})$. The magnitude of physical higgsino($\tilde H_L$)-lepton (${e_R}$)-slepton(${\tilde {e_R}}$) vertex after giving VEV to ${\cal Z}_I$  is given as :
\begin{eqnarray}
\label{eq:eReRZi}
& &  |C_{{\tilde H}_L {e_R} \tilde {e_R}}| \equiv\frac{{\cal V}^{-\frac{13}{18}}\langle{\cal Z}_i\rangle}{{\sqrt{\hat{K}^{2}_{{\cal Z}_1{\bar{\cal Z}}_1}{\hat{K}_{{\cal A}_3{\bar {\cal A}}_3}}{\hat{K}_{{\cal A}_3{\bar {\cal A}}_3}}}}}{\tilde {\cal A}_3}\chi^{Z_I}{\chi^{ {\cal A}_3}} \equiv {\cal V}^{-\frac{10}{9}}.
\end{eqnarray}
 The contribution of higgsino-quark($u_R$)-squark(${\tilde u_R}$) vertex has already been evaluated in \cite{gravitino_DM} by expanding  $\frac{e^{\frac{K}{2}}}{2}\left({\cal D}_{{\cal Z}_I}D_{{\cal A}_4}W\right)\chi^{{\cal Z}_I}{\chi^{ {{\cal A}}_4}}$ in the fluctuations linear in ${\cal Z}_I$ as well as ${\cal A}_4$ about their stabilized VEV's. The magnitude of the same is given as:
 \begin{eqnarray}
\label{eq:uRuRZi}
& &  |C_{{\tilde H}_L {u_R} \tilde {u_R}}|\equiv\frac{{\cal V}^{\frac{5}{18}}\langle{\cal Z}_i\rangle}{{\sqrt{\hat{K}^{2}_{{\cal Z}_1{\bar{\cal Z}}_1}{\hat{K}_{{\cal A}_4{\bar {\cal A}}_4}}{\hat{K}_{{\cal A}_4{\bar {\cal A}}_4}}}}}{\tilde {\cal A}_4}\chi^{Z_I}{\chi^{ {\cal A}_4}} \equiv {\cal V}^{-\frac{10}{9}}.
\end{eqnarray}
 The results of coefficients of both slepton(squark)-lepton(quark)-higgsino as  given in set of eqs.~(\ref{eq:elelZi})-(\ref{eq:uRuRZi}) are as follows:
\beqn
\label{eq:Higgs+gaug}
&& |C_{{\tilde H^{c}}_L {e_L} \tilde {e_L}}|\equiv {\cal V}^{-\frac{3}{2}},|C_{{\tilde H^{c}}_L {e_L} \tilde {e_R}}| \equiv |C_{{\tilde H}_L {e_R} \tilde {e_L}}|\equiv {\cal V}^{-\frac{9}{5}}, |C_{{\tilde H}_L {e_R} \tilde {e_R}}|\equiv {\cal V}^{-\frac{10}{9}},\nonumber\\
&& |C_{{\tilde H^{c}}_L {u_L} \tilde {u_L}}|\equiv {\cal V}^{-\frac{4}{5}}, |C_{{\tilde H^{c}}_L {u_L} \tilde {u_R}}| \equiv |C_{{\tilde H}_L {u_R} \tilde {u_L}}|\equiv{\cal V}^{-\frac{5}{3}}, |C_{{\tilde H}_L {u_R}|\tilde {u_R}}|\equiv {\cal V}^{-\frac{10}{9}}.
\eeqn
Utilizing the aforementioned results and the results of various gaugino-fermion-sfermion vertices as given in equation (\ref{verticesgaugino}), and  by adding the contribution of same as according to equation (\ref{eq:neutralinos_I}), the volume suppression factors coming from the neutralino-lepton-slepton vertices are given as:
\beqn
\label{eq:chiLR}
&& |C_{\chi^0_1 {e_L}\tilde {e_L}}|=|C_{\chi^0_2 {e_L}\tilde {e_L}}|\equiv {\cal V}^{-\frac{3}{2}}, |C_{\chi^0_3 {e_L}\tilde {e_L}}|\equiv {\tilde f}{\cal V}^{-1}, |C_{\chi^0_1 {e_L}\tilde {e_R}}|=|C_{\chi^0_2 {e_L}\tilde {e_R}}|\equiv {\cal V}^{-\frac{9}{5}},  \nonumber\\
&&|C_{\chi^0_3 {e_L}\tilde {e_R}}|\equiv  {\tilde f} {\cal V}^{-\frac{15}{9}}, | C_{\chi^0_1 {e_R}\tilde {e_L}}|=|C_{\chi^0_2 {e_R}\tilde {e_L}}|\equiv {\cal V}^{-\frac{9}{5}}, |C_{\chi^0_3 {e_R}\tilde {e_L}}|\equiv{\tilde f} {\cal V}^{-\frac{15}{9}}, \nonumber\\
&& | C_{\chi^0_1 {e_R}\tilde {e_R}}|=|C_{\chi^0_2 {e_R}\tilde {e_R}}|\equiv {\cal V}^{-\frac{10}{9}}, |C_{\chi^0_3 {e_R}\tilde {e_R}}|\equiv {\tilde f} {\cal V}^{-\frac{3}{5}}.
\eeqn
The volume suppression factors coming from the neutralino-quark-squark vertices are given as:
\beqn
\label{eq:chiLRu}
&& |C_{\chi^0_1 {u_L}\tilde {u_L}}|=|C_{\chi^0_2 {u_L}\tilde {u_L}}|\equiv {\cal V}^{-\frac{4}{5}}, |C_{\chi^0_3 {u_L}\tilde {u_L}}|\equiv {\tilde f}{\cal V}^{-\frac{4}{5}}, |C_{\chi^0_1 {u_L}\tilde {u_R}}|=|C_{\chi^0_2 {u_L}\tilde {u_R}}|\equiv{\cal V}^{-\frac{5}{3}}, \nonumber\\
&&|C_{\chi^0_3 {u_L}\tilde {u_R}}|\equiv  {\tilde f} {\cal V}^{-\frac{14}{9}}  |C_{\chi^0_1 {u_R}\tilde {u_L}}|=|C_{\chi^0_2 {u_R}\tilde {u_L}}|\equiv {\cal V}^{-\frac{5}{3}}, |C_{\chi^0_3 {u_R}\tilde {u_L}}|\equiv {\tilde f} {\cal V}^{-\frac{14}{9}}, \nonumber\\
&&| C_{\chi^0_1 {u_R}\tilde {u_R}}|=|C_{\chi^0_2 {u_R}\tilde {u_R}}|\equiv {\cal V}^{-\frac{10}{9}}, |C_{\chi^0_3 {u_R}\tilde {u_R}}|\equiv {\tilde f} {\cal V}^{-\frac{3}{5}}.
\eeqn
The interaction Lagrangian governing the neutralino-slepton(squark)-lepton(quark) interaction can be written as:
\begin{eqnarray}
\label{eq:LagLR1}
{\cal L}= \sum_{i=1,3} C_{\chi^0_i {f_L}\tilde {f_L}} f_{L}\tilde{f_{L}}\chi^0_i + C_{\chi^0_i {f_L}\tilde {f_R}} f_{L}\tilde{f_{R}}\chi^0_i +C_{\chi^0_i {f_R}\tilde {f_L}} f_{R}\tilde{f_{L}}\chi^0_i +C_{\chi^0_i {f_R}\tilde {f_R}}f_{ R}\tilde{f_{R}}\chi^0_i,
\end{eqnarray}
 where f=(e,u). Rewriting $f_L$ as well as  $f_R$ in term of diagonalized basis states  $f_1$ and $f_2$,  the equation takes the form as of equation (\ref{eq:eff1loop2}):
\beqn
\label{eq:eff1loop1}
&& {\cal L}_{int}= \bar{\chi}_{f}
                \left(( C_{\chi^0_i {f_L}\tilde {f_L}} D_{f_{11}}+ C_{\chi^0_i {f_L}\tilde {f_R}}  D_{f_{21}}) \PR +(C_{\chi^0_i {f_R}\tilde {f_L}}  D_{f_{11}}+ C_{\chi^0_i {f_R}\tilde {f_R}} D_{f_{21}}) \PL\right) {\phi_{f_1}}\chi^0_i \nonumber\\
                && +\bar{\chi}_{f}
                \left((C_{\chi^0_i {f_L}\tilde {f_L}}  D_{f_{12}}+ C_{\chi^0_i {f_L}\tilde {f_R}} D_{f_{22}}) \PR +( C_{\chi^0_i {f_R}\tilde {f_L}} D_{f_{12}}+ C_{\chi^0_i {f_R}\tilde {f_R}}D_{f_{22}}) \PL \right)   {\phi_{f_2}}\chi^0_i.  \eeqn
Using equation (\ref{eq:EDM}), dipole moment contribution will follow:
\beqn
&&\frac{d_f}{e}|_{\chi_i}=   \sum_{i=1,3}\frac{m_{\tilde{\chi}_i^0}}{{(4\pi)^2 }} \Bigl[ \frac{1}{m^{2}_{\tilde {f_1}}} {\rm Im}\left(C_{\chi^0_i {f_L}\tilde {f_L}}C_{\chi^0_i {f_R}\tilde {f_R}} D_{f_{11}} D_{f_{21}}^*+ C_{\chi^0_i {f_L}\tilde {f_R}}C_{\chi^0_i {f_R}\tilde {f_L}} D_{f_{21}}D_{f_{11}}^*\right)  Q'_{\tilde {f_1}} B\Bigl(\frac{m_{\tilde{\chi}_i^0}^2}{m_{{\tilde f_1}^2}}\Bigr)\nonumber\\
&&+ {\frac{1}{m^{2}_{\tilde {f_2}}}} {\rm Im}\left(C_{\chi^0_i {f_L}\tilde {f_L}}C_{\chi^0_i {f_R}\tilde {f_R}}  D_{f_{12}}D_{f_{22}}^*+ C_{\chi^0_i {f_L}\tilde {f_R}}C_{\chi^0_i {f_R}\tilde {f_L}} D_{f_{22}}D_{f_{12}}^*\right)   Q'_{\tilde {f_2}} B\Bigl(\frac{m_{\tilde{\chi}_i^0}^2}{m_{{\tilde f_2}^2}}\Bigr)\Bigr].
\eeqn
 Using the  values of first generation scalar/slepton mass $m_{{\tilde f_1}} ={\cal V}^{\frac{1}{2}}m_{\frac{3}{2}}$ and $m_{\tilde{\chi}_1^0}=m_{\tilde{\chi}_2^0}={\cal V}^{\frac{59}{72}}m_{\frac{3}{2}}$, $m_{\tilde{\chi}_3^0}={\cal V}^{\frac{2}{3}}m_{\frac{3}{2}}$; one gets:
  \beqn
&& B\Bigl(\frac{m_{\tilde{\chi}_2^0}^2}{m_{{\tilde f_i}^2}}\Bigr)= B\Bigl(\frac{m_{\tilde{\chi}_1^0}^2}{m_{{\tilde f_i}^2}}\Bigr)=\frac{1}{2\Bigl(\frac{m_{\tilde{\chi}_1^0}^2}{m_{{\tilde f_i}^2}}-1\Bigr)^2}\Bigl(1+\frac{m_{\tilde{\chi}_1^0}^2}{m_{{\tilde f_i}^2}}+\frac{m_{\tilde{\chi}_1^0}^2}{m_{{\tilde f_i}^2}}ln \Bigl(\frac{m_{\tilde{\chi}_1^0}^2}{m_{{\tilde f_i}^2}}\Bigr)\Bigr)\Bigl({1-\frac{m_{\tilde{\chi}_1^0}^2}{m_{{\tilde f_i}^2}}}\Bigr)  \sim\frac{1}{{\cal V}^{\frac{23}{36}}}, \\
&& B\Bigl(\frac{m_{\tilde{\chi}_3^0}^2}{m_{{\tilde f_i}^2}}\Bigr)=\frac{1}{2\Bigl(\frac{m_{\tilde{\chi}_3^0}^2}{m_{{\tilde f_i}^2}}-1\Bigr)^2}\Bigl(1+\frac{m_{\tilde{\chi}_3^0}^2}{m_{{\tilde f_i}^2}}+\frac{m_{\tilde{\chi}_3^0}^2}{m_{{\tilde f_i}^2}}ln \Bigl(\frac{m_{\tilde{\chi}_3^0}^2}{m_{{\tilde f_i}^2}}\Bigr)\Bigr)\Bigl({1-\frac{m_{\tilde{\chi}_3^0}^2}{m_{{\tilde f_i}^2}}}\Bigr)    \sim \frac{m_{{\tilde f_i}^2}}{m_{\tilde{\chi}_3^0}^2}={\cal V}^{-\frac{1}{3}}.
\eeqn
Utilizing above and the results of $C_{\chi^0_i {e_{L/R}}\tilde {e_{L/R}}}$ as given in  equation (\ref{eq:chiLR}), and further simplifying,  dominant contribution of EDM of electron will be given as\footnote{We use the assumption that the complex phases appearing in effective Yukawa couplings are of O(1).}:
\beqn
&& \frac{d_e}{e}|_{\chi_i}\equiv \frac{{\cal V}^{\frac{59}{72}}m_{\frac{3}{2}}\left(  {\cal V}^{-\frac{8}{3}} \sin {\theta_e} \sin {\phi_e} \right)}{(4\pi)^2 {\cal V}^{\frac{23}{36}} } \left[\frac{C_{{\tilde {e_2}}{\tilde {e_2}^*}{\gamma}}}{m_{\tilde {e_2}}^2} -\frac{C_{{\tilde {e_1}}{\tilde {e_1}^*}{\gamma}}}{m_{\tilde {e_1}}^2} \right].
\eeqn
Similarly, using results of $C^{\chi^0_i {u_{L/R}}\tilde {u_{L/R}}}$ as given in  equation (\ref{eq:chiLRu}), the  dominant contribution of EDM of quark will be given as:
\beqn
&& \frac{d_u}{e}|_{\chi_i}\equiv \frac{{\cal V}^{\frac{59}{72}}m_{\frac{3}{2}}\left( {\cal V}^{-\frac{17}{9}} \sin {\theta_e} \sin {\phi_e} \right)}{(4\pi)^2  {\cal V}^{\frac{23}{36}}} \left[\frac{C_{{\tilde {u_2}}{\tilde {u_2}^*}{\gamma}}}{m_{\tilde {u_2}}^2} -\frac{C_{{\tilde {u_1}}{\tilde {u_1}^*}{\gamma}}}{m_{\tilde {u_1}}^2}  \right].
\eeqn
Incorporating the value of $ C_{{\tilde {e_i}}{\tilde {e_i}}{\gamma}}$ from equation no (\ref{eq:Ce1e1gamma1}), one gets
\beqn
&&\frac{d_e}{e}|_{\chi_i}\equiv \frac{{\cal V}^{\frac{59}{72}}m_{\frac{3}{2}}\left({\cal V}^{-\frac{8}{3}} \sin {\theta_e} \sin {\phi_e}\right)}{(4\pi)^2  {\cal V}^{\frac{23}{36}}}{\cal V}^{\frac{62}{45}}{\tilde f}\left[  \frac{\cos^2 {\theta_e}}{m_{\tilde{e_2}}^2} -\frac{\sin^2  {\theta_e} }{m_{\tilde {e_1}}^2} \right],\nonumber\\
&& {\hskip -0.4in}{\rm and}~~\frac{d_u}{e}|_{\chi_i}\equiv \frac{{\cal V}^{\frac{59}{72}}m_{\frac{3}{2}}\left({\cal V}^{-\frac{17}{9}} \sin {\theta_u} \sin {\phi_u}\right)}{(4\pi)^2  {\cal V}^{\frac{23}{36}} }{\cal V}^{\frac{62}{45}}{\tilde f}\left[  \frac{\cos^2 {\theta_u}}{m_{\tilde{u_2}}^2} -\frac{\sin^2  {\theta_u} }{m_{\tilde {u_1}}^2} \right].
\eeqn
Incorporating value of $\sin {\theta_{e}}=\sin {\theta_{u}}= 1$, $\sin{\phi_e}=\sin{\phi_u}= (0,1]$, ${\tilde f}\sim {\cal V}^{-\frac{23}{30}}$, and value of scalar masses $m_{{\tilde e_i}}=m_{{\tilde u_i}}= {\cal V}^{\frac{1}{2}}m_{\frac{3}{2}}$, the numerical value of EDM of electron for this case will be:
 \beqn
\hskip -0.2in\frac{d_e}{e}|_{\chi_i}\equiv \frac{{\cal V}^{\frac{59}{72}}m_{\frac{3}{2}}}{(4\pi)^2{\cal V}^{\frac{23}{36}}  }({\tilde f}{\cal V}^{\frac{62}{45}})\times{\cal V}^{-\frac{8}{3}}\left( \frac{1}{{\cal V} m^{2}_{\frac{3}{2}}}\right) \equiv \frac{{\tilde f} {\cal V}^{\frac{59}{72}+ \frac{62}{45}-\frac{8}{3}-\frac{23}{36}-1}}{(4\pi)^2 m_{\frac{3}{2}}} \equiv 10^{-37} cm.
\eeqn
and the numerical value of EDM of neutron/quark will be:
 \beqn
\hskip -0.2in \frac{d_n}{e}|_{\chi_i}\equiv \frac{{\cal V}^{\frac{59}{72}}m_{\frac{3}{2}}}{(4\pi)^2{\cal V}^{\frac{23}{36}}  }({\tilde f} {\cal V}^{\frac{62}{45}})\times{\cal V}^{-\frac{17}{9}}\left( \frac{1}{{\cal V} m^{2}_{\frac{3}{2}}}\right) \sim \frac{{\tilde f} {\cal V}^{\frac{59}{72}+ \frac{62}{45}-\frac{17}{9}-\frac{23}{36}-1}}{(4\pi)^2 m_{\frac{3}{2}}} \equiv 10^{-34} cm.
\eeqn

{{\bf R-parity violating vertices contribution:}}
 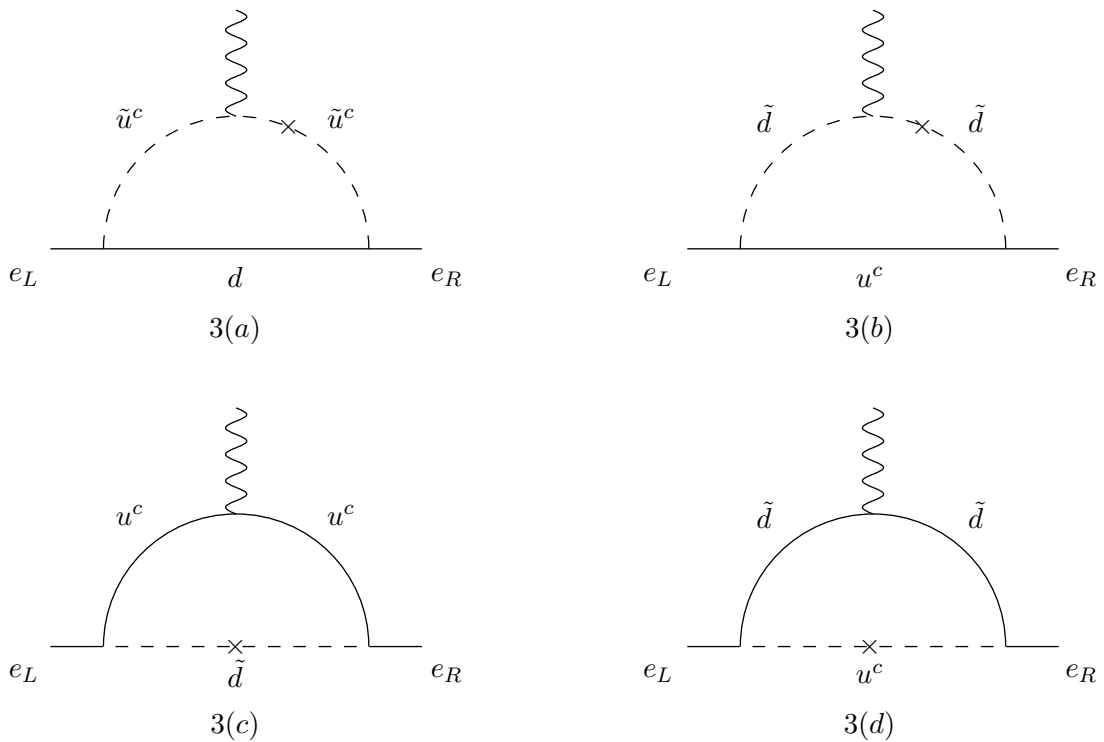
\begin{figure}
\begin{center}
\begin{picture}(250,270)(70,0)
    \Line(10,50)(25,50)
   \DashLine(25,50)(135,50){5}
    \Line(135,50)(150,50)
   \CArc(80,50)(50,0,180)
   \Photon(80,100)(80,140){4}{4}
   \Text(80,50)[]{$\times$}
   \Text(40,100)[]{$u^{c}$}
   \Text(120,100)[]{$u^{c}$}
   \Text(0,40)[]{{$e_{L}$}}
   \Text(160,40)[]{{$e_{R}$}}
   \Text(80,40)[]{{{${\tilde d}$}}}
   \Text(80,20)[]{$3(c)$}
     \Line(250,50)(265,50)
   \DashLine(265,50)(375,50){5}
    \Line(375,50)(390,50)
   \CArc(320,50)(50,0,180)
   \Photon(320,100)(320,140){4}{4}
   \Text(320,50)[]{$\times$}
    \Text(280,100)[]{${\tilde d}$}
   \Text(360,100)[]{${\tilde d} $}
   \Text(240,40)[]{{$e_{L}$}}
   \Text(400,40)[]{{$e_{R}$}}
   \Text(320,40)[]{{{$u^c$}}}
    \Text(320,20)[]{$3(d)$}
    \Line(10,200)(150,200)
   \DashCArc(80,200)(50,0,180){5}
   \Photon(80,250)(80,290){4}{4}
   \Text(100,247)[]{$\times$}
   \Text(40,250)[]{${\tilde u}^{c} $}
   \Text(120,250)[]{${\tilde u}^{c} $}
   \Text(0,190)[]{{$e_{L}$}}
   \Text(160,190)[]{{$e_{R}$}}
   \Text(80,190)[]{{{$d$}}}
   \Text(80,170)[]{$3(a)$}
   \Line(250,200)(390,200)
    \DashCArc(320,200)(50,0,180){5}
   \Photon(320,250)(320,290){4}{4}
   \Text(340,247)[]{$\times$}
    \Text(280,250)[]{${\tilde d}$}
   \Text(360,250)[]{${\tilde d} $}
   \Text(240,190)[]{{$e_{L}$}}
   \Text(400,190)[]{{$e_{R}$}}
   \Text(320,190)[]{{{$u^c$}}}
    \Text(320,170)[]{$3(b)$}
   \end{picture}
\end{center}
\caption{One-loop diagrams involving R-parity violating couplings.}
 \end{figure}
 We have explicitly taken into account the contribution of R-parity violating couplings in the context of ${\cal N}=1$ gauged supergravity limit of $\mu$-split SUSY set-up discussed in \cite{gravitino_DM}. Though one would certainly expect a very suppressed value of EDM because of presence of heavy sfermions as well as vanishing contribution of R-parity violating vertices, we discuss the effect of the same on EDM of electron/neutron  just to compare the order of magnitude  of EDM with respect to R-parity conserving loop diagrams. Though the R-parity violating interaction vertices are complex but due to presence of both R-parity violating vertex as well as its conjugate in the one-loop diagrams as given in Figure~3,  the complex phase  disappears and therefore, contribution of the same to EDM will vanish. However, as similar to neutralino and gaugino one-loop diagrams, the non-zero phase corresponding to CP-violating effect can appear only by considering the chirality flip between slepton(squark) fields appearing as a propagators in the one-loop diagram. Due to chirality flip, the matrix amplitude depends on the off-diagonal component of slepton(squark) mass matrix, the contribution of which further depends on complex trilinear coupling ${\cal A}_{IJK}$ as well as supersymmetric mass parameter $\mu$.

The one-loop Feynman diagrams for electron EDM mediated by R-parity violating interaction vertices are given in Figure~3. Using the analytical results as given in \cite{frank_hamidian} to get the numerical estimate of EDM of electron, we have
\beqn
\label{EDMRPV}
&& {\hskip -0.4in} \frac{d_e}{e}|_{RPV}=-|C_{e_{L}{\tilde u}^{c}_R d_L}|^2 C_{{\tilde u}^{c}_{R} {\tilde u^{c *}_{R}} \gamma}\ \frac{2 e}{3}\ |{{\cal A}_{u_j}}| \frac{m_{d_{k}}}{m^{3}_{\tilde u}}\ \sin {\theta_u}\ \sin {\phi_{A_u}}B(r_{d_k})-\nonumber\\
&& {\hskip -0.5in} |C_{e_{L}{\tilde d_L} u^{c}_R }|^2 C_{{\tilde d_L}^{*} \tilde d_L \gamma} \ \frac{ e}{3}\ |{{\cal A}_{d_j}}| \frac{m_{u_{k}}}{m^{3}_{\tilde d}}\ \sin {\theta_d}\ \sin {\phi_{A_d}}B(r_{u_k}) -|C_{e_{L}{\tilde u}^{c}_R d_L}|^2 C_{{ u}^{c}_R {u^{c *}_R} \gamma}\ \frac{2 e}{3}\ |{{\cal A}_{d_j}}| \frac{m_{u_{j}}}{m^{3}_{\tilde d}}\ \sin {\theta_d}\ \sin {\phi_{A_d}}A(r_{u_k}) \nonumber\\
&&  {\hskip -0.5in} -|C_{e_{L}{\tilde d_L} u^{c}_R }|^2 C_{{d}^{*}_L d_L \gamma} \ \frac{e}{3}\ |{{\cal A}_{u_j}}| \frac{m_{d_{j}}}{m^{3}_{\tilde u}}\ \sin {\theta_u}\ \sin {\phi_{A_u}}A(r_{d_k}).
\eeqn
where $r_{(u_k/d_k)}=  {m^{2}_{(u_k/d_k)}}/{m^{2}_{\tilde f}}$ and form of one-loop functions $A(r)$ and $B(r)$ is defined in (\ref{eq:A}).

One can draw the similar R-parity violating one-loop diagram to calculate quark EDM ($u$) by replacing $(e_L,e_R) \leftrightarrow(u_L,u_R)$ and ${\tilde u}^{c}\leftrightarrow {\tilde e}^{c}$. The analytical expression in case of quark EDM will  be of the following form:
\beqn
\label{EDMRPV}
&& {\hskip -0.4in} \frac{d_u}{e}|_{RPV}=-|C_{u_{L}{\tilde e}^{c}_R d_L}|^2 C_{{\tilde e}^{c}_{R} {\tilde e^{c *}_{R}} \gamma}\ e \ |{{\cal A}_{e_j}}| \frac{m_{d_{k}}}{m^{3}_{\tilde e}}\ \sin {\theta_e}\ \sin {\phi_{A_e}}B(r_{d_k})-\nonumber\\
&&  {\hskip -0.4in} |C_{u_{L}{\tilde d_L} e^{c}_R }|^2 C_{{\tilde d}^{*}_L \tilde d_L \gamma} \ \frac{ e}{3}\ |{{\cal A}_{d_j}}| \frac{m_{e_{k}}}{m^{3}_{\tilde d}}\ \sin {\theta_d}\ \sin {\phi_{A_d}}B(r_{e_k}) -|C_{u_{L}{\tilde e}^{c}_R d_L}|^2 C_{{ e}^{c}_{R} {e^{c *}_{R}} \gamma}\ \frac{ e}{3}\ |{{\cal A}_{d_j}}| \frac{m_{e_{j}}}{m^{3}_{\tilde d}}\ \sin {\theta_d}\ \sin {\phi_{A_d}}A(r_{e_k}) \nonumber\\
&& {\hskip -0.4in} -|C_{u_{L}{\tilde d_L} e^{c}_R }|^2 C_{{d}^{*}_L d_L \gamma} \ e \ |{{\cal A}_{e_j}}| \frac{m_{d_{j}}}{m^{3}_{\tilde e}}\ \sin {\theta_e}\ \sin {\phi_{A_u}}A(r_{d_k}).
\eeqn
The magnitude of the coefficient of interaction vertices $C_{e_{L}{\tilde u}^c d}$, $C_{e_{L}{\tilde d} u^c }$, $C_{u_{L}{\tilde e}^{c}_R d_L}$ and $C_{u_{L}{\tilde d_L} e^{c}_R }$  have already been obtained in \cite{gravitino_DM} and given as:
\beqn
C_{e_{L}{\tilde u}^{c}_R d_L} = C_{e_{L}{\tilde d}u^{c}_R }=C_{u_{L}{\tilde e}^{c}_R d_L}=C_{u_{L}{\tilde d_L} e^{c}_R } \equiv {\cal V}^{\frac{5}{3}}e^{i \phi_{y_\alpha}},
\eeqn
where $\phi_{y_\alpha}$ is the phase factor associated with complex R-parity violating interaction vertices.

The volume suppression factors coming from $C_{{\tilde u}^{c}_R {\tilde u^{c *}_R} \gamma}$, $C_{{\tilde e}^{c}_R {\tilde e^{c *}_R} \gamma}$ and  $C_{{\tilde d_L} {\tilde d_L}^{*}\gamma}$ vertices have already been obtained in the case of gaugino one-loop diagrams and given as:
\beqn
C_{{\tilde u}^{c}_R {\tilde u^{c *}_R} \gamma}= C_{{\tilde e}^{c}_R {\tilde e^{c *}_R} \gamma}\equiv {\cal V}^{\frac{62}{45}}{\tilde f}, C_{{\tilde d_L}{\tilde d_L}^{*}\gamma} \equiv {\cal V}^{\frac{53}{45}}{\tilde f}.
\eeqn
We set $C_{f f^{*}\gamma}|_{\rm EW}$ to be the charge of the quark $d_L$. The reason for the same is as follows. Consider the following kinetic-term-like term contributing to the quark-quark-photon vertex in ${\cal N}=1$ gauged supergravity action of Wess and Bagger: $g_{YM} g_{a_2 {\bar a}_2}\frac{{\bar\chi}^{{\bar a}_2}_L\Gamma^{a_2}_{T_B a_2}X^B\slashed{A}\chi^{a_2}}{\left(\sqrt{K_{{a_2}{\bar a}_2}}\right)^2}\in
\frac{g_{a_2{\bar a}_2}{\bar\chi}^{{\bar a}_2}\slashed{D}\chi^{a_2}}{\left(\sqrt{K_{a_2{\bar a}_2}}\right)^2}$. For the purpose of demonstrating the possibility of obtaining a SM-like quark-quark-photon coupling at the EW scale, let us assume that all moduli save $T_B, a_2$ have been stabilized at values indicated earlier and $\left(n^0_\beta\right)_{\rm max}\sim{\cal V}$ and consequently we take the K\"{a}hler potential to be:
\begin{eqnarray}
\label{K_TBa2}
K&\sim&-2 ln \Bigl[\left(T_B + {\bar T}_B - a_2\left\{C_{2{\bar 2}} {\bar a}_2 + C_{2{\bar 1}}\langle {\bar a}_1\rangle + C_{2{\bar 3}}\langle{\bar a}_3\rangle + C_{2{\bar 4}}\langle{\bar a}_4\rangle\right\} +  c.c. + {\cal V}^{\frac{2}{3}}\right)^{\frac{3}{2}} + {\cal V}\biggr]\nonumber\\
& \equiv & - 2 ln\biggl[\left(T_B + {\bar T}_B - C_{2{\bar 2}}|a_2|^2 - a_2{\bar \Sigma}_2 + h.c. + {\cal V}^{\frac{2}{3}}\right)^{\frac{3}{2}} + {\cal V}\Bigr].
\end{eqnarray}
Consider having frozen all moduli save $T_B$ and $a_2$. Then from:
$$\left(\begin{array}{cc}g_{T_B{\bar T}_B} & g_{T_B{\bar a}_2} \\ g_{a_2{\bar T}_B} & g_{a_2{\bar a}_2}\end{array}\right)^{-1}=\frac{1}{g_{T_B{\bar T}_B} g_{a_2{\bar a}_2} - |g_{T_B{\bar a}_2}|^2}\left(\begin{array}{cc} g_{a_2{\bar a}_2} & - g_{T_B{\bar a}_2} \\
- g_{a_2{\bar T_B}} & g_{T_B{\bar T}_B}\end{array}\right),$$ if $g_{T_B{\bar a}_2}|_{EW}$ is small such that
\begin{equation}
\label{metric_componets_small_large}
|g_{T_B{\bar a}_2}|^2_{EW} > g_{T_B{\bar T}_B} g_{a_2{\bar a}_2}|_{EW}
\end{equation}
 then:
\begin{eqnarray}
\label{inverse_EW}
& & g^{T_B{\bar T}_B}|_{EW}\sim \frac{g_{a_2{\bar a}_2}}{|g_{T_B{\bar a}_2}|^2},\
g^{a_2{\bar a}_2}|_{EW}\sim \frac{g_{T_B{\bar T}_B}}{|g_{T_B{\bar a}_2}|^2}|_{EW}, g^{T_B{\bar a}_2}|_{EW}\sim \frac{1}{g_{T_B{\bar a}_2}}|_{EW}\equiv{\rm large}.
\end{eqnarray}
Using (\ref{K_TBa2}), we evaluate  ${\bar\partial}_{{\bar a}_2}\partial_{a_2}K
, {\bar\partial}_{{\bar T}_B}\partial_{T_B}K, {\bar\partial}_{{\bar T}_B}\partial_{a_2}K \partial_{T_B}{\bar\partial}$ and ${\partial}_{{\bar T}_B}\partial_{a_2}K$. If ${\bar\partial}_{{\bar T}_B}\partial_{a_2}K|_{EW}\sim\delta<<1$ such that (\ref{metric_componets_small_large}) is satisfied then:
{
\begin{eqnarray}
\label{smallBbara2_EW}
&& {\hskip -1.0in} 3\langle\left(T_B + {\bar T}_B - C_{2{\bar 2}}|a_2|^2 - a_2{\bar \Sigma}_2 + h.c. + {\cal V}^{\frac{2}{3}}\right)\rangle_{EW}^{\frac{3}{2}}  \sim \langle\biggl[\left(T_B + {\bar T}_B - C_{2{\bar 2}}|a_2|^2 - a_2{\bar \Sigma}_2 + h.c. + {\cal V}^{\frac{2}{3}}\right)^{\frac{3}{2}} + {\cal V}\biggr]\rangle_{EW}.
\end{eqnarray}}
Using (\ref{smallBbara2_EW}), one sees that:
\begin{equation}
\label{dTBga2Tbbar}
\partial_{T_B}g_{a_2{\bar T}_B}|_{EW\ {\rm near}\ (\ref{smallBbara2_EW})}\sim\frac{9\left(C_{2{\bar 2}}{\bar a}_2 + {\bar\Sigma}_2\right)}{\biggl[\left(T_B + {\bar T}_B - C_{2{\bar 2}}|a_2|^2 - a_2{\bar \Sigma}_2 + h.c. + {\cal V}^{\frac{2}{3}}\right)^{\frac{3}{2}} + {\cal V}\biggr]^2}\sim {\cal V}^{-\frac{29}{18}},
\end{equation}
assuming $\langle a_{1,2,3,4}\rangle|_{EW}\sim{\cal O}(1)\langle a_{1,2,3,4}\rangle|_{M_s}$. If $g_{a_2{\bar a}_2}|_{M_s}\sim g_{a_2{\bar a}_2}|_{EW}\sim10^{-2}$, then from (\ref{metric_componets_small_large}), one sees:
\begin{equation}
\label{gBBbar_EW}
g_{T_B{\bar T_B}}|_{EW}\sim\delta^\prime<10^2\delta^2.
\end{equation}
Noting that:
\begin{equation}
\label{aff-con}
\Gamma^{a_2}_{T_ba_2}=\frac{g^{a_2{\bar T}_B}}{2}\left(\partial_{T_B}g_{a_2{\bar T}_B} +
\partial_{a_2}g_{T_B{\bar T}_B}\right) + \frac{g^{a_2{\bar a}_2}}{2}\left(\partial_{T_B}g_{a_2{\bar a}_2} + \partial_{a_2}g_{T_B{\bar a}_2}\right),
\end{equation}
we see that one can get a large contribution to (\ref{aff-con}) from $g^{a_2{\bar T}_B}\partial_{T_B}g_{a_2{\bar T}_B}|_{EW}$ given by:
\begin{equation}
\label{ginva2TbdTbga2Tb}
g^{a_2{\bar T}_B}\partial_{T_B}g_{a_2{\bar T}_B}|_{EW,\ {\cal V}\sim10^4}\sim\frac{10^{-6.5}}{\delta}.
\end{equation}
Let us look at implementation of (\ref{gBBbar_EW}) and its consequences. From above calculations, one notes that
(\ref{gBBbar_EW}) is identically satisfied if (\ref{smallBbara2_EW}) is satisfied. Consider working with
$\tau_{S,B},z^i,a_I,...$ instead of $T_{S,B},z^i,a_I,...$ having frozen $G^a$ and other open-string moduli. Noting then that:
\begin{equation}
K^{\alpha{\bar\beta}}=\frac{1}{K_{\tau_S{\bar\tau}_S}K_{\tau_B{\bar\tau}_B} - |K_{\tau_S{\bar\tau}_B}|^2}\left(\begin{array}{cc}K_{\tau_B{\bar\tau}_B} & - K_{\tau_B{\bar\tau}_S} \\
- K_{\tau_S{\bar\tau}_B} & K_{\tau_S{\bar\tau}_S}\end{array}
\right),
\end{equation}
and assuming $K_{\tau_B{\bar\tau}_B}|_{EW}\sim\delta^\prime<<1, K_{\tau_S{\bar\tau}_S}|_{M_s}\sim K_{\tau_S{\bar\tau}_S}|_{EW}\sim{\cal V}^{-1}, K_{\tau_S{\bar\tau}_B}|_{M_s}\sim
K_{\tau_S{\bar\tau}_B}|_{EW}\sim{\cal V}^{-\frac{5}{3}}$ implying $|K_{\tau_S{\bar\tau}_B}|^2>K_{\tau_S{\bar\tau}_S}K_{\tau_B{\bar\tau}_B}$, one obtains:
\begin{eqnarray}
\label{metric_bulk_EW}
& & {\hskip -0.5in} K^{\tau_S{\bar\tau}_S}|_{EW}\sim\frac{K_{\tau_B{\bar\tau}_B}|_{EW}}{|K_{\tau_S{\bar\tau}_B}|^2_{EW}}\sim\delta^\prime {\cal V}^{\frac{10}{3}},  K^{\tau_B{\bar\tau}_B}|_{EW}\sim\frac{K_{\tau_S{\bar\tau}_S}|_{EW}}{|K_{\tau_S{\bar\tau}_B}|^2_{EW}}\sim{\cal V}^{\frac{7}{3}},  K^{\tau_S{\bar\tau}_B}|_{EW}\sim\frac{1}{K_{\tau_S{\bar\tau}_B}|_{EW}}\sim{\cal V}^{\frac{5}{3}}.
\end{eqnarray}
Equation (\ref{metric_bulk_EW}) implies:
\begin{eqnarray}
\label{F-nonrenormalization}
& & {\hskip -0.5in} {\bar F}^{\tau_S}|_{M_s}\sim K^{\tau_S{\bar\tau}_S}D_{{\tau}_S}{W}  \sim {\bar F}^{\tau_S}|_{EW}=e^{\frac{K}{2}}\left(K^{\tau_S{\bar\tau}_S}D_{\tau_S}{ W} + K^{\tau_B{\bar\tau}_S}D_{{\tau}_B}{W}\right)\sim\left(\delta^\prime {\cal V}^{\frac{10}{3}} + {\cal V}\right)m_{3/2}\sim\frac{1}{\cal V}.
\end{eqnarray}
So, the $F^{\tau_S}$-term (potential $||F^{\tau_S}||^2$) is 1-loop RG-invariant! Further, the complete $F$-term potential:
\begin{eqnarray}
\label{F_term_1L_RG_inv}
& & {\hskip -0.5in} V|_{M_s}\sim e^K K^{\tau_S{\bar\tau}_S}|D_{\tau_S}W|^2\sim {\cal V}m_{3/2}^2    \sim e^K\left(K^{\tau_S{\bar\tau}_S}|D_{\tau_S}W|^2 + K^{\tau_B{\bar\tau}_B}|D_{\tau_B}W|^2 + K^{\tau_S{\bar\tau}_B}D_{\tau_S}D_{{\bar\tau}_B}{\bar W} + {\rm h.c.}\right)_{EW}\nonumber\\
&& \sim\left(\delta^\prime {\cal V}^{\frac{10}{3}} + {\cal V}\right)m_{3/2}^2\sim{\cal V}m_{3/2}^2,
 \end{eqnarray}
is also 1-loop RG-invariant. So, the quark-quark-photon vertex  can be made to be of ${\cal O}(1)$ for $\delta\sim10^{-13}$, i.e., one can hope that the coupling $C_{ff^{*}\gamma}\sim{\cal O}(1)$ for $f({\rm fermion})\equiv e, u$.

For $r_{(u_k/d_k/e_k)}=  {m^{2}_{(u_k/d_k/e_k)}}/{m^{2}_{\tilde f_i}}$, $A(r_{u_k/d_k/e_k})= B(r_{u_k/d_k/e_k})=1$.
As mentioned in equation (\ref{me21}), $|{\cal A}'_{e}|= |{\cal A}_{e}^{*}-{\mu}\cot{\beta}| \equiv {\cal V}m_{\frac{3}{2}}$, $|{\cal A}'_{u}|= |{\cal A}_{u}^{*}-{\mu}\cot{\beta}| \equiv {\cal V}m_{\frac{3}{2}}$. Using the these results and results of coefficient of interaction vertices as given above; considering $\sin{\phi_u}=\sin{\phi_d}= (0,1]$, $\sin{\theta_e}=\sin{\theta_u}=1$, the magnitude of dominant contribution of EDM of electron will be given as:
\beqn
\frac{d_e}{e}|_{\rm RPV}\sim  \frac{2}{3} \frac{{\cal V}^{-\frac{10}{3}+1 +\frac{62}{45}}}{{\cal V}^{\frac{3}{2}} m^{2}_{\frac{3}{2}}}m_{u_k} \equiv 10^{-31}GeV^{-1} \equiv 10^{-45}cm,
\eeqn
and the magnitude of dominant contribution of EDM of neutron/quark will be given as follows:
\beqn
\frac{d_n}{e}|_{\rm RPV}\sim  \frac{{\cal V}^{-\frac{10}{3}+1 +\frac{62}{45}}}{{\cal V}^{\frac{3}{2}} m^{2}_{\frac{3}{2}}}m_{e_k} \equiv 10^{-31}GeV^{-1} \equiv 10^{-45}cm.
\eeqn

\subsection{One-Loop Diagrams involving Neutral Scalar (Higgs) in the Loop}
In this subsection, we estimate the contribution of one-loop diagrams involving fermions and Higgs as propagators to EDM of fermion. The fine-tuning argument given by N.~Arkani-Hamed and S.~Dimopoulos in \cite{HamidSplitSUSY} is not just able to provide a light Higgs by diagonalising the Higgs mass matrix, it is important to give a reasonable order of magnitude of EDM by considering diagonalised Higgs mass eigenstates (light Higgs as one of the eigenstate of Higgs mass matrix) as scalar propagator in the one-loop.  In the discussion so far, we have argued that CP-violating phases in the one-loop diagrams contribution to EDM of electron/neutron are accomplished by considering off-diagonal contribution of sfermion mass matrix at electroweak scale. In this subsection, we will discuss the one-loop diagrams in which non-zero CP-violating phases appears through mixing between Higgs doublet in Higgs mass matrix. Using the same approach, we have already calculated the mass of one of the Higgs formed by linear combination of two Higgs doublets $H_{u,d}$ to be light (identified with Position moduli ${\cal Z}_{1,2}$ in our set-up (see \cite{Dhuria+Misra_mu_Split SUSY},\cite{gravitino_DM})). Now, we implement this approach to calculate non-zero EDM of electron/neutron by considering eigenstates of Higgs mass matrix as propagators in the one-loop diagram.

{{\bf SM-like Yukawa coupling contribution}}: The one-loop diagram mediated by SM-like Yukawa coupling is given in Figure 4. The effective one-loop operator given in equation (\ref{eq:eff1loop}) can be recasted in the following form:
\beqn
\label{eq:eff1loop1H}
{\cal L}_{int}=\sum_{i} \bar{\chi}_{f}
                (C_{f^{*}_{L}f_{R}{H_i}} \PL+ C_{f^{*}_{L}f_{R}{H_i}} \PR) {\phi_{H_i}}{ \chi_f}+H.c.
\eeqn
For $i=1,2$, above equation can be expanded as:
 \beqn
\label{eq:eff1loop2H}
{\cal L}_{int}= \bar{\chi}_{f}
                (C_{f^{*}_{L}f_{R}{H_1}} \PL +C_{f^{*}_{L}f_{R}{H_1}}  \PR){\phi_{H_1}}{ \chi_f}+\bar{\chi}_{e}
                (C_{f^{*}_{L}f_{R}{H_2}}  \PL +C_{f^{*}_{L}f_{R}{H_2}}  \PR) {\phi_{H_2}}{ \chi_f}+ H.c.\nonumber
\eeqn
where ${\phi_{H_1}}$ and ${\phi_{H_2}}$ correspond to eigenstates of mass matrix of Higgs doublet and ${\chi_f}$ corresponding to fermion. Using equation (\ref{massmat}), the aforementioned vertices can be expressed in terms of undiagonalized ($H_u,H_d$) basis as follows:
\begin{eqnarray}
\label{Celerh1}
&& C_{f^{*}_{L}f_{R}{H_1}}= D_{h_{11}}C_{f^{*}_{L}f_{R}{H_u}}+D_{h_{12}}C_{f^{*}_{L}f_{R}{H_d}}, C_{f^{*}_{L}f_{R}{H_2}}= D_{h_{21}}C_{f^{*}_{L}f_{R}{H_u}}+D_{h_{22}}C_{f^{*}_{L}f_{R}{H_d}}.
\end{eqnarray}
\begin{figure}
\begin{center}
\begin{picture}(145,97) (130,40)
   \ArrowLine(100,50)(135,50)
    \ArrowLine(215,50)(135,50)
    \Text(215,50)[]{$\times$}
    \ArrowLine(215,50)(295,50)
     \ArrowLine(330,50)(295,50)
     \DashCArc(215,50)(80,0,180){5}
   \Photon(185,50)(240,20){4}{4}
   \Text(150,110)[]{$H^{0}_{i}$}
   \Text(290,100)[]{$H^{0}_{i}$}
   \Text(90,50)[]{{$f_{L}$}}
   \Text(340,50)[]{{$f^{c}_{R}$}}
   \Text(170,40)[]{{{$f^{c}_{R}$}}}
   \Text(250,40)[]{{{$f_{L}$}}}  \end{picture}
\end{center}
\caption{One-loop diagram involving scalar(Higgs) and SM-like fermions.}
 \end{figure}
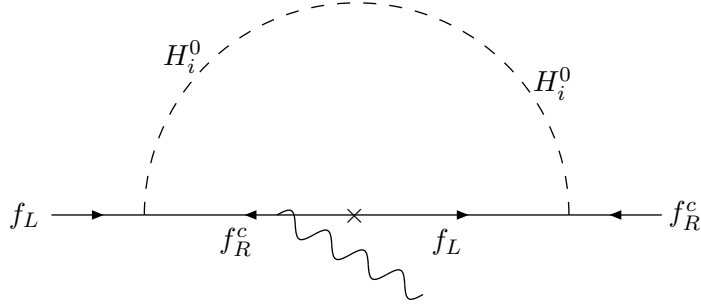
 In ${\cal N}=1$ gauged supergravity, the interaction vertices $C_{e^{*}_{L}e_{R}{H_u}/{H_d}}$ and $C_{u^{*}_{L}u_{R}{H_u}/{H_d}}$ will be given by expanding $e^{\frac{K}{2}}{\cal D}_{{\cal A}_1}D_{{\cal A}_3}W$ and $e^{\frac{K}{2}}{\cal D}_{{\cal A}_2}D_{{\cal A}_4}W$ respectively in the fluctuations linear in ${\cal Z}_{i}$ about its stabilized VEV. The values of the same have already been obtained in \cite{gravitino_DM} and given as:
\beqn
\label{eq:Yhatz1a1a3}
&&{\rm for}~f=e,~ C_{e^{*}_{L}e_{R}{H_u}/{H_d}}= \hat{Y}^{\rm eff}_{{\cal Z}_I{\cal A}_1{\cal A}_3} =\frac{{\cal O}({\cal Z}_I-{\cal V}^{\frac{1}{36}})\ {\rm term\ in}\
e^{\frac{K}{2}}{\cal D}_{{\cal A}_1}D_{{\cal A}_3}W}{\sqrt{K_{{\cal Z}_i\bar{\cal Z}_i}K_{{\cal A}_1\bar{\cal A}_1}K_{{\cal A}_3\bar{\cal A}_3}}}\equiv   {\cal V}^{-\frac{47}{45}} e^{i \phi_{{\cal Y}_e}}\nonumber\\
&&
{\rm for}~f=u,~ C_{u^{*}_{L}u_{R}{H_u}/{H_d}}= \hat{Y}^{\rm eff}_{{\cal Z}_I{\cal A}_2{\cal A}_4} =\frac{{\cal O}({\cal Z}_I-{\cal V}^{\frac{1}{36}})\ {\rm term\ in}\
e^{\frac{K}{2}}{\cal D}_{{\cal A}_2}D_{{\cal A}_4}W}{\sqrt{K_{{\cal Z}_1\bar{\cal Z}_1}K_{{\cal A}_1\bar{\cal A}_1}K_{{\cal A}_3\bar{\cal A}_3}}}\equiv {\cal V}^{-\frac{19}{18}} e^{i \phi_{{\cal Y}_u}},
\eeqn
where $e^{i \phi_{{\cal Y}_e}}$ and $e^{i \phi_{{\cal Y}_e}}$ are the phase factors associated with complex effective Yukawa couplings.

Going back to equation (\ref{Celerh1}),
\begin{eqnarray}
\label{Celerh2}
&& C_{e^{*}_{L}e_{R}{H_1}}\equiv  {\cal V}^{-\frac{47}{45}}e^{i \phi_{{\cal Y}_e}}(D_{h_{11}} +D_{h_{21}}), C_{e^{*}_{L}e_{R}{H_2}}\equiv{\cal V}^{-\frac{47}{45}}e^{i \phi_{{\cal Y}_e}}( D_{h_{12}} +D_{h_{22}}),\nonumber\\
&&  C_{u^{*}_{L}u_{R}{H_1}}\equiv{\cal V}^{-\frac{19}{18}}e^{i \phi_{{\cal Y}_u}}(D_{h_{11}} +D_{h_{21}}), C_{u^{*}_{L}u_{R}{H_2}}\equiv{\cal V}^{-\frac{19}{18}}e^{i \phi_{{\cal Y}_u}}( D_{h_{12}} +D_{h_{22}})
\end{eqnarray}
Now, the one-loop EDM of the electron(quark) in this  case will be given as \cite{keum+kong}:
\beqn
\label{eq:EDMh}
{\frac{d}{e}}|_{H_{1,2}}=\frac{m_f Q_f}{(4\pi)^2} \left( \frac{1}{m^{2}_{H_1}}{\rm Im}(C_{f^{*}_{L}f_{R}{H_1}}C_{f^{*}_{L}f_{R}{H_1}}^*) A\left(\frac{m^{2}_{f}}{m^{2}_{H_1}}\right) + \frac{1}{m^{2}_{H_2}} {\rm Im}(C_{f^{*}_{L}f_{R}{H_2}} C_{f^{*}_{L}f_{R}{H_2}}^*)
        A\left(\frac{m^{2}_{f}}{m^{2}_{H_2}}\right)\right),
\eeqn
where $m_{f}$ corresponds to fermion mass and $m_{H_{1,2}}$ correspond to eigenstates of Higgs mass matrix. Since  we are considering only first generation fermions in our $D3/D7$ $\mu$-split SUSY set up, physical mass eigenstate of fermion is same as usual Dirac mass term corresponding to first generation lepton/quark only. Using the fact that phase factors associated with Wilson line modulus ${\cal A}_{1/2}$ ( identified with first generation L-handed lepton/quark), Wilson line modulus ${\cal A}_{3/4} $( identified with first generation R-handed lepton/quark) and position modulus ( identified with Higgs doublet) are distinct and the effective Yukawa couplings  also produce non-zero phase factor, the masses of SM-fermions can be complex. Therefore, we assume that overall phase formed by adding all phase factors associated with fields as all coefficient of Yukawa coupling add up in such a way that  overall phase vanishes and fermion mass is real.

Using (\ref{Celerh2}),
\beqn
&& {\rm Im}(C_{e^{*}_{L}e_{R}{H_2}}C_{e^{*}_{L}e_{R}{H_2}}^*) = - {\rm Im}(C_{e^{*}_{L}e_{R}{H_1}}C_{e^{*}_{L}e_{R}{H_1}}^*) \equiv \frac{1}{2}{\cal V}^{-\frac{94}{45}}\sin{\theta_h}\sin{\phi_h}\nonumber\\
&& {\rm Im}(C_{u^{*}_{L}u_{R}{H_2}}C_{u^{*}_{L}u_{R}{H_2}}^*) = - {\rm Im}(C_{u^{*}_{L}u_{R}{H_1}}C_{u^{*}_{L}u_{R}{H_1}}^*) \equiv \frac{1}{2}{\cal V}^{-\frac{19}{18}}\sin{\theta_h}\sin{\phi_h}.\nonumber
 \eeqn
 Given that $\sin \theta_h=
\frac{2|{\hat \mu} {\cal B}|}{\sqrt{\left(M_{H_u}^2-M_{H_d}^2\right)^2+4 ( {{\hat \mu} {\cal B}})^2}}$. Using the values given above, $\sin{\theta_h} \in[0,1]$. We also make an assumption that $\phi_h~ \exists ~(0,\frac{\pi}{2}]$. Using equation (\ref{eq:A}) and value of $m_{e}= 0.5 MeV$, $m_{H_1}\equiv  125 GeV$ and $m_{H_2}\equiv  {\cal V}^{\frac{59}{72}}m_{\frac{3}{2}}$,
$A\left(\frac{m^{2}_{e}}{m^{2}_{H_1}}\right)=  A\left(\frac{m^{2}_{e}}{m^{2}_{H_2}}\right)\equiv 1$.
using (\ref{eq:A}), the dominant contribution of electron EDM in this case will be given as
\beq
\label{eq:EDMh1}
{\frac{d_e}{e}}|_{H_{1,2}}=\frac{10^{-3}}{4(4\pi)^2} {\cal V}^{-\frac{94}{45}} \left(\frac{1}{m^{2}_{H_1}}-\frac{1}{m^{2}_{H_2}}\right)\equiv 10^{-20} GeV^{-1}\equiv{\cal O}(10^{-34}) cm.
\eeq
The  numerical estimate of neutron/quark EDM will be given as:
\beq
\label{eq:EDMh2}
{\frac{d_n}{e}}|_{H_{1,2}}=\frac{10^{-3}}{2 (4\pi)^2} {\cal V}^{-\frac{19}{9}} \left(\frac{1}{m^{2}_{H_1}}-\frac{1}{m^{2}_{H_2}}\right)\equiv10^{-29} GeV^{-1}\equiv{\cal O}(10^{-33}) cm.
\eeq

{{\bf Chargino contribution }}: The one-loop diagram corresponding to electron EDM mediated via Higgs and chargino exchange is given in Figure~5. Due to presence of heavy fermions and  light as well as heavy scalars (eigenvalues of Higgs mass matrix) existing as propagators in the loop, using analytical expression of one-loop EDM as given in equation (\ref{eq:EDM}), one can expect an enhancement in the order of magnitude of EDM. We explicitly analysis the contribution of this loop diagram to EDM at one loop in the context of ${\cal N}=1$ gauged supergravity action. One can not have similar diagram for quark because of violation of charge conservation. So we use the loop diagram given in Figure~5 to get the analysis of EDM of electron only. The effective one-loop operator will be of the following form:
\beqn
\label{eq:eff1loop1}
{\cal L}_{int}=\sum_{i,j} \bar{\chi}_{f}
                (C_{f^{*}_{L}\chi^{+}_{j}{H_i^{0}}} \PL){\phi_{H_i^{0}}}{\tilde \chi}^{+}_{j}+\bar{\chi}_{f}( C_{q^{*}_{R}\chi^{-}_{j}{H_i^{0}}} \PR) {\phi_{H_i^{0}}}{\tilde \chi}^{-}_{j}+H.c... \ i,j=1,2.
\eeqn
\begin{figure}
\begin{center}
\begin{picture}(145,97) (130,40)
   \ArrowLine(100,50)(135,50)
    \ArrowLine(215,50)(135,50)
    \Text(215,50)[]{$\times$}
    \ArrowLine(215,50)(295,50)
     \ArrowLine(330,50)(295,50)
     \DashCArc(215,50)(80,0,180){5}
   \Photon(185,50)(240,20){4}{4}
   \Text(150,110)[]{$H^{0}_{i}$}
   \Text(290,100)[]{$H^{0}_{i}$}
   \Text(90,50)[]{{$e_{L}$}}
   \Text(340,50)[]{{$e_{R}$}}
   \Text(170,40)[]{{{$\chi^{+}_{i}$}}}
   \Text(250,40)[]{{{$\chi^{-}_{i}$}}}  \end{picture}
\end{center}
\caption{One-loop diagram involving Higgs and charginos.}
 \end{figure}
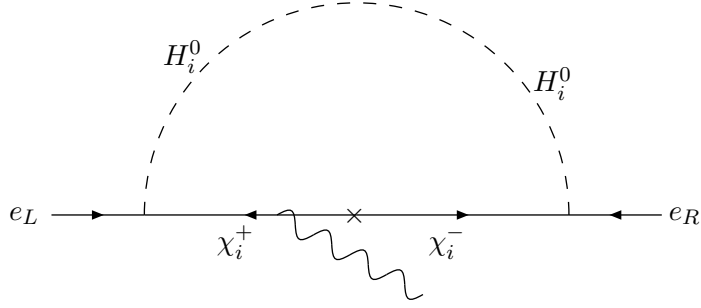
Using equation (\ref{massmat}), one can represent coefficient of interaction vertices in terms of undiagonalized basis of Higgs mass matrix as follows:
\begin{eqnarray}
\label{Celerchi}
&&C_{e^{*}_{L}\chi^{+}_{j}{H_1^{0}}} = D_{h_{11}}C_{e^{*}_{L}\chi^{+}_{j}{H_u^{0}}} +D_{h_{12}}C_{e^{*}_{L}\chi^{+}_{j}{H_d^{0}}}, C_{e^{*}_{L}\chi^{+}_{j}{H_2^{0}}}= D_{h_{21}}C_{e^{*}_{L}\chi^{+}_{j}{H_u^{0}}}+D_{h_{22}}C_{e^{*}_{L}\chi^{+}_{j}{H_u^{0}}} \nonumber\\
&&C_{e^{*}_{R}\chi^{-}_{j}{H_1^{0}}} = D_{h_{11}}C_{e^{*}_{R}\chi^{-}_{j}{H_u^{0}}} +D_{h_{12}}C_{e^{*}_{R}\chi^{-}_{j}{H_d^{0}}}, C_{e^{*}_{R}\chi^{-}_{j}{H_2^{0}}}= D_{h_{21}}C_{e^{*}_{R}\chi^{-}_{j}{H_u^{0}}}+D_{h_{22}}C_{e^{*}_{R}\chi^{-}_{j}{H_u^{0}}}.
\end{eqnarray}
As given in appendix,
\beqn
&& {\tilde \chi}^{+}_{1}= -{\tilde H}^{+}_{u}+\left(\frac{v}{M_P}{\tilde f}{\cal V}^{\frac{5}{6}}\right){\tilde \lambda}^{+}_{i},{\tilde \chi}^{-}_{1}=  -{\tilde H}^{-}_{d}+\left(\frac{v}{M_P}{\tilde f}{\cal V}^{\frac{5}{6}}\right){\tilde \lambda}^{-}_{i},~ {\rm and}~m_{{\tilde \chi}^{\pm}_{1}}\equiv  {\cal V}^{\frac{59}{72}}m_{\frac{3}{2}}\nonumber\\
&&  {\tilde \chi}^{+}_{2}= {\tilde \lambda}^{+}_{i} +\left(\frac{v}{M_P}{\tilde f}{\cal V}^{\frac{5}{6}}\right){\tilde H}^{+}_{u},{\tilde \chi}^{-}_{2}= {\tilde \lambda}^{-}_{i} +\left(\frac{v}{M_P}{\tilde f}{\cal V}^{\frac{5}{6}}\right) {\tilde H}^{-}_{d},~ {\rm and}~m_{{\tilde \chi}^{\pm}_{2}}\equiv  {\cal V}^{\frac{2}{3}}m_{\frac{3}{2}}.\nonumber
\eeqn
Using the above,
\beqn
\label{eq:Ceechi}
&&C_{e^{*}_{L}\chi^{+}_{1}{H_u^{0}}/{H_d^{0}}}= -C_{e^{*}_{L}{\tilde H^{+}_u}{H_u^{0}}/H_d^{0}}+ \left(\frac{v}{M_P}{\tilde f}{\cal V}^{\frac{5}{6}}\right)C_{e^{*}_{L}{\tilde {\lambda}^{+}_i}{H_u^{0}}/H_d^{0}}\nonumber\\
&& C_{e^{*}_{L}\chi^{+}_{2}{H_u^{0}}/{H_d^{0}}}= C_{e^{*}_{L}{\tilde {\lambda}^{+}_i}{H_u^{0}}/H_d^{0}}+\left(\frac{v}{M_P}{\tilde f}{\cal V}^{\frac{5}{6}}\right) C_{e^{*}_{L}{\tilde H^{+}_u}{H_u^{0}}/H_d^{0}}\nonumber\\
&& C_{e^{*}_{R}\chi^{-}_{1}{H_u^{0}}/{H_d^{0}}}= -C_{e^{*}_{R}{\tilde H^{-}_d}{H_u^{0}}/H_d^{0}}+ \left(\frac{v}{M_P}{\tilde f}{\cal V}^{\frac{5}{6}}\right)C_{e^{*}_{R}{\tilde {\lambda}^{+}_i}{H_u^{0}}/H_d^{0}}\nonumber\\
&&C_{e^{*}_{R}\chi^{-}_{2}{H_u^{0}}/{H_d^{0}}}= C_{e^{*}_{R}{\tilde {\lambda}^{+}_i}{H_u^{0}}/H_d^{0}}+\left(\frac{v}{M_P}{\tilde f}{\cal V}^{\frac{5}{6}}\right) C_{e^{*}_{R}{\tilde H^{-}_d}{H_u^{0}}/H_d^{0}}.
\eeqn
 The interaction vertices $C_{e^{*}_{L}{\tilde H^{+}_u}{H_u^{0}/H_d^{0}}}$ and $ C_{e^{*}_{R}{\tilde H^{-}_d}{H_u^{0}}/H_d^{0}}$ corresponding to Figure 5 will be given by expanding the $e^{\frac{K}{2}}{\cal D}_{{\cal Z}_1}D_{{\cal A}_1}W$  and $e^{\frac{K}{2}}{\cal D}_{{\cal Z}_1}D_{{\cal A}_3}W$ in the fluctuations linear in ${\cal Z}_{i}$ about its stabilized VEV. The contributions of $e^{\frac{K}{2}}{\cal D}_{z_1}D_{a_1}W$  as well as $e^{\frac{K}{2}}{\cal D}_{z_1}D_{a_3}W$ have been given  in terms of undiagonalized $(z_i,a_i)$  basis in \cite{gravitino_DM}. We assume that $ e^{\frac{K}{2}}{\cal D}_i D_{\bar {{\cal A}_1}}W \sim {\cal O}(1) e^{\frac{K}{2}}{\cal D}_i D_{{\bar a}_1}W$. Since the EW symmetry gets broken for the higgsino(${\tilde H^{+}_u}$)-lepton (${e_L}$)-Higgs(${H^{0}_{u}/H^{0}_d}$) vertex, we evaluate the contribution of the same by expanding $e^{\frac{K}{2}}{\cal D}_{z_1}D_{a_1}W$ in the fluctuations linear in $z_1$ as well as $(z_i- {\cal V}^{\frac{1}{36}})$, and then giving  VEV to $z_i$. Doing so, the magnitude of coefficient of this vertex  will be given as :
\beqn
\label{eq:Yhatz1z1a1}
|C_{e^{*}_{L}{\tilde H^{+}_u}{H_u^{0}/H_d^{0}}}|\sim  \frac{\langle{\cal Z}_i\rangle {\cal O}({\cal Z}_i-{\cal V}^{\frac{1}{36}})\ {\rm term\ in}\
e^{\frac{K}{2}}{\cal D}_{{\cal Z}_1}D_{{\cal A}_1}W}{\sqrt{(K_{{\cal Z}_1\bar{\cal Z}_1})^3 K_{{\cal A}_1\bar{\cal A}_1}}}\equiv {\cal V}^{-\frac{1}{10}},~{\rm for}~{\cal V}=10^5.
\eeqn
Similarly, the contribution of physical  higgsino(${\tilde H^{-}_d}$)-lepton(${e_R}$)-Higgs(${H^{0}_{u}/H^{0}_d}$) vertex  will be given as
\beqn
\label{eq:Yhatz1z1a3}
|C_{e^{*}_{R}{\tilde H^{-}_d}{H_u^{0}/H_d^{0}}}| \sim \frac{{\cal O}({\cal Z}_i-{\cal V}^{\frac{1}{36}})\ {\rm term\ in}\
e^{\frac{K}{2}}{\cal D}_{{\cal Z}_1}D_{{\cal A}_3}W}{\sqrt{K_{{\cal Z}_1\bar{\cal Z}_1}K_{{\cal Z}_1\bar{\cal Z}_1}K_{{\cal A}_3\bar{\cal A}_3}}}\equiv   {\cal V}^{\frac{1}{10}},~{\rm for}~{\cal V}=10^5.
\eeqn
 The coefficient of interaction vertex $e^{-}_{L}-H^{0}_{u}- {\tilde \lambda^{+}_i}$ corresponding to Figure~5 will be given by $
{\cal L}_{e^{-}_{L}-H^{0}_{u}-{\tilde \lambda^{+}_i}}= g_{YM}g_{ {{\cal A}_1}\bar{T}_B}X^{{*}B}{\bar\chi}^{\bar {\cal A}_1}\tilde{\lambda^{+}_i}+\partial_{{\cal A}_1}T_B D^{B} {\bar\chi}^{\bar {\cal A}_1}\tilde{\lambda^{+}_i}$.
Since $\partial_{{\cal A}_1}T_B$ does not give any term which is linear in ${\cal Z}_i$, so the second term contributes zero to the given vertex. By expanding $g_{{\cal A}_1\bar{T}_B}$ in the fluctuation linear in ${\cal Z}_1$ around its stabilized VEV, in terms of undiagonalized basis, we have: $ g_{{T_B} {\bar a}_1}\rightarrow-{\cal V}^{-\frac{13}{12}} (z_1-{\cal V}^{-\frac{1}{36}}),~{\rm and}~g_{YM}\sim{\cal V}^{-\frac{1}{36}}$.
 Considering $g_{YM}g_{{T_B} {\bar a}_1} \sim {\cal O}(1)g_{YM}g_{{T_B} {\bar {\cal A}}_1}$ as shown in \cite{gravitino_DM}; incorporating values of  $X^{B}=-6i\kappa_4^2\mu_7Q_{T_B}$, $\kappa_4^2\mu_7\sim \frac{1}{\cal V}$ and  $Q_{T_B}\sim{\cal V}^{\frac{1}{3}} (2\pi\alpha^\prime)^2\tilde{f}$, we get the contribution of physical gaugino(${\tilde \lambda^{+}_i}$)-lepton$(e_{L})$-Higgs($H^{0}_{u}$) interaction vertex given as follows:
\begin{equation}
\label{eq:eLHambda}
|C_{e_{L}{\tilde \lambda^{+}_i}{H_u^{0}}/H_d^{0}}|\equiv \frac{g_{YM}g_{{T_B} {\bar {\cal A}_1}}X^{T_B}\sim {\cal V}^{-\frac{47}{36}} {\tilde f}}{{ \sqrt{\hat{K}_{{\cal A}_1{\bar {\cal A}}_1}\hat{K}_{{\cal Z}_1{\bar {\cal Z}}_1}} }}{\cal Z}_1{\bar\chi}^{\bar {\cal A}_1}\tilde{\lambda^{0}}
 \equiv\tilde{f}\left({\cal V}^{-\frac{3}{2}}\right).
\end{equation}
To calculate the coefficient of  interaction vertex $e^{*}_{R}-H^{0}_{u}-\tilde{\lambda^{-}_i}$, we need to expand $g_{{\cal A}_3\bar{T}_B}$ in the fluctuation quadratic in ${\cal Z}_1$ to first conserve $SU(2)_L$ symmetry and after giving  VEV to one of the ${\cal Z}_i$, we get the required contribution
\begin{equation}
\label{eq:eRHlambda}
|C_{e^{*}_{R}{\tilde \lambda^{+}_i}{H_u^{0}}/H_d^{0}}|\equiv \frac{g_{YM}g_{{T_B} {\bar {\cal A}_3}}X^{T_B}\sim {\cal V}^{-\frac{16}{9}}\langle{\cal Z}\rangle {\tilde f}}{{ \sqrt{\hat{K}_{{\cal A}_3{\bar {\cal A}}_3}\hat{K}^{2}_{{\cal Z}_1{\bar {\cal Z}}_1}} }}{\cal Z}_1{\bar\chi}^{\bar {\cal A}_3}\tilde{\lambda^{0}}
 \equiv\tilde{f}\left({\cal V}^{-\frac{15}{9}}\frac{\langle{\cal Z}_i\rangle}{M_P}\right).
\end{equation}
Incorporating the results given in eqs.~(\ref{eq:Yhatz1z1a1})-(\ref{eq:eRHlambda}) in equation (\ref{eq:Ceechi}),
we have,
\beqn
\label{eq:results}
 && |C_{e^{*}_{L}\chi^{+}_{1}{H_u^{0}}/{H_d^{0}}}|\equiv{\cal V}^{-\frac{1}{10}}, |C_{e^{*}_{L}\chi^{+}_{2}{H_u^{0}}/{H_d^{0}}}|\equiv{\cal V}^{\frac{1}{10}}, \nonumber\\
 && |C_{e^{*}_{R}\chi^{-}_{1}{H_u^{0}}/{H_d^{0}}}|\equiv\tilde{f}{\cal V}^{-\frac{3}{2}},
 |C_{e^{*}_{R}\chi^{-}_{2}{H_u^{0}}/{H_d^{0}}}|\equiv \tilde{f}{\cal V}^{-\frac{15}{9}}\frac{\langle{\cal Z}_i\rangle}{M_P}.
 \eeqn
Now, the one-loop EDM of the electron in this  case will be given as \cite{keum+kong}:
\beqn
\label{eq:EDMchi}
{\frac{d}{e}}|_{\chi^{\pm}_i}=\sum_{i}\frac{m_{\chi^{\pm}_j} Q'_{{e}(i)}}{(4\pi)^2} \left[ \frac{1}{m^{2}_{H^{0}_i}}{\rm Im}\left(\left(C_{e^{*}_{L}\chi^{+}_{i}{H_i^{0}}}C_{e^{*}_{R}\chi^{-}_{j}{H_i^{0}}}^*\right) A\left(\frac{m^{2}_{\chi^{\pm}_i}}{m^{2}_{H^{0}_i}}\right) \right)\right].
\eeqn
where $m_{\chi^{\pm}_j}$ and $m^{2}_{H^{0}_i}$ corresponds to masses eigenstates of chargino and Higgs mass matrix. The effective charge for this loop diagram will be $Q'_{{e}(i)}= Q_e C_{{\chi^{+}_i}{\chi^{-}_i}{\gamma}}$ where $C_{{\chi^{+}_1}{\chi^{-}_1}{\gamma}}=C_{{{\tilde H}^{+}_i}{{\tilde H}^{-}_i}{\gamma}},C_{{\chi^{+}_2}{\chi^{-}_2}{\gamma}}=C_{{{\tilde \lambda}^{+}_{i}{{\tilde  \lambda}^{-}_i}{\gamma}}}$.
The contributions of both higgsino-higgsino-gauge boson vertex and gaugino-gaugino-gauge boson have already obtained in the context of ${\cal N}=1$ gauged  supergravity in \cite{gravitino_DM}. Using the same,
\begin{equation}
C_{{\chi^{+}_1}{\chi^{-}_1}{\gamma}}\equiv {\tilde f}{\cal V}^{-\frac{5}{18}}, C_{{\chi^{+}_2}{\chi^{-}_2}{\gamma}}\equiv {\tilde f}{\cal V}^{-\frac{11}{18}}.
\end{equation}
Utilizing the results of $C_{e^{*}_{L/R}\chi^{\pm}_{i}{H_i^{0}}}$ vertices given in (\ref{eq:results}) and the assumption that value of phase factor associated with these couplings are of ${\cal O}(1)$; $m_{\chi^{\pm}_1}=m_{H_2}= {\cal V}^{\frac{59}{72}}m_{\frac{3}{2}}$,$m_{\chi^{\pm}_2} = {\cal V}^{\frac{2}{3}}m_{\frac{3}{2}}$ and $m_{H_1}\sim 125 GeV$ as given in section {{\bf 2}}, $\sin \theta_h=
 (0,1]$, $\phi_e= (0,\frac{\pi}{2}]$, and
$A(\frac{m^{2}_{\chi^{\pm}_i}}{m^{2}_{H^{0}_i}})\equiv \frac{m^{2}_{H^{0}_i}}{m^{2}_{\chi^{\pm}_i}}$ by using (\ref{eq:A}),  we have:
\beq
\label{eq:EDMchi1}
{\frac{d}{e}}|_{\chi^{\pm}_i}\equiv \frac{1}{\sqrt{2}(4\pi)^2}({\cal V}^{-\frac{1}{10}+\frac{1}{10}})\times\frac{{\tilde f}{\cal V}^{-\frac{5}{18}}} {{\cal V}^{\frac{59}{72}}m_{\frac{3}{2}}} \equiv {\cal O}(10^{-32}) cm,~ {\rm for} ~{\cal V}={\cal O}(1)\times 10^4.
\eeq
\subsection{One-Loop Diagrams involving Gravitino and Sgoldstino in the Loop}

{{\bf Gravitino contribution:}} In this section, we estimate the EDM of electron(quark) by considering the gravitino as a propagator in one-loop diagrams despite the fact that these are logarithmically divergent. The loop diagrams are given in Figure 6. To get the numerical estimate of EDM corresponding to these diagrams, we first need to determine the contribution of relevant vertices in ${\cal N}=1$ gauged supergravity. The same are evaluated as follows:
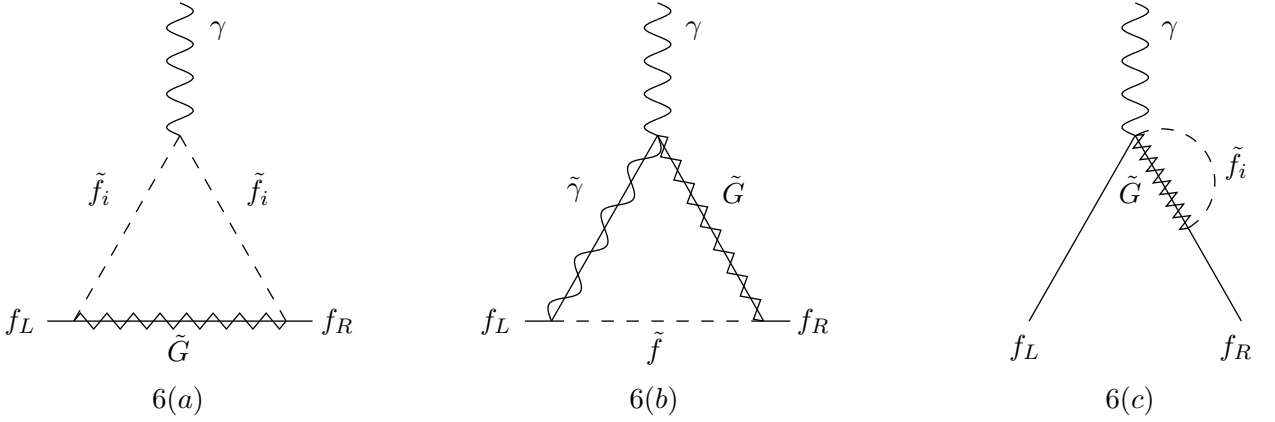
\begin{figure}
\begin{center}
\begin{picture}(100,137) (280,20)
   \Line(100,50)(110,50)
   \Line(110,50)(190,50)
 \ZigZag(110,50)(190,50){3}{8}
   \Line(190,50)(200,50)
   \DashLine(110,50)(150,120){5}
   \Photon(150,120)(150,170){5}{4}
   \DashLine(150,120)(190,50){5}
   \Text(165,160)[]{$\gamma$}
   \Text(120,100)[]{${\tilde f}_{i}$}
   \Text(180,100)[]{${\tilde f}_{i}$}
   \Text(90,50)[]{{$f_{L}$}}
   \Text(210,50)[]{{$f_{R}$}}
   \Text(150,40)[]{{{${\tilde G}$}}}
   \Text(150,20)[]{{{$6(a)$}}}
      \Line(280,50)(290,50)
      \DashLine(290,50)(370,50){5}
      \Line(370,50)(380,50)
   \Line(290,50)(330,120)
   \Photon(290,50)(330,120){4}{4}
   \Photon(330,120)(330,170){5}{4}
   \Line(330,120)(370,50)
   \ZigZag(330,120)(370,50){3}{8}
   \Text(345,160)[]{$\gamma$}
   \Text(300,100)[]{${\tilde \gamma}$}
   \Text(360,100)[]{${\tilde G}$}
   \Text(270,50)[]{{$f_{L}$}}
   \Text(390,50)[]{{$f_{R}$}}
   \Text(330,40)[]{{{${\tilde f}$}}}
   \Text(330,20)[]{{{$6(b)$}}}
   \Line(470,50)(510,120)
   \Photon(510,120)(510,170){5}{4}
   \DashCArc(520,102.5)(20,300,118){5}
   \ZigZag(510,120)(530,85){3}{8}
   \Line(510,120)(530,85)
   \Line(530,85)(550,50)
   \Text(525,160)[]{$\gamma$}
   \Text(510,100)[]{${\tilde G}$}
   \Text(550,110)[]{${\tilde f}_{i}$}
   \Text(470,40)[]{{$f_{L}$}}
   \Text(550,40)[]{{$f_{R}$}}
   \Text(510,20)[]{{$6(c)$}}
   \end{picture}
\end{center}
\caption{One-loop diagram involving gravitino.}
 \end{figure}
 In ${\cal N}=1$ gauged supergravity, the gravitino-fermion-sfermion vertex will be given as: $
 {\cal L}_{\tilde G-f-{\tilde f}}=-\frac{1}{2}\sqrt{2}eg_{ij}\partial_{\mu}{\phi}^i \chi^{j}{\gamma}^{\mu}{\gamma}^{\nu}\psi_{\mu}$.  The physical ${\psi_{\mu}}$- lepton(quark)-slepton(squark) vertex will be given as:
 \beqn
 \label{Cpsilltilde}
&& |C_{\tilde G\ e_L \ {\tilde e_L}}| \equiv \frac{g_{{\cal A}_1 {\bar {\cal A}}_1}}{\sqrt{{\cal K}_{{\cal A}_1 {\bar {\cal A}_1}}{\cal K}_{{\cal A}_1 {\bar {\cal A}_1}}}}\partial_{\mu}{\cal A}_1 \chi^{{\cal A}_1}{\gamma}^{\mu}{\gamma}^{\nu}\psi_{\mu}\equiv\partial_{\mu}{\cal A}_1 \chi^{{\cal A}_1}{\gamma}^{\mu}{\gamma}^{\nu}\psi_{\mu},\nonumber\\
 && |C_{\tilde G\ u_L \ {\tilde u_L}}| \equiv \frac{g_{{\cal A}_2 {\bar {\cal A}}_2}}{\sqrt{{\cal K}_{{\cal A}_2 {\bar {\cal A}_2}}{\cal K}_{{\cal A}_2 {\bar {\cal A}_2}}}}\partial_{\mu}{\cal A}_2 \chi^{{\cal A}_2}{\gamma}^{\mu}{\gamma}^{\nu}\psi_{\mu}\equiv \partial_{\mu}{\cal A}_2 \chi^{{\cal A}_2}{\gamma}^{\mu}{\gamma}^{\nu}\psi_{\mu}.
 \eeqn
 The contribution of physical sfermion -sfermion -photon vertices have already been obtained in subsection {\bf 3.1} and the values of the same are given as:
 $|C_{\tilde e_L \tilde e_L \gamma}|\equiv {\cal V}^{\frac{44}{45}}{\tilde {\cal A}}_1 \partial_{\mu}{\tilde {\cal A}}_1 A^{\mu},  |C_{\tilde u_L \tilde u_L \psi_{\mu}}| \equiv {\cal V}^{\frac{53}{45}}{\tilde {\cal A}}_2 \partial_{\mu}{\tilde {\cal A}}_2 A^{\mu}.$
 The contribution of the fermion-sfermion-photino($\gamma$) vertex in the context of ${\cal N}=1 $ gauged supergravity action will given by $
{\cal L}_{f {\tilde f} {\tilde \gamma}}= g_{YM}g_{{\cal A}_I \bar{T}_B}X^{{*}B}{\bar\chi}^{\bar {\cal A}_I}\tilde{\gamma}+\partial_{{\cal A}_1}T_B D^{B} {\bar\chi}^{\bar {\cal A}_I}\tilde{\gamma}$.
 For $f={e}$,  by expanding $g_{a_1 \bar{T}_B}$ in the fluctuations linear in $a_1$ around its stabilized VEV, we have: $ g_{\bar {a_1}T_B}= {\cal V}^{-\frac{2}{9}}(a_1- {\cal V}^{-\frac{2}{9}}), ~{\rm and}~ \partial_{{\cal A}_1}T_B \rightarrow {\cal V}^{\frac{10}{9}}({\cal A}_1- {\cal V}^{-\frac{2}{9}})$.
Assuming that $g_{\bar { {\cal A}_1}T_B}={\cal O}(1)g_{\bar {a_1} T_B}$ and using  $X^{{*}B}={\kappa}^{2}_4 {\mu}_7 Q_B, D^B=\frac{4\pi\alpha^\prime\kappa_4^2\mu_7Q_Bv^B}{\cal V}$ where $ Q_B \sim {\cal V}^{\frac{1}{3}}{\tilde f}$ and ${\kappa}^{2}_4 {\mu}_7\sim \frac{1}{\cal V}$, we get:
 $ |C_{e_L {\tilde e_L} {\tilde \gamma}}|\sim  \frac{{\cal V}^{-\frac{2}{9}}{\tilde f}}{\sqrt{{\cal K}_{{\cal A}_1 {\bar {\cal A}_1}}{\cal K}_{{\cal A}_1 {\bar {\cal A}_1}}}}{\tilde {\cal A}}_1 {\bar\chi}^{\bar {\cal A}_1}\tilde{\gamma}\equiv {\tilde f}{\cal V}^{-1}{\tilde {\cal A}}_1 {\bar\chi}^{\bar {\cal A}_1}\tilde{\gamma}.$
For $f={u}$,  using  $ g_{\bar {{\cal A}_2}T_B} \sim {\cal O}(1) g_{\bar {a_2}T_B}= {\cal V}^{-\frac{5}{4}}(a_2- {\cal V}^{-\frac{1}{3}})$ and $\partial_{{\cal A}_2}T_B \rightarrow {\cal V}^{\frac{1}{9}}({\cal A}_2- {\cal V}^{-\frac{1}{3}})$, we have:
   \beqn
  |C_{u_L {\tilde u_L} {\tilde \gamma}}|\sim  \frac{{\cal V}^{-\frac{11}{9}}{\tilde f}}{\sqrt{{\cal K}_{{\cal A}_2 {\bar {\cal A}_2}}{\cal K}_{{\cal A}_2 {\bar {\cal A}_2}}}}{\tilde {\cal A}}_2 {\bar\chi}^{\bar {\cal A}_2}\tilde{\gamma}\equiv {\tilde f}{\cal V}^{-\frac{4}{5}}{\tilde {\cal A}}_2 {\bar\chi}^{\bar {\cal A}_2}\tilde{\gamma}.
  \eeqn
 The contribution of the gravitino-fermion-sfermion-photon vertex in the context of ${\cal N}=1 $ gauged supergravity action will be given as: $
  {\cal L}=-\frac{1}{2}{\sqrt 2}e g_{{\cal A}_I T^{*}_B}X^{B}A_{\mu}{\bar\chi}^{{\cal A}_I}{\gamma}^{\mu}{\gamma}^{\nu}{\psi_{\mu}}$.  Using the above-mentioned value of $g_{{\cal A}_1 T^{*}_B}$, $g_{{\cal A}_2 T^{*}_B}$  and $X^{B}$, the coefficient of physical  gravitino-lepton(quark)-slepton(squark)-photon vertex will be given as:
  \beqn
  \label{Gelel}
  && | C_{\tilde G e_L {\tilde e_L}\gamma}|\sim \frac{{\cal V}^{-\frac{8}{9}}{\tilde f}}{\sqrt{{\cal K}_{{\cal A}_1 {\bar {\cal A}_1}}{\cal K}_{{\cal A}_1 {\bar {\cal A}_1}}}}A_{\mu}{\bar\chi}^{{\cal A}_1}{\gamma}^{\mu}{\gamma}^{\nu}{\psi_{\mu}}\equiv {\tilde f}{\cal V}^{-\frac{5}{3}} A_{\mu}{\bar\chi}^{{\cal A}_1}{\gamma}^{\mu}{\gamma}^{\nu}{\psi_{\mu}},\nonumber\\
  &&| C_{\tilde G u_L {\tilde u_L}\gamma}|\sim \frac{{\cal V}^{-\frac{35}{18}}{\tilde f}}{\sqrt{{\cal K}_{{\cal A}_2 {\bar {\cal A}_2}}{\cal K}_{{\cal A}_2 {\bar {\cal A}_2}}}}A_{\mu}{\bar\chi}^{{\cal A}_2}{\gamma}^{\mu}{\gamma}^{\nu}{\psi_{\mu}}\equiv {\tilde f}{\cal V}^{-\frac{5}{3}} A_{\mu}{\bar\chi}^{{\cal A}_2}{\gamma}^{\mu}{\gamma}^{\nu}{\psi_{\mu}}.
  \eeqn
The contribution of photon ($\gamma$)-photino (${\tilde \gamma}$)-gravitino($\gamma$) vertex will be given $
{\cal L}= \frac{i}{4}e\bar {\gamma}^{\mu}{\lambda}[\slashed{\partial},\slashed{\cal A}]{\psi}_{\mu}$. We notice that there is no moduli space-dependent factor coming from this vertex.

The above Feynman diagrams involving gravitino in a loop have been explicitly worked out in \cite{mandez+orteu} to calculate the magnetic moment of muon in the context of spontaneously broken minimal ${\cal N}=1 $ gauged supergravity. We explicitly utilize their results in a modified form to get the estimate of EDM of electron/quark in the ${\cal N}=1$ gauged supergravity. The modified  results of magnetic moment of electron after multiplying with volume suppression factors coming from relevant vertices as calculated in equations (\ref{Cpsilltilde})-(\ref{Gelel}) are as follows:

For Figure~6(a):
  \beqn
&& {\hskip -1.2in} {a}^{\rm div}_{f}|_{6(a)}\equiv{\tilde f}{\cal V}^{a}(G_N m^{2}_{f}/{\pi})\sum_{j=1,2}\Bigl[\Gamma(\epsilon-1)[-\frac{1}{90}{\mu}^2+\frac{1}{18}{\mu}^{2}_{j}]+
 \Gamma(\epsilon)[\frac{2}{45}{\mu}^2+\frac{2}{9}] +(-1)^j \sin{\theta} \Gamma(\epsilon-1)[-{\mu}^{2}_{j}/3{\mu}]\Bigr],
 \eeqn
 where ${\cal V}^{a}$ is the Calabi-Yau volume-suppression factor.
 Here $\mu={m_f}/m_{\frac{3}{2}}$ and $\mu_j=m_{{\tilde f}_j}/m_{3/2}$, $j=1,2$; m is lepton mass, $m_{\frac{3}{2}}$ is gravitino mass. $m_{{\tilde f}_1}$ and $m_{{\tilde f}_2}$ are eigenvalues of diagonalized slepton(squark) mass matrix. In our set-up $\sin{\theta}=1$. Using $m_{{\tilde f}_1}=m_{{\tilde f}_2}={\cal V}^{\frac{1}{2}}m_{\frac{3}{2}}$, $m_{\frac{3}{2}}={\cal V}^{-2}M_P$  and $m_e={\cal O}(1) MeV$, we have
 $\mu_1=\mu_2= \frac{1}{\cal V}~~{\rm and}~~\mu= 10^{-11}~{\rm for}~{\cal V}=10^5$.

 For $f=e$, incorporating these values, dominant contribution will be of the form:
 \beqn
&&  {\hskip -1.0in}  {a}^{\rm div}_{e}|_{6(a)}\equiv {\tilde f}{\cal V}^{\frac{44}{45}}(G_N m^{2}_{e}/{\pi}) \Bigl[\frac{1}{18 {\cal V}^2}\Gamma(\epsilon-1) + \frac{2}{9}  \Gamma(\epsilon)\Bigr]  \equiv {\tilde f}{\cal V}^{\frac{44}{45}}(G_N m^{2}_{e}/{\pi})   \Bigl[\frac{1}{18 {\cal V}^2}\Gamma(\epsilon-1)+\frac{1}{18 {\cal V}^2}\Gamma(\epsilon) + a'\Bigr].
 \eeqn
 where $ a'=(\frac{2}{9} -\frac{1}{18 {\cal V}^2}) \Gamma(\epsilon)$ is divergent piece.
 Using $-\Gamma(\epsilon-1)=\Gamma(\epsilon)(1+\epsilon)$, the finite contribution will be given as:
 ${a}^{\rm finite}_{e}|_{6(a)}\equiv \frac{1}{18}{\tilde f}{\cal V}^{-\frac{46}{45}}(G_N m^{2}_{e}/{\pi})$. Similarly, using the volume suppression factor coming from quark-quark photon vertex, we get:
${a}^{\rm finite}_{u}|_{6(a)}\equiv \frac{1}{18}{\tilde f}{\cal V}^{-\frac{37}{45}}(G_N m^{2}_{u}/{\pi})$. Now we use the relation between anomalous magnetic moment and electric dipole moment to get the numerical estimate of EDM of electron in this case. As given in \cite{Graesser+thomas}, $
a_f= \frac{2 |m_f|}{e Q_f}|d_f|\cos{\phi}$,
 where $m_f$ and $Q_f$ correspond to mass and charge of fermion; $d_f$ is electric dipole moment of fermion and $\phi$ is defined as $
\phi\equiv {\rm Arg}(d_f m^{*}_f)$. We consider that in the loop diagrams involving sfermion as propagators, the non-trivial phase responsible to generate EDM  appears from eigenstates of sfermion mass matrix (off-diagonal component of slepton mass matrix) and we assume the value of same as ${\phi_{d_f}}~ \exists~ (0,\frac{\pi}{2}]$. The first generation electron/quark mass has been calculated from complex effective Yukawa coupling(${\cal Y}^{\rm eff}_{{\cal Z}_I {\cal A}_{1/3}{\cal A}_{2/4}}$) in ${\cal N}=1$ gauged supergravity and there is a distinct phase factor ${\phi_{y_e/y_u}}$ associated with the same. Using the fact that ${\phi_{d_f}}\neq {\phi_{y_e/y_u}}$  the relative phase between two will be in the interval $\phi~ \exists~(0,\frac{\pi}{2})\sim {\cal O}(1)$. Hence,
\beqn
&& \frac{d_e}{e}|_{6(a)}= 2{|m_e|}\ {a}^{\rm finite}_{e}|_{6(a)}\equiv  \frac{1}{18}{\tilde f}{\cal V}^{-\frac{46}{45}}(G_N m_{e}/{\pi})\equiv 10^{-67}cm, \nonumber\\
&& \frac{d_u}{e}|_{6(a)}= 2{|m_u|}\ {a}^{\rm finite}_{u}|_{6(a)}\equiv  \frac{1}{18}{\tilde f}{\cal V}^{-\frac{37}{45}}(G_N m_{u}/{\pi})\equiv 10^{-67}cm.
\eeqn
For Figure~6(b):
 \beqn
 &&  {\hskip -0.9in}  {a}^{\rm div}_{f}|_{6(b)}\equiv{\tilde f}{\cal V}^{-a}(G_N m^{2}_{f}/{\pi})\sum_{j=1,2}\Bigl[\Gamma(\epsilon-1)[\frac{1}{20}{\mu}^2-\frac{1}{6}{\mu}^{2}_{j}]+
 \Gamma(\epsilon)[-\frac{7}{60}{\mu}^{2}]  +(-1)^j \sin{2\alpha}\Gamma(\epsilon-1)[{\mu}^{2}_{j}/{\mu}]\Bigr]; ~{\rm where}~ {f=e,u}. \nonumber
 \eeqn
 For $f=e$, incorporating the values of masses and simplifying, now we will have
  \beqn
 &&  {\hskip -1.3in} {a}^{\rm div}_{e}|_{6(b)}\equiv{\tilde f}{\cal V}^{-1}(G_N m^{2}_{e}/{\pi})\Bigl[-\frac{1}{6 {\cal V}^2}\Gamma(\epsilon-1) -\frac{7}{60}{\mu}^{2}  \Gamma(\epsilon)\Bigr]\equiv {\tilde f}{\cal V}^{-1}(G_N m^{2}_{f}/{\pi}) \Bigl[ -\frac{7}{60}{\mu}^{2}  \Gamma(\epsilon-1)-\frac{7}{60}{\mu}^{2}  \Gamma(\epsilon)+a'\Bigr].
 \eeqn
 where $a'=(-\frac{1}{6 {\cal V}^2}+\frac{7}{60}{\mu}^{2})\Gamma(\epsilon-1)$ is divergent piece.
 Picking up the finite contribution, we get $  {a}^{\rm finite}_{e}|_{6(b)}\equiv  \frac{7}{60}{\tilde f}{\cal V}^{-1}(G_N m^{2}_{f}/{\pi}){\mu}^{2}$, and therefore,
\beq
\frac{d_e}{e}|_{6(b)} \equiv2 |m_e| \ {{a}^{finite}_{e}|_{6(b)}}\equiv 10^{-65}GeV^{-1} \equiv 10^{-79}cm.
\eeq
Similarly, using volume suppression factor coming from quark-quark photon vertex,
 \beqn
 \frac{d_u}{e}|_{6(b)}\equiv 10^{-64}GeV^{-1} \equiv 10^{-78}cm.
 \eeqn
For Figure 6(c):
 \beqn
 && {\hskip -0.9in} {a}^{\rm div}_{f}|_{6(c)}\equiv {\tilde f}{\cal V}^{-\frac{5}{3}}(G_N m^{2}_{f}/{\pi})\sum_{j=1,2}\Bigl[\Gamma(\epsilon-1)[-\frac{1}{90}{\mu}^2+\frac{1}{9}{\mu}^{2}_{j}]+
 \Gamma(\epsilon)[\frac{1}{10}{\mu}^2-\frac{2}{9}]  +(-1)^j \sin{2\alpha}\Gamma(\epsilon-1)[-2{\mu}^{2}_{j}/3{\mu}]\Bigr]. \nonumber
  \eeqn
As Similar to the above, incorporating the value of masses and further simplifying, dominant contribution is given by:
 \beqn
 &&  {\hskip -0.9in} {a}^{\rm div}_{f}|_{6(c)}\equiv{\tilde f}{\cal V}^{-\frac{5}{3}}(G_N m^{2}_{f}/{\pi}) \Bigl[\frac{1}{9 {\cal V}^2}\Gamma(\epsilon-1)-\frac{2}{9}
 \Gamma(\epsilon)\Bigr]\equiv {\tilde f}{\cal V}^{-\frac{5}{3}}(G_N m^{2}_{f}/{\pi}) \Bigl[\frac{1}{9 {\cal V}^2} \Gamma(\epsilon-1)+\frac{1}{9 {\cal V}^2} \Gamma(\epsilon)+a'\Bigr].
  \eeqn
  where $a'= (-\frac{2}{9}-\frac{1}{9 {\cal V}^2} )\Gamma(\epsilon)$ is divergent piece.
  Considering the finite piece,  ${a}^{\rm finite}_{q}|_{6(c)}= \frac{1}{9}{\tilde f}{\cal V}^{-\frac{11}{3}}(G_N m^{2}_{f}/{\pi})$. Again using $\frac{d_f}{e}|_{6(c)}= 2 |m_f| \ {{a}^{\rm finite}_{e}|_{6(c)}}$, we get
 \beq
\frac{d_e}{e}|_{6(c)}= \frac{d_u}{e}|_{6(c)}\equiv 10^{-66} GeV^{-1} \equiv 10^{-80} cm.
\eeq
Hence, the overall contribution of EDM of electron as well as neutron/quark in case of one-loop Feynman diagrams involving gravitino is:
\beq
\frac{d_{e}}{e}|_{\tilde G}=\frac{d_{n}}{e}|_{\tilde G}=\frac{d_{e/u}}{e}|_{6(a)}+\frac{d_{e/u}}{e}|_{6(b)}+\frac{d_{e/u}}{e}|_{6(c)}\equiv 10^{-67}cm.
\eeq
{{\bf Sgoldstino contribution:}} In supersymmetric models, the sgoldstino is the bosonic component of the superfield corresponding to which there is an $F$-term (D-term) supersymmetry breaking. In our set up, supersymmetry is broken in the bulk sector and the scale of the same is governed by $F$-term (assuming that in dilute flux approximation $V_{D} << V_{F}$) corresponding to bulk fields ($F^{\tau_{S}},F^{\tau_{B}},{\cal G}^a$) where $\tau_{S}$ and ${\tau_{B}}$ correspond to `small' and `big' divisor volume moduli and ${\cal G}^a$ correspond to complexified NS-NS and RR axions. It was shown in \cite{gravitino_DM}, at $M_s$,  $|F^{\tau_{S}}|>|F^{{\cal G}^a}|, |F^{\tau_B}|$. From {\bf 4.1}, the requirement of the quark-quark-photon coupling to be the SM at the EW scale, we see that $|F^{\tau_B}|$ is the most dominant $F$-term at the EW scale. To obtain an estimate of the off-shell goldstino multiplet, we consider the same to be: $(\tau_B, \chi_{B},F^B)$, where $\tau_B$ is a complex scalar field. Here, we identify $\sigma_B$ with scalar(sgoldstino)field and $\rho_B$ with pseudo-scalar(sgoldstino) field.

{\emph {Mass of sgoldstino}}: The dominant contribution to $F$-term potential, at the string scale $M_s$, is given by $V = ||F^{\tau_S}||^2$, where $F^{{\bar\tau}_S}= e^{K/2} {\bar\partial}^{{\bar\tau}_S}\partial^{\beta}K D_{\beta}W$\footnote{We note that $e^{K({\tau}_{S,B},G^a,z^i, a_I,...)}
{\bar\partial}^{\bar{\cal I}}\partial^{\cal J}K({\tau}_{S,B},G^a,z^i, a_I,...) D_{\cal J}W D_{\bar{\cal I}}{\bar W}({\cal I}\equiv T_{s,b},G^a,z^i,a_I,...)=e^{K(\tau_{z,b},G^a,z^i,a_I,...)}{\bar\partial}^{\bar{\alpha}}\partial^\beta K(\tau_{S,B},G^a,z^i,a_I,...) D_\beta W D_{\bar\alpha}{\bar W}(\beta=\tau_{S,B},G^a,z^i,a_I)$; however $G_{{\cal I}{\bar{\cal J}}}=\partial_{\cal I}{\bar\partial}_{\bar{\cal J}} K(T_{S,B},G^a,z^i, a_I,...),
G_{\alpha{\bar\beta}}\neq\partial_\alpha{\bar\partial}_{\bar{\beta}}K(\tau_{S,B},G^a,z^i,a_I,...)$ as $\tau_{S,B}$ is not an ${\cal N}=1$ chiral coordinate.}. At the EW scale the $F$-term potential receives the dominant contribution
from the $||D_{\tau_B}W||^2$ term and is estimated to be: $V(n_s=2)|_{EW}\sim e^KK^{\tau_S{\bar\tau}_B}D_{\tau_S}W D_{{\bar\tau}_B}{\bar W} + e^KK^{\tau_B{\bar\tau}_B}|D_{\tau_B}W|^2$, near $\langle\sigma_S\rangle\sim\frac{ln {\cal V}}{({\cal O}(1))_{\sigma_S}^4}, \langle\sigma_B\rangle\equiv  \frac{{\cal V}^{\frac{2}{3}}}{({\cal O}(1)_{\sigma_B})^4}$ yields $$\frac{\partial^2V}{\partial\sigma_B^2}\biggr|_{EW}\equiv  {\cal V}^{-\frac{1}{3}}m_{3/2}^2({\cal O}(1)_{\sigma_B})^2\left(({\cal O}(1)_{\sigma_S})^2 + \frac{({\cal O}(1)_{\sigma_S})^6}{ln {\cal V}}\right).$$ For the aforementioned ${\cal O}(1)_{\sigma_B}=\frac{{\cal O}(1)_{\sigma_S}}{2}\equiv 3.5$ for ${\cal V}\sim10^4,
 \left(\frac{\partial^2V}{\partial\sigma_B^2}\right)_{EW}\equiv {\cal V}^{\frac{4}{3}}$ and the canonically normalized coefficient quadratic in the fluctuations, yields the sgoldstino mass estimate:
$$m_{\tau_B} \sim \sqrt{\frac{\left(\frac{\partial^2V}{\partial\sigma_B^2}\right)_{EW}}{\kappa^2_4\mu_7 K^{\tau_B{\bar\tau}_B}_{EW}}}\sim {\cal O}(1)m_{3/2}.$$
It will be interesting to get the contribution of the same to electron/neutron EDM.

\begin{figure}
\begin{center}
\begin{picture}(100,137) (220,20)
   \Line(100,50)(110,50)
   \DashLine(110,50)(190,50){5}
   \Line(190,50)(200,50)
   \Line(110,50)(150,120)
   \Photon(150,120)(150,170){5}{4}
   \Line(150,120)(190,50)
   \Text(165,160)[]{$\gamma$}
   \Text(125,100)[]{${f}_{i}$}
   \Text(175,100)[]{${f}_{i}$}
   \Text(90,50)[]{{$f_{L}$}}
   \Text(210,50)[]{{$f_{R}$}}
   \Text(150,40)[]{{{${{\tau}_B}$}}}
   \Text(150,20)[]{{{$7(a)$}}}
      \Line(280,50)(290,50)
      \Line(290,50)(370,50)
      \Line(370,50)(380,50)
   \Photon(290,50)(330,120){5}{4}
   \Photon(330,120)(330,170){5}{4}
   \DashLine(330,120)(370,50){5}
   \Text(345,160)[]{{$\gamma$}}
   \Text(305,100)[]{${\tilde \gamma}$}
   \Text(355,100)[]{${\tau}_B$}
   \Text(270,50)[]{{$f_{L}$}}
   \Text(390,50)[]{{$f_{R}$}}
   \Text(330,40)[]{{{${f_L}$}}}
   \Text(330,20)[]{{{$7(b)$}}}
   \end{picture}
\end{center}
\caption{One-loop diagram involving sgoldstino.}
 \end{figure}
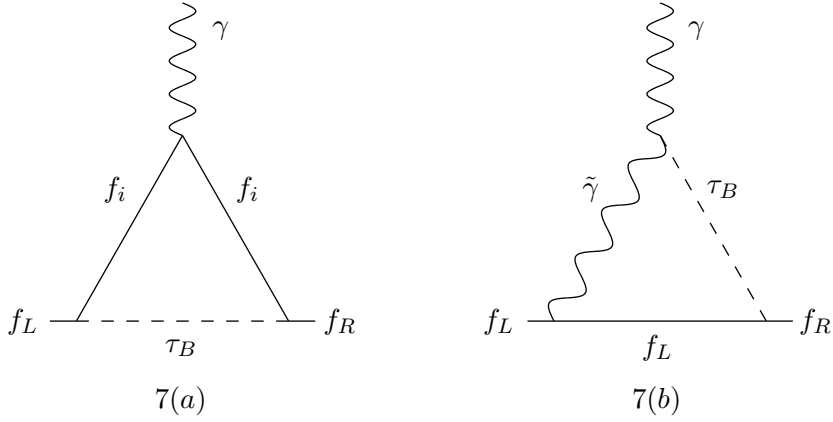

 To get the analysis of one-loop diagrams involving sgoldstino,  we consider only scalar sgoldstino field ($\sigma_s$) and first calculate the contribution of vertices involving sgoldstino in the context of ${\cal N}=1$ gauged supergravity. The coefficient of lepton($e_L$)-scalar(sgoldstino($\sigma_B$))-lepton($e_R$) vertex has been calculated by expanding $\frac{e^{\frac{K}{2}}}{2}\left({\cal D}_{{\cal A}_1} D_{{\cal A}_3}W\right){\bar \chi}^{{\cal A}_1}\chi^{{\cal A}_3}$ in the fluctuations linear in $\sigma_B$ in ${\cal N}=1$ gauged supergravity. By expanding above in the fluctuations linear in $\sigma_B\rightarrow \sigma_B+{\cal V}^{\frac{2}{3}}{M_P}$ , on simplifying,
we have: $\frac{e^{\frac{K}{2}}}{2} {\cal D}_{{\cal A}_1}D_{{\cal A}_3}W\chi^{{\cal A}_1}{\bar\chi}^{{\cal A}_3} \sim {\cal V}^{-\frac{13}{3}}\delta{\sigma_B}\chi^{{\cal A}_1}{\bar\chi}^{{\cal A}_3}$. The physical lepton($e_L$)-sgoldstino($\sigma_B$)-lepton($e_R$) vertex will be given as
\begin{eqnarray}
\label{eq:2a2}
|C_{{\delta \sigma_B} {e_{L}} {e^{c}_{R}}}| \equiv \frac{{\cal V}^{-\frac{13}{3}}}{{\sqrt{k^{2}_4 \mu_7 G^{\tau_B \bar {\tau_B}}\hat{K}_{{\cal A}_1{\bar{\cal A}}_1}{\hat{K}_{{\cal A}_3{\bar{\cal A}}_3}}}}}\equiv {\cal V}^{-\frac{92}{15}}, {\rm for}~{\cal V}\sim {10^5}.
\end{eqnarray}
Similarly,  the coefficient of quark($u_L$)-scalar(sgoldstino($\sigma_B$))-lepton ($u_R$) vertex can be calculated by expanding $\frac{e^{\frac{K}{2}}}{2}\left({\cal D}_{{\cal A}_2} D_{{\cal A}_4}W\right){\bar \chi}^{{\cal A}_2}\chi^{{\cal A}_4}$ in the fluctuations linear in $\sigma_B$ in $N=1$ gauged supergravity. Using the similar procedure, we get: $\frac{e^{\frac{K}{2}}}{2} {\cal D}_{{\cal A}_2}D_{{\cal A}_4}W\chi^{{\cal A}_2}{\bar\chi}^{{\cal A}_4}
\sim{\cal V}^{-4}\delta{\sigma_B}\chi^{{\cal A}_2}{\bar\chi}^{{\cal A}_4}$.
Therefore, the physical quark($u_L$)-sgoldstino ($\sigma_B$)-quark($u_R$)vertex will be given as
\begin{eqnarray}
\label{eq:2a2u}
|C_{{\delta \sigma_B} {u_{L}} {u^{c}_{R}}}|\sim \frac{{\cal V}^{-5}}{{\sqrt{k^{2}_4 \mu_7 G^{\tau_B \bar {\tau_B}}\hat{K}_{{\cal A}_2{\bar{\cal A}}_2}{\hat{K}_{{\cal A}_4{\bar{\cal A}}_4}}}}}\equiv {\cal V}^{-\frac{33}{5}},~{\rm for}~{\cal V}\sim {10^5}.
\end{eqnarray}
In ${\cal N}=1$ supergravity, the contribution of photon-sgoldstino(scalar)-photon will be accommodated by gauge kinetic term ${\cal L}= Re(T_B) F\wedge *_{4}F$,
where $Re(T_B)=  \sigma_B - C_{i \bar j}a_i a_{\bar j}$.  Considering $\sigma_B \rightarrow \langle \sigma_B \rangle+\delta \sigma_B $, coefficient of the physical vertex will be given as:
$|C_{\gamma\gamma \delta\sigma_B}| \equiv \frac{1/M_{p}}{\sqrt{k^{2}_4 \mu_7 G^{\tau_B \bar {\tau_B}}}}\sim \frac{{\cal V}^{-\frac{2}{3}}}{M_{p}}.$
The possibility of getting fermion-fermion-photon vertex $C_{f f^{*} \gamma}\equiv {\cal O}(1)$ has been shown in subsection {\bf 3.1}.

Now we use the values of coefficients of relevant vertices to evaluate the estimate of EDM for loop diagrams given in Figure~7(a) and 7(b). The diagrams have been evaluated in \cite{brignole et al} to determine the estimate of magnetic moment of muon in ${\cal N}=1$  global supersymmetry. Utilizing their results in a modified form in the context of ${\cal N}=1$ gauged SUGRA and the relation between magnetic moment and EDM as given above, for Figure~7(a), the magnitude of electric dipole moment will be
\beqn
|\frac{d_f}{e}|_{7(a)}=\frac{m_f}{16 \pi^2}\cos{\phi}\Bigl[ (C_{\delta\sigma_B f_L f^{c}_R})^2 \int^{1}_{0} dx \frac{x^{2}(2-x)}{m^{2}_{\sigma_B}(1-x) + m^{2}_{f}x^2} \Bigr].
\eeqn
Putting the value of $|C_{\delta\sigma_B e_L e^{c}_R}|\equiv {\cal V}^{-\frac{92}{15}}, |C_{\delta\sigma_B u_L u^{c}_R}|\equiv {\cal V}^{-\frac{33}{6}}$, and the value of masses $m_{\sigma_B}= m_{\frac{3}{2}}, m_{e}=0.5 MeV$, we get
\beqn
|\frac{d_e}{e}|_{7(a)}\equiv 10^{-95}cm,~ {\rm and}~|\frac{d_n}{e}|_{7(a)}\equiv 10^{-89}cm.
\eeqn
For Figure~7(b):
\beqn
&& {\hskip -0.25in} |\frac{d_f}{e}|_{7(b)}= \frac{C_{\delta\sigma_B f_L f^{c}_R}C_{\gamma\gamma \delta\sigma_B}}{8 \pi^2} \Bigl[ \Delta_{UV} - \frac{1}{2}-  \int^{1}_{0}dx \int^{1-x}_{0} dy  \log \Bigl[\frac {m^{2}_{\sigma_B}y + m^{2}_{f} x^2}{\mu^2} \Bigr] \Bigr].\nonumber\\
\eeqn
where $\Delta_{UV}= \log[\frac{\Delta^{2}_{\rm UV}}{\mu^2}]-1$.
Incorporating values of relevant inputs and considering the finite piece,
\beqn
&& |\frac{d_e}{e}|_{7(b)}\equiv 10^{-72}cm,  |\frac{d_n}{e}|_{7(b)}\equiv 10^{-68}cm.
\eeqn
Hence, the overall contribution of sgoldstino to EDM of electron/neutron is
\beqn
&&|\frac{d_{e}}{e}|_{sgoldstino} = |\frac{d_{e}}{e}|_{7(a)}+ |\frac{d_{e}}{e}|_{7(b)}\equiv10^{-72}cm,\\
&& \hskip -0.5in{\rm and}\nonumber\\
&&
|\frac{d_{n}}{e}|_{sgoldstino} = |\frac{d_{n}}{e}|_{7(a)}+ |\frac{d_{n}}{e}|_{7(b)}\equiv10^{-68}cm.
\eeqn
The results of all possible one-loop diagrams contributing to EDM of electron/neutron are summarized in a table given below:
\begin{table}[h]
\label{table:decay_lifetime}
\caption{Results of EDM of electron/neutron for all possible one-loop diagrams}  
\centering  
{\hskip -0.1in}
\begin{tabular}{l c c rrrrrrr}  
\hline\hline                       
 One-loop particle exchange  & origin of complex phase & $d_{e}$(e cm)  & $d_n$(e cm)
\\ [1.0ex]
\hline
{$\lambda^{0}{\tilde f}$} &  Diagonalized sfermion mass eigenstates & $10^{-39}$& $10^{-38}$    \\[1ex]
{$\chi^{0}_i \tilde f$} & ''& $10^{-37}$ & $ 10^{-34} $ & \\[1ex]
{$f \tilde f$}& ''& $10^{-45}$& $ 10^{-45}  $   \\[1ex]
{$f h^{0}_i$} &  Digonalized Higgs mass eigenstates  & $10^{-34}$ & $ 10^{-33}  $   \\[1ex]
{$\chi^{\pm} h^{0}_i$}  &  ''  & $10^{-32}$ & $ -  $   \\[1ex]
{${\rm gravitino}(\tilde G)\ {\tilde f}$}    &  Diagonalized sfermion mass eigenstates & $10^{-67}$ & $10^{-67}$   \\[1ex]
{${\rm sgoldstino}\ {\tilde f}$}   &Diagonalized sfermion mass eigenstates  & $10^{-72}$ & $10^{-68}$   \\[1ex]
 \hline                          

\end{tabular}
\label{tab:PPer}
\end{table}

\section{Two-loop Level Barr-Zee Type Contribution to Electric Dipole Moment}

In the two-loop diagrams discussed in this section, the CP-violating effects are mainly demonstarted by complex effective Yukawa couplings which include R-parity violating couplings, SM-like Yukawa couplings as well as couplings involving higgsino, and complex scalar trilinear couplings in the context of ${\cal N}=1$ gauged supergravity. In the subsection given below, we present the contribution of individual Barr-Zee type diagrams formed by including an internal fermion loop generated by R-parity violating interactions, SM-like Yukawa interactions and gaugino(gaugino)-higgsino(higgsino)-Higgs couplings. The two-loop diagrams are shown in Figure 8.

\subsection{Two-loop Level Barr-Zee Feynman Diagrams involving internal Fermion Loop}

\begin{figure}
\begin{center}
\begin{picture}(100,275) (290,-130)
   \ArrowLine(110,60)(300,60)
   \DashLine(130,60)(175,110){4}
   \Text(186,112)[]{${P_L}$}
   \CArc(200,120)(25,0,180)
   \ArrowArc(200,120)(25,180,0)
   \Photon(225,110)(280,60){3}{7}
   \Photon(200,145)(200,190){2}{4}
    \Text(150,100)[]{$h^{0}_i$}
   \Text(215,165)[]{$\gamma^0$}
    \Text(165,130)[]{$u(e)$}
   \Text(110,53)[]{{$e_{L}(u_L)$}}
   \Text(300,53)[]{{$e_{R}(u_R)$}}
   \Text(260,100)[]{{$\gamma^0$}}
   \Text(200,40)[]{{$8(a)$}}
 \ArrowLine(360,60)(550,60)
  \DashLine(380,60)(425,110){4}
  \CArc(450,120)(25,0,180)
   \ArrowArc(450,120)(25,180,0)
    \Photon(475,110)(530,60){3}{7}
   \Photon(450,145)(450,190){2}{4}
    \Text(400,100)[]{$h^{0}_i$}
   \Text(465,165)[]{$\gamma^0$}
    \Text(420,130)[]{$\chi^{\pm}_i$}
   \Text(360,53)[]{{$e_{L}(u_L)$}}
   \Text(550,53)[]{{$e_{R}(u_R)$}}
   \Text(510,100)[]{{$\gamma^0$}}
   \Text(450,40)[]{{$8(b)$}}
    \Line(235,-120)(425,-120)
    \DashLine(245,-120)(300,-60){4}
     \Text(310,-58)[]{${P_L}$}
     \CArc(325,-60)(25,0,180)
     \ArrowArc(325,-60)(25,180,0)
    \Photon(350,-60)(415,-120){3}{7}
    \Photon(325,-35)(325,10){2}{4}
    \Text(265,-80)[]{${\tilde \nu}_{iL}$}
   \Text(340,-15)[]{$\gamma^0$}
    \Text(290,-50)[]{$u(e)$}
   \Text(235,-127)[]{{$e_{L}(u_L)$}}
   \Text(425,-127)[]{{$e_{R}(u_R)$}}
   \Text(390,-80)[]{{$\gamma^0$}}
   \Text(325,-140)[]{{$8(c)$}}
\end{picture}
\end{center}
\caption{two-loop diagram involving fermions in the internal loop.}
 \end{figure}
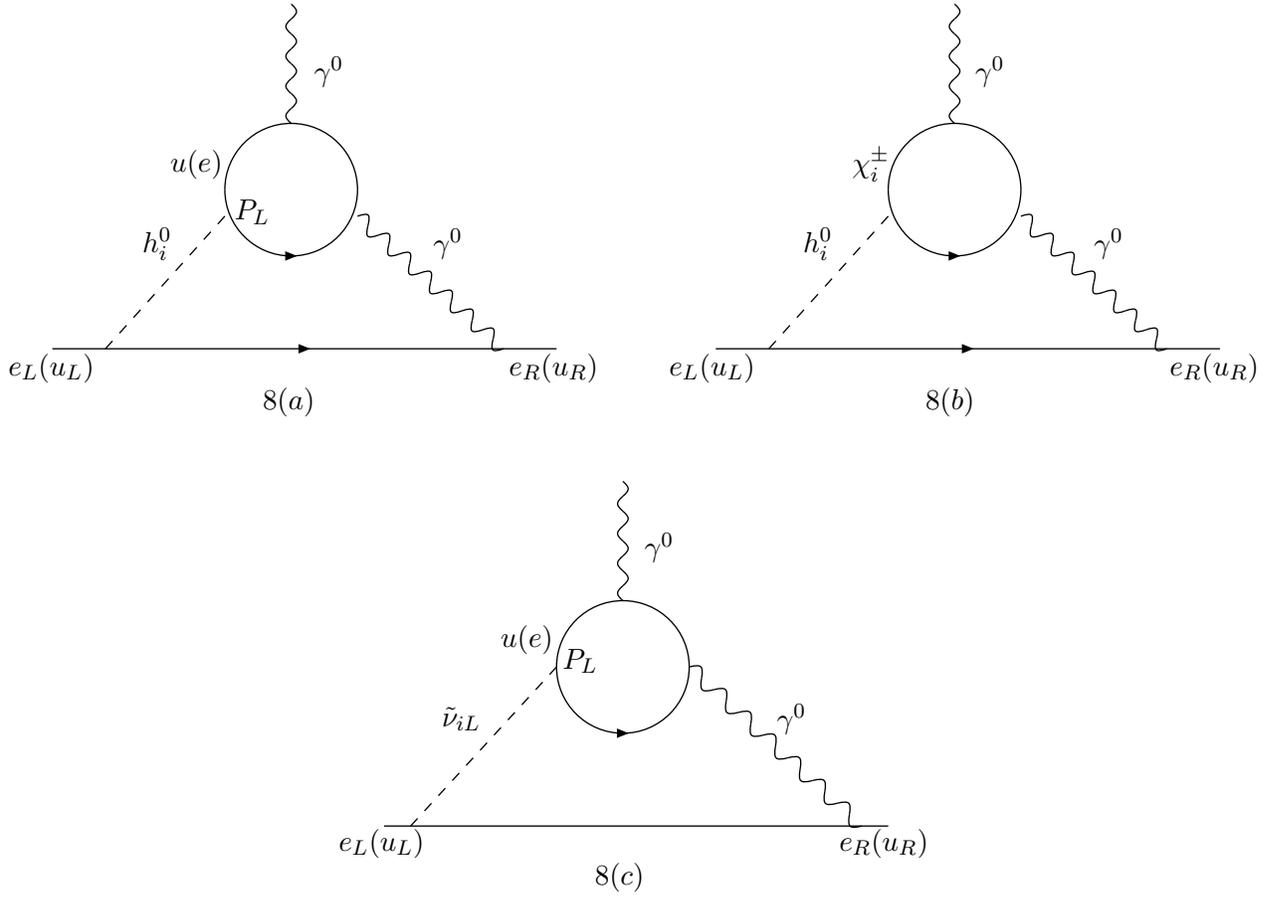
  {{\bf Higgs contribution:}}
For a two-loop diagram given in Figure~8(a), the interaction Lagrangian is governed by Yukawa couplings given as:
\beqn
{\cal L} \supset  {\hat Y}_{H^{0}_i u_{L} u^{c}_{R}}{H}^{0}_{i} u_{kL} u^{c}_{kR}+ {\hat Y}^{*}_{H^{0}_i e_{L} e^{c}_{R}} H^{0}_{i} e_{jL} e^{c}_{jR}+h.c..
\eeqn
We have already given the estimate of effective Yukawa couplings for first generation of leptons and quarks in \cite{gravitino_DM} in the context of ${\cal N}=1$ gauged supergravity. Using those results, we have
\beqn
 {\hat Y}_{H^{0}_i e_{L} e^{c}_{R}}\sim {\hat Y}^{\rm eff}_{{\cal Z}_i {\cal A}_1 {\cal A}_3}\equiv   {\cal V}^{-\frac{47}{45}}e^{i \phi_{y_e}}, {\hat Y}_{H^{0}_i u_{L} u^{c}_{R}}\sim {\hat Y}^{\rm eff}_{{\cal Z}_i {\cal A}_2 {\cal A}_4}\equiv  {\cal V}^{-\frac{17}{18}}e^{i \phi_{y_u}}; {\rm for~ {\cal V}=10^5},
\eeqn
where $e^{i \phi_{y_e}}$ and $e^{i \phi_{y_u}}$ are non-zero phases of aforementioned Yukawa couplings.

For a two-loop Barr-Zee diagram involving internal fermion loop and taking into account the chirality flip between internal loop and external line, the analytical expression has been derived in \cite{Pilaftsis, Yamanaka_fermion}. Using the same, the electric dipole moment of electron for a loop diagram given in Figure 8(a) will be\footnote{We consider $Q^{'}_{e}= C_{e e^{*}\gamma} Q_{e} \sim Q_e$ because $ C_{e e^{*}\gamma}\sim {\cal O}(1)$ as shown in {{\bf subsection 4.3}}. Similarly $Q^{'}_{u}\sim Q_{u}$.}:
\beqn
\label{eq:2loopdH1}
\frac{d}{e}|_{H}= \sum_{i=1,2} Im \left({\hat Y}_{H^{0}_i e_{L} e^{c}_{R}}{\hat Y}_{H^{0}_i u_{L} u^{c}_{R}} \right)\frac{\alpha_{em}Q^{2}_{u}Q_{e}}{16 \pi^3 m_{e j}} \left(f(z_1)- g(z_1)\right).
\eeqn
and EDM of neutron will be given as:
\beqn
\label{eq:2loopdH1n}
\frac{d}{n}|_{H}= \sum_{i=1,2} Im \left({\hat Y}_{H^{0}_i e_{L} e^{c}_{R}}{\hat Y}_{H^{0}_i u_{L} u^{c}_{R}} \right)\frac{\alpha_{em}Q^{2}_{e}Q_{u}}{16 \pi^3 m_{e j}} \left(f(z_2)- g(z_2)\right),
\eeqn
where $z_1= \frac{m^{2}_{e}}{m^{2}_{H^{0}_i}}; z_2= \frac{m^{2}_{u}}{m^{2}_{H^{0}_i}}$ and
\beqn
{\hskip -0.2in} && f(z)=\frac{z}{2}\int^{1}_{0} dx \frac{1- 2x(1-x)}{x(1-x)-z}{\rm ln} \left(\frac{x(1-x)}{z} \right), g(z)=\frac{z}{2}\int^{1}_{0} dx \frac{1}{x(1-x)-z}{\rm ln} \left(\frac{x(1-x)}{z} \right).
\eeqn
Using the value of masses $m_{H^{0}_1}=125 GeV$,  $m_{H^{0}_2}={\cal V}^{\frac{59}{72}}m_{\frac{3}{2}}$ and $m_{e}=0.5GeV$, $
f\left( \frac{m^{2}_{e}/m^{2}_{u}}{m^{2}_{H^{0}_1}}\right)= g\left( \frac{m^{2}_{e}/m^{2}_{u}}{m^{2}_{H^{0}_1}}\right)= 10^{-10}; f\left( \frac{m^{2}_{e}/m^{2}_{u}}{m^{2}_{H^{0}_2}}\right)= g\left( \frac{m^{2}_{e}/m^{2}_{u}}{m^{2}_{H^{0}_2}}\right)= 10^{-23}$,
Utilizing the same and assuming $e^{i (\phi_{y_e}-\phi_{y_u})}=(0,1]$, equation (\ref{eq:2loopdH1}) and (\ref{eq:2loopdH1n}) reduces to give EDM result as follows:
\beqn
\label{eq:2loopdH12}
\frac{d}{e}|_{H}=\frac{d}{n}|_{H}\sim {\cal V}^{-2} \times 10^{-2} \times 10^{-10}= 10^{-22}GeV^{-1} \equiv 10^{-36} cm.
\eeqn
{{\bf Chargino contribution:}} In a loop diagram 8(b), the general Lagrangian governing the interaction of chargino's will be as:
\beqn
{\cal L} \supset  {C}_{ikk}{H}^{0}_{i} {\chi}^{+}_{kL} {\chi}^{-}_{kR}+ {\hat Y}^{*}_{H^{0}_i{e_L}{e^{c}_R}}H^{0}_{i} e_{jL} e^{c}_{jR}+ {\hat Y}^{*}_{H^{0}_i{u_L}{u^{c}_R}}H^{0}_{i} u_{jL} u^{c}_{jR}+ h.c.
\eeqn
We evaluate the the contribution of chargino(${\chi}^{\pm}_i$)-Higgs-chargino (${\chi}^{\pm}_1$)  vertex in ${\cal N}=1 $ gauged supergravity. As described in appendix, ${\chi}^{\pm}_1$ and ${\chi}^{\pm}_2$  correspond to a higgsino (${\tilde H}^{\pm}_{i}$) with a very small admixture of gaugino (${\lambda}^{\pm}_{i}$) and vice versa. So $C_{\chi^{+}_i \chi^{-}_1 H^{0}_i}\equiv C_{{\tilde H^{+}_{i}{\tilde H^{-}_{i} H^{0}_i}}}$; and $C_{\chi^{+}_2 \chi^{-}_2 H^{0}_i} \equiv C_{{{\tilde \lambda}^{+}_i {\lambda}^{-}_{i}H^{0}_i}}$.

  \underline{Higgsino($\chi^{-}_{kL}$)-Higgs-higgsino($\chi^{+}_{kR}$) vertex}:
Given that higgsino is a majorana particle, therefore $\chi^{+}_{kR}= (\chi^{-}_{kL})^c$. In our model,  higgsino has been identified with position moduli ${\cal Z}_i$, the contribution of this vertex in ${\cal N}=1$ gauged sypergravity will be given by expanding $e^{\frac{K}{2}} {\cal D}_{{\cal Z}_i} {\cal D}_{\bar {\cal Z}_i}W $ in the fluctuations linear in ${\cal Z}_i$ about its stabilized VEV. Since $SU(2)_L$ symmetry is not conserved for this vertex, we will expand the above in the fluctuations quadratic in ${\cal Z}_i$; giving  VEV to one of the ${\cal Z}_i$.  Considering ${z_i} \rightarrow {\cal V}^{\frac{1}{18}}+ \delta {z_i}$; we have
${\cal D}_{z_i} {\cal D}_{\bar z_i}W= {\cal V}^{-\frac{16}{9}}z_i \langle z_i\rangle$. Using  ${\cal D}_{{\cal Z}_i} {\cal D}_{\bar {\cal Z}_i}W \sim {\cal D}_{ \bar {z_1}}D_{z_i}W$, the physical vertex will be given as:
$C_{\chi^{+}_i \chi^{-}_1 H^{0}_i} \equiv C_{{\tilde H^{+}_{i}{\tilde H^{-}_{i} H^{0}_i}}} =\frac{{\cal V}^{-\frac{7}{4}}}{(\sqrt{\hat{K}_{{\cal Z}_1{\bar{\cal Z}}_1}})^4}= {\cal V}^{\frac{1}{4}}e^{i{ \phi_{\chi_1}}}$
where ${{ \phi_{\chi_1}}}$ correspond to non-zero phase associated with the aforementioned coupling.

\underline{Gaugino($\lambda^{+}_{kR}$)-Higgs- gaugino ($\lambda^{+}_{kL}$) vertex}: The coefficient of this vertex will be given from the kinetic term of gaugino. The interaction term corresponding to this coupling will be given by considering term $ {\cal L}= i{\bar {\lambda_L}}{\gamma}^m {\frac{1}{4}(K_{{\cal Z}_i}{\partial}_m {{\cal Z}_i}- c.c.)}{\lambda}_{L}+(\partial_{{\cal Z}_i} T_B){\bar {\lambda_L}}{\gamma}^m {\frac{1}{4}(K_{{\cal Z}_i}{\partial}_m {{\cal Z}_i}\\-
 c.c.)}{\lambda}_{L}$, where  ${\lambda}_{L}$ corresponds to gaugino. Given that charged(gaugino's) are either $SU(2)_L$ singlets or triplets, the aforementioned vertex  does not preserve $SU(2)_L$ symmetry - one has to obtain the term bilinear in ${\cal Z}_i$ such that we give  VEV to one of the ${\cal Z}_i$. Since $(\partial_{{\cal Z}_i} T_B)$ does not contain terms bilinear in ${\cal Z}_i$ which is needed to ensure $SU(2)_L$ symmetry, second term contributes zero to the given vertex. In terms of undiagonalized basis,
$\partial_{z_i}K\sim{\cal V}^{-\frac{2}{3}}\langle z_i\rangle$, and using
$\partial_{{\cal Z}_i}K\sim  {\cal O}(1) \partial_{z_i}K$, we have:
$\partial_{{\cal Z}_i}K\sim {\cal V}^{-\frac{2}{3}}\langle{\cal Z}_i\rangle$, incorporating the same, we get
\begin{eqnarray}
\label{eq:Cggh}
 & & {\cal L}=  \frac{{\cal V}^{-\frac{2}{3}}\langle{\cal Z}_i\rangle{\bar {\lambda_L}}\frac{\slashed{\partial}{{\cal Z}_i}}{M_P}{\lambda_L}}{{\sqrt{(\hat{K}_{{\cal Z}_1{\bar {\cal Z}}_1}})^2}}\sim   {\cal V}^{\frac{13}{36}}h {\bar {\lambda_L}}\frac{\slashed{p}_h}{M_P}{\lambda_L}
 \sim {\cal V}^{\frac{13}{36}}{h}{\bar {\lambda_L}}\frac{{\gamma}\cdot({p_{e_L} + p_{e_R}})}{M_P}{\lambda}_{L}.
 \end{eqnarray}
Therefore,
$ C_{\chi^{+}_2 \chi^{-}_2 H^{0}_i} \equiv C_{H^{0}_i{{\lambda^{-}_{R}}}{\lambda}^{+}_{L}}\sim {\cal V}^{\frac{13}{36}}\frac{m_{e}}{M_P}e^{i{{ \phi_{{\tilde\lambda}^{0}_1}}}}$
where ${{ \phi_{{\tilde\lambda}^{0}_1}}}$ correspond to non-zero phase associated with the aforementioned coupling.

 The contribution of gaugino-gaugino-gauge boson as well as higgsino-higgsino-gauge boson have been already evaluated in the context of ${\cal N}=1$ gauged supergravity. The volume suppression factors corresponding to these vertices are as follows:
\beqn
\label{eq:vertices}
&& |C_{\chi^{+}_{1}\chi^{-}_{1}\gamma}|\equiv |C_{{\tilde H}^{+}_{u}{\tilde H}^{-}_{d}\gamma}|\equiv {\tilde f} {\cal V}^{-\frac{5}{18}};  |C_{\chi^{+}_{2}\chi^{-}_{2}\gamma}|\equiv  |C_{{\tilde \lambda}^{+}_{i}{\tilde \lambda}^{-}_{i}\gamma}|\equiv {\tilde f} {\cal V}^{-\frac{11}{18}}
\eeqn

Now, EDM of electron for a loop diagram given in Figure 8(b) will be given as:
\beqn
\label{eq:2loopdchi}
\frac{d_e}{e}|_{\chi^{\pm}_k}= \sum_{i= {1,2}}\sum_{k=1,2} Im \left( {\hat Y}_{H^{0}_i{e_L}{e^{c}_{R}}}{C}_{\chi^{+}_{k}\chi^{-}_{k}h} \right)\left({C}_{\chi^{+}_{k}\chi^{-}_{k}\gamma} \right)^2\frac{\alpha_{em}Q^{2}_{\chi_i}Q_{e}}{16 \pi^3 m_{\chi^{\pm}_k}} \left(f(z)- g(z)\right)\nonumber
\eeqn
and EDM of neutron for a loop diagram given in Figure 8(b) will be given as:
\beqn
\label{eq:2loopdchin}
\frac{d_{u/n}}{e}|_{\chi^{\pm}_k}= \sum_{i= {1,2}}\sum_{k=1,2} Im \left( {\hat Y}_{H^{0}_i{e_L}{e^{c}_{R}}}{C}_{\chi^{+}_{k}\chi^{-}_{k}h} \right)\left({C}_{\chi^{+}_{k}\chi^{-}_{k}\gamma} \right)^2\frac{\alpha_{em}Q^{2}_{\chi_i}Q_{u}}{16 \pi^3 m_{\chi^{\pm}_k}} \left(f(z)- g(z)\right)\nonumber
\eeqn
where $z= \frac{m^{2}_{\chi^{\pm}_k}}{m^{2}_{H^{0}_i}}; f\Bigl( \frac{m^{2}_{\chi^{\pm}_1}}{m^{2}_{H^{0}_1}}\Bigr)- g\Bigl( \frac{m^{2}_{\chi^{\pm}_1}}{m^{2}_{H^{0}_1}}\Bigr)= 10; f\Bigl( \frac{m^{2}_{\chi^{\pm}_1}}{m^{2}_{H^{0}_2}}\Bigr)- g\Bigl( \frac{m^{2}_{\chi^{\pm}_1}}{m^{2}_{H^{0}_2}}\Bigr)= 1, f\Bigl( \frac{m^{2}_{\chi^{\pm}_2}}{m^{2}_{H^{0}_1}}\Bigr)- g\Bigl( \frac{m^{2}_{\chi^{\pm}_2}}{m^{2}_{H^{0}_1}}\Bigr)= 10 ; f\Bigl( \frac{m^{2}_{\chi^{\pm}_2}}{m^{2}_{H^{0}_2}}\Bigr)- g\Bigl( \frac{m^{2}_{\chi^{\pm}_2}}{m^{2}_{H^{0}_2}}\Bigr)= 0.1$.
Considering  $({ \phi_{\chi_i}}-{\phi_{y_e}})= ({ \phi_{\chi_i}}-{\phi_{{\tilde\lambda}^{0}_1}})\sim (0,\frac{\pi}{2}]$; for $m_{\chi^{\pm}_1}= {\cal V}^{\frac{59}{72}}m_{\frac{3}{2}}$, $m_{\chi ^{\pm}_2}= {\cal V}^{\frac{2}{3}}m_{\frac{3}{2}}$, ${\hat Y}_{H^{0}_i{e_L}{e^{c}_{R}}}={\cal V}^{-\frac{47}{45}}$, ${\hat Y}_{H^{0}_i{e_L}{e^{c}_{R}}}={\cal V}^{-\frac{17}{18}}$; the value of EDM of electron and neutron will be given as:
\beqn
\label{eq:2loopdchi2}
\frac{d_e}{e}|_{\chi_i}=\frac{d_n}{e}|_{\chi_i}\sim  \frac{ {\tilde f}^2{\cal V}^{-\frac{8}{3}}}{m_{\frac{3}{2}}} \times 10^{-5} \times 10^{1}\equiv10^{-33}GeV^{-1} \equiv 10^{-47} cm.
\eeqn
{{\bf R Parity violating contribution:}}
For a loop diagram given in Figure 8(c), the Lagrangian governing the interaction of neutrino will correspond to R-parity violating interactions given as:
\beqn
{\cal L} \supset  {\tilde \lambda }_{{\tilde \nu}_L u_L u^{c}_R}{\nu}_{iL} u_{kL} u^{c}_{kR}+ {\tilde \lambda }_{{\tilde \nu}_L e_L e^{c}_R}{\nu}_{i L} e_{j L} e^{c}_{j R}+h.c..
\eeqn
The contribution of R-parity violating interaction terms ${\hat \lambda}_{ikk}$ and ${\hat \lambda}^{*}_{ijj}$
are given by expanding ${\cal D}_{{\cal A}_1}{\cal D}_{{\cal A}_3} W$ and ${\cal D}_{{\cal A}_2}{\cal D}_{{\cal A}_4} W$ in the fluctuations linear in ${\cal A}_{1}$  around its stabilized VEV. The values of the same have already been calculated in the context of ${\cal N}=1$ gauged supergravity action and given as follows:
\beqn
&& {\tilde \lambda }_{{\tilde \nu}_L e_L e^{c}_R} \equiv{\cal V}^{-\frac{5}{3}}e^{i \phi_{\lambda_e}}, {\tilde \lambda }_{{\tilde \nu}_L u_L u^{c}_R} \equiv{\cal V}^{-\frac{5}{3}}e^{i \phi_{\lambda_u}},
 \eeqn
 where $e^{i \phi_{\lambda_e}}$ and $e^{i \phi_{\lambda_u}}$ are non-zero phases corresponding to above-mentioned complex R-parity violating couplings. The EDM of electron in this case will be:
\beqn
\label{eq:2loopdnu}
\frac{d}{e}|_{RPV}= Im \left( {\tilde \lambda }_{{\tilde \nu}_L e_L e^{c}_R}{\tilde \lambda }_{{\tilde \nu}_L u_L u^{c}_R} \right)\frac{\alpha_{em}Q^{2}_{u}Q_{e}}{16 \pi^3 m_{e j}} \left(f(z_1)- g(z_1)\right),
\eeqn
and EDM of neutron will be given as:
\beqn
\label{eq:2loopdnun}
\frac{d}{u/n}|_{RPV}= Im \left( {\tilde \lambda }_{{\tilde \nu}_L e_L e^{c}_R}{\tilde \lambda }_{{\tilde \nu}_L u_L u^{c}_R} \right)\frac{\alpha_{em}Q^{2}_{e}Q_{u}}{16 \pi^3 m_{e j}} \left(f(z_2)- g(z_2)\right).
\eeqn
where $z_1= \frac{m^{2}_{e}}{m^{2}_{\nu_{iL}}}; z_2= \frac{m^{2}_{u}}{m^{2}_{\nu_{iL}}}$. Using the value of masses $m_{{\nu}_{iL}}= {\cal V}^{\frac{1}{2}}m_{\frac{3}{2}}$, $m_{e}=0.5GeV$ and $m_{u}={\cal O}(1)$, $f\left( \frac{m^{2}_{e}/m^{2}_{u}}{m^{2}_{{\nu}_{iL}}}\right)= g\left( \frac{m^{2}_{e}/m^{2}_{u}}{m^{2}_{{\nu}_{iL}}}\right)= 10^{-27}$,  and assuming $( \phi_{\lambda_e}-\phi_{\lambda_u})= (0,\frac{\pi}{2}]$; equation (\ref{eq:2loopdnu}) and (\ref{eq:2loopdnun}) reduce to give EDM result as follows:
\beqn
\frac{d}{e}|_{RPV}= \frac{d}{n}|_{RPV} \sim {\cal V}^{-\frac{10}{3}} \times 10^{-2} \times 10^{-27}\equiv 10^{-55}GeV^{-1} \equiv 10^{-70} cm.
\eeqn

\subsection{Two-loop Level Barr-Zee Feynman Diagrams involving Internal Sfermion Loop}

In this subsection, we evaluate the contribution of heavy sfermion loop generated by trilinear scalar interactions including Higgs. The loop diagrams are mediated by $\gamma h$ exchange. Unlike one-loop diagrams, here we do not have to consider the mixing sleptons(squarks) because of the fact that non-zero phase associated with complex scalar trilinear interaction is sufficient to generate non-zero EDM of elctron/neutron. We first evaluate the contribution of relevant vertices in the context of ${\cal N}=1$ gauged supergravity for two-loop diagrams shown in Figure 9.

\begin{figure}
\begin{center}
\begin{picture}(100,200) (290,20)
   \ArrowLine(110,60)(300,60)
   \DashLine(130,60)(175,110){4}
   \DashCArc(200,120)(25,0,180){5}
   \DashCArc(200,120)(25,180,0){5}
   \Photon(225,110)(280,60){3}{7}
   \Photon(200,145)(200,190){2}{4}
    \Text(150,100)[]{$h^{0}_i$}
   \Text(215,165)[]{$\gamma^0$}
    \Text(160,130)[]{${\tilde u_j}({\tilde e_j})$}
   \Text(110,53)[]{{$e_{L k}(u_{L k})$}}
   \Text(295,53)[]{{$e_{R k}(u_{R k})$}}
   \Text(255,100)[]{{$\gamma^0$}}
   \Text(200,30)[]{{$9(a)$}}
 \ArrowLine(360,60)(550,60)
  \DashLine(380,60)(425,110){4}
  \DashCArc(450,120)(25,0,180){5}
   \DashArrowArc(450,120)(25,180,0){5}
    \Photon(475,110)(510,60){3}{6}
   \Photon(475,110)(500,160){3}{5}
    \Text(400,100)[]{$h^{0}_i$}
   \Text(515,165)[]{$\gamma^0$}
    \Text(410,130)[]{${\tilde u_j}({\tilde e_j})$}
   \Text(367,53)[]{{$e_{L k}(u_{L k})$}}
   \Text(550,53)[]{{$e_{R k}(u_{R k})$}}
   \Text(500,100)[]{{$\gamma^0$}}
   \Text(450,30)[]{{$9(b)$}}
\end{picture}
\end{center}
\caption{Two-loop diagram involving sfermions in the internal loop.}
\end{figure}
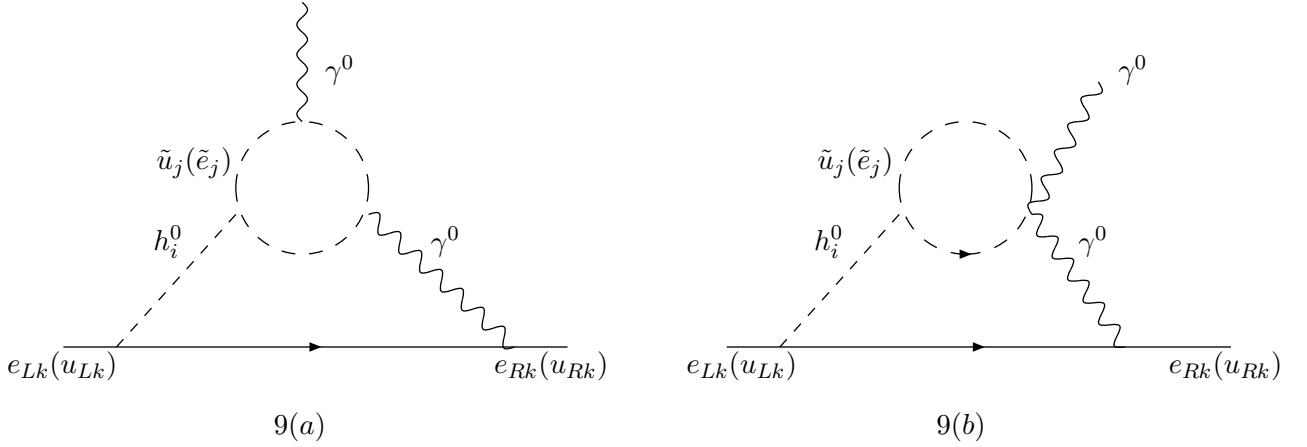
\underline{Slepton(${\tilde e}_{jR}$)-slepton(${\tilde e}_{j R}$)-Higgs vertex:} By expanding effective supergravity potential
$V|_{EW}\sim e^KK^{\tau_S{\bar\tau}_B}D_{\tau_S}W D_{{\bar\tau}_B}{\bar W} + e^KK^{\tau_B{\bar\tau}_B}|D_{\tau_B}W|^2$ in the fluctuations linear in ${\cal Z}_i \rightarrow {\cal Z}_i +{\cal V}^{\frac{1}{36}}M_P$ , ${\cal A}_3 \rightarrow {\cal A}_3 +{\cal V}^{-\frac{13}{18}}M_P$,  contribution of the term quadratic in ${\cal A}_3$ as well as ${\cal Z}_i$ is of the order $ {\cal V}^{-\frac{59}{36}}\langle{\cal Z}_i\rangle $, which after giving VEV to  one of the ${\cal Z}_i$, will be given as:

\begin{equation}
\label{eq:CeReRh}
 C_{{\tilde e_R}{\tilde e_R}^{*} H^{0}_i}\equiv \frac{1}{{\sqrt{(\hat{K}_{{\cal Z}_i{\bar{\cal Z}}_i})^2 (\hat{K}_{{\cal A}_3{\bar{\cal A}}_3})^2}}}\left[{\cal V}^{-\frac{59}{36}}\langle {\cal Z}_i\rangle\right]\equiv ( {\cal V}^{-2}M_P) e^{i\phi_{{\tilde e_R}}}.
\end{equation}
where $\phi_{{\tilde e_R}}$ is non-zero phase corresponding to aforementioned complex scalar 3-point interaction vertex. Using the similar procedure, the coefficient of slepton(${\tilde e}_{jL}$)-slepton(${\tilde e}_{j L}$)-Higgs vertex will be given as:
    \beqn
    C_{{\tilde e_L}{\tilde e_L}^{*} H^{0}_i}\equiv   \frac{1}{{\sqrt{(\hat{K}_{{\cal Z}_i{\bar{\cal Z}}_i})^2 (\hat{K}_{{\cal A}_1{\bar{\cal A}}_1})^2}}}\left[{\cal V}^{-\frac{95}{36}}\langle {\cal Z}_i\rangle\right]\equiv( {\cal V}^{-\frac{12}{5}}M_P) e^{i\phi_{{\tilde e_L}}}.
    \eeqn
$\phi_{{\tilde e_L}}$ is non-zero phase corresponding to this particular complex scalar 3-point interaction vertex.

 \underline{Squark(${\tilde u}_{jR}$)-squark(${\tilde u}_{j R}$)-Higgs vertex:} By expanding $V|_{EW}$ in the fluctuations around  ${\cal Z}_i \rightarrow {\cal Z}_i +{\cal V}^{\frac{1}{36}}M_P$ , ${\cal A}_4 \rightarrow {\cal A}_4 +{\cal V}^{-\frac{11}{9}}M_P$,  contribution of term quadratic in ${\cal A}_4$ as well as ${\cal Z}_i$ is of the order $ {\cal V}^{-\frac{23}{36}}\langle{\cal Z}_i\rangle $, which after giving VEV to  one of the ${\cal Z}_i$, will be given as:
\begin{equation}
\label{eq:CuRuRh}
 C_{{\tilde u_R}{\tilde u_R}^{*} H^{0}_i}\equiv \frac{1}{{\sqrt{(\hat{K}_{{\cal Z}_i{\bar{\cal Z}}_i})^2 (\hat{K}_{{\cal A}_4{\bar{\cal A}}_4})^2}}}\left[{\cal V}^{-\frac{23}{36}}\langle {\cal Z}_i\rangle\right]\equiv( {\cal V}^{-2}M_P) e^{i\phi_{{\tilde u_R}}}.
\end{equation}
where $\phi_{{\tilde u_R}}$ is non-zero phase corresponding to aforementioned complex scalar 3-point interaction vertex.

\underline{Squark(${\tilde u}_{jL}$)-squark(${\tilde u}_{j L}$)-Higgs vertex:} By expanding $V|_{EW}$ in the fluctuations around  ${\cal Z}_i \rightarrow {\cal Z}_i +{\cal V}^{\frac{1}{36}}M_P$ , ${\cal A}_2 \rightarrow {\cal A}_2 +{\cal V}^{-\frac{1}{3}}M_P$,  contribution of term quadratic in ${\cal A}_2$ as well as ${\cal Z}_i$ is of the order $ {\cal V}^{-\frac{131}{36}}\langle{\cal Z}_i\rangle $, which after giving VEV to  one of the ${\cal Z}_i$, will be given as:
\begin{equation}
\label{eq:CuLuLh}
 C_{{\tilde u_L}{\tilde u_L}^{*} H^{0}_i}\equiv \frac{1}{{\sqrt{(\hat{K}_{{\cal Z}_i{\bar{\cal Z}}_i})^2 (\hat{K}_{{\cal A}_2{\bar{\cal A}}_2})^2}}}\left[{\cal V}^{-\frac{131}{36}}\langle {\cal Z}_i\rangle\right]\equiv ( {\cal V}^{-\frac{20}{9}}M_P) e^{i\phi_{{\tilde u_L}}}.
\end{equation}
where $\phi_{{\tilde u_L}}$ is non-zero phase corresponding to aforementioned complex scalar 3-point interaction vertex.

The contribution of  slepton (${\tilde e}_{j R}$)-slepton (${\tilde e}_{j R}$)-photon($\gamma$)- photon($\gamma$) vertex  will be given by:
${\bar{\partial}}_{{\bar {\cal A}}_{3}}\partial_{{\cal A}_3}G_{T_B{\bar T}_B} X^{T_B} X^{\bar {T_B}} {A}^{\mu} A_{\nu}$. On solving: $
{\bar{\partial}}_{{\bar {\cal A}}_{3}}\partial_{{\cal A}_3}G_{T_B{\bar T}_B} \sim {\cal V}^{\frac{1}{9}}{\cal A}^{*}_{1}{\cal A}_{1}$, Incorporating values of $X^B$ as mentioned earlier, the real physical slepton(${\tilde e}_{jR}$)-slepton(${\tilde e}_{j R}$)-photon($\gamma$)-photon($\gamma$) vertex is proportional to
\begin{equation}
\label{eq:CllZZ}
C_{{\tilde e}_R {\tilde e}^{*}_R \gamma \gamma}\equiv \frac{{\cal V}^{\frac{1}{9}}{\tilde f}^2 {\cal V}^{-\frac{4}{3}}}{\sqrt{(K_{{\cal A}_3 {\cal A}_3})^2}}\equiv  {\tilde f}^2 {\cal V}^{-\frac{13}{5}}.
\end{equation}
   The coefficient of  real physical ${\tilde e}_{j L}$ - ${\tilde e}_{j L}$ $\gamma$- $\gamma$ vertex has been obtained in \cite{gravitino_DM}. The value of the same is given by $C_{{\tilde e}_L {\tilde e}^{*}_L \gamma \gamma}\equiv {\tilde f}^2 {\cal V}^{-3}$.
 Similarly, the coefficient of real  physical ${\tilde u}_{j R}-{\tilde u}_{j R}-\gamma-\gamma$ vertex  will be given by:
${\bar{\partial}}_{{\bar {\cal A}}_{4}}\partial_{{\cal A}_4}G_{T_B{\bar T}_B} X^{T_B} X^{\bar {T_B}} {A}^{\mu} A_{\nu}$. On solving, the volume suppression factor corresponding to this  vertex will be given as:
\begin{equation}
C_{{\tilde u}_R {\tilde u}^{*}_R \gamma \gamma}\sim \frac{{\rm coefficient~of}~{\bar{\partial}}_{{\bar {\cal A}}_{4}}\partial_{{\cal A}_4}G_{T_B{\bar T}_B}}{\sqrt{(K_{{\cal A}_4 {\cal A}_4})^2}}\equiv {\tilde f}^2 {\cal V}^{-\frac{118}{45}}.
\end{equation}
The coefficient of  real physical (${\tilde u}_{jL}-{\tilde u}_{j L}-\gamma-\gamma$) vertex  will be given by:
${\bar{\partial}}_{{\bar {\cal A}}_{2}}\partial_{{\cal A}_2}G_{T_B{\bar T}_B}
X^{T_B} X^{\bar {T_B}} {A}^{\mu} A_{\nu}$. On solving, the volume suppression factor corresponding to this  vertex will be given as:
\begin{equation}
C_{{\tilde u}_L {\tilde u}^{*}_L \gamma \gamma}\sim \frac{{\rm coefficient~of}~{\bar{\partial}}_{{\bar {\cal A}}_{2}}\partial_{{\cal A}_2}G_{T_B{\bar T}_B}}{\sqrt{(K_{{\cal A}_2 {\cal A}_2})^2}}\equiv {\tilde f}^2{\cal V}^{-\frac{127}{45}}.
\end{equation}
The contribution of  real scalar-scalar-photon vertices have already been obtained in section {\bf 2}, and given as:
\beqn
&& C_{{\tilde e}_L {\tilde e}^{*}_L \gamma}\equiv ({\tilde f} {\cal V}^{\frac{44}{45}}){\tilde {\cal A}_1} A^\mu\partial_\mu {\tilde {\cal A}}_{1}, C_{{\tilde e}_R {\tilde e}^{*}_R \gamma} \equiv ({\tilde f}{\cal V}^{\frac{53}{45}}){\cal A}_3 A^\mu\partial_\mu {\bar {\cal A}}_{3}, \nonumber\\
&& C_{{\tilde u}_L {\tilde u}^{*}_L \gamma}\equiv({\tilde f}{\cal V}^{\frac{53}{45}}){\tilde {\cal A}_2} A^\mu\partial_\mu {\tilde {\cal A}}_{2}, C_{{\tilde u}_R {\tilde u}^{*}_R \gamma}\equiv ({\tilde f}{\cal V}^{\frac{62}{45}}){\tilde {\cal A}_4} A^\mu\partial_\mu {\tilde {\cal A}}_{4}.
\eeqn
The analytical expression for the EDM involving sfermion/scalar in an internal loop has been provided in \cite{Yamanaka_sfermion}. Using the same, for Figure~9(a), EDM of electron will be given as:
\beqn
\label{eq:deksfermion}
{\hskip -0.1in} \frac{d_{e}}{e}|^{\rm sfermion}_{4.9(a)}= \sum_{i=1,2} \ \sum_{{j={{\tilde u}_L, {\tilde u}_R}}} Im ({\hat Y}_{H^{0}_ie_L e^{c}_R}{C_{H^{0}_i j j^{*}}})(C_{j j^{*}\gamma})^2\times \frac{\alpha_{em} \eta_c Q_{e_j}q^{2}_{j}}{32 \pi^3 m^{2}_{H^{0}_i}}F({\tilde z}),
\eeqn
and EDM of neutron/quark will be given as:
\beqn
\label{eq:deksfermionn}
{\hskip -0.1in}  \frac{d_{n}}{e}|^{\rm sfermion}_{4.9(a)}= \sum_{i=1,2} \ \sum_{{j={{\tilde e}_L, {\tilde e}_R}}} Im ({\hat Y}_{H^{0}_iu_L u^{c}_R}{C_{H^{0}_i j j^{*}}})(C_{j j^{*}\gamma})^2\times \frac{\alpha_{em} \eta_c Q_{u_j}q^{2}_{j}}{32 \pi^3 m^{2}_{H^{0}_i}}F({\tilde z}).
\eeqn
where $z=\frac{m^{2}_{j}}{m^{2}_{H^{0}_i}};
F(z)=-\int^{1}_{0} dx \frac{x(1-x)}{x(1-x)-z}{\rm ln} \left(\frac{x(1-x)}{z} \right)$.
Considering $(\phi_{{\tilde u_{L/R}}}-\Phi_{y_e})=(\phi_{{\tilde e_{L/R}}}-\Phi_{y_u})=(0,\frac{\pi}{2}]$; $|{\hat Y}_{H^{0}_ie_L e^{c}_R}|\equiv {\cal V}^{-\frac{47}{45}}, |{\hat Y}_{H^{0}_iu_L u^{c}_R}|\equiv {\cal V}^{-\frac{19}{18}}$ and using the value of masses, $m_{{\tilde e}_L}=m_{{\tilde e}_R}= m_{{\tilde u}_L}=m_{{\tilde u}_R}={\cal V}^{\frac{1}{2}}m_{\frac{3}{2}}$, $m_{H^{0}_1}=125 GeV$ and $m_{H^{0}_2}={\cal V}^{\frac{59}{72}}m_{\frac{3}{2}}$ , we have $F\Bigl(\frac{m^{2}_{j}}{m^{2}_{H^{0}_1}}\Bigr)= 10^{-17} , F\Bigl(\frac{m^{2}_{j}}{m^{2}_{H^{0}_2}}\Bigr)= 1$. Incorporating the value of interaction vertices, equation (\ref{eq:deksfermion}) and (\ref{eq:deksfermionn}) reduce to give EDM results as follows:
\beqn
\label{eq:deksfermion1}
&&\frac{d_{e}}{e}|^{\rm sfermion}_{9(a)}= 10^{-8}\times {\cal V}^{-\frac{4}{15}}{\tilde f}^2 \equiv10^{-15} GeV^{-1}\equiv 10^{-29}cm;~{\rm for~{\cal V}=10^4},\nonumber\\
&&
\frac{d_{n}}{e}|^{\rm sfermion}_{9(a)}= 10^{-8}\times {\cal V}^{-\frac{3}{10}}{\tilde f}^2\equiv 10^{-15} GeV^{-1}\equiv 10^{-29}cm;~{\rm for~{\cal V}= 10^4}.
\eeqn
For a loop diagram given in Figure 9(b), EDM of electron will be given as:
\beqn
\frac{d_{e}}{e}|^{\rm sfermion}_{9(b)}= \sum_{i=1,2} \ \sum_{{j={{\tilde u}_L, {\tilde u}_R}}} Im ({\hat Y}_{H^{0}_ie_L e^{c}_R}{C_{H^{0}_i j j^{*}}})(C_{j j^{*}\gamma \gamma})\times \frac{\alpha_{em}Q_{e_j}q^{2}_{j}}{32 \pi^3 m^{2}_{H^{0}_i}}F({\tilde z}),
\eeqn
and EDM of neutron will be given as:
\beqn
{\hskip -0.1in} \frac{d_{n/u}}{e}|^{\rm sfermion}_{9(b)}= \sum_{i=1,2} \ \sum_{{j={{\tilde e}_L, {\tilde e}_R}}} Im ({\hat Y}_{H^{0}_iu_L u^{c}_R}{C_{H^{0}_i j j^{*}}})(C_{j j^{*}\gamma \gamma})\times \frac{\alpha_{em}Q_{u_j}q^{2}_{j}}{32 \pi^3 m^{2}_{H^{0}_i}}F({\tilde z}).
\eeqn
where $F\left(\frac{m^{2}_{j}}{m^{2}_{H^{0}_1}}\right)= 10^{-17}, F\left(\frac{m^{2}_{j}}{m^{2}_{H^{0}_1}}\right)= 1 $.
Incorporating the value of masses and estimate of relevant coupling veretx, EDM of electron will be
 \beqn
\frac{d_{e}}{e}|^{\rm sfermion}_{9(b)}= 10^{-9} \times {\cal V}^{-\frac{17}{3}}{\tilde f}^2 \equiv 10^{-43} GeV^{-1}\equiv 10^{-57}cm.
\eeqn
The EDM of neutron in this case will be given as:
 \beqn
\frac{d_{n}}{e}|^{\rm sfermion}_{9(b)}= 10^{-9} \times {\cal V}^{-\frac{91}{18}}{\tilde f}^2 \equiv 10^{-40} GeV^{-1}\equiv 10^{-54}cm.
\eeqn
The overall contribution of EDM of electron as well as neutron corresponding to 2-loop diagram involving sfermions is:
\beqn
\frac{d_{e/n}}{e}|^{\rm sfermion}= \frac{d_{e/n}}{e}|^{\rm sfermion}_{9(a)} +\frac{d_{e/n}}{e}|^{\rm sfermion}_{9(b)}\equiv10^{-29}cm.
\eeqn
\subsection{Two-loop level Barr-Zee Feynman diagram involving $W^{\pm}$ boson in the internal loop}

In this subsection, we discuss the important contribution of Barr-Zee diagram involving W boson as an internal loop. In the 1-loop as well as 2-loop diagrams discussed so far, we have discussed the contribution mediated by Higgs exchange. The non-zero phases in 1-loop diagram are affected by considering a mixing between Higgs doublet in $\mu$ split SUSY model while in 2-loop diagrams, the phases are affected through complex effective Yukawa coupling.  It has been found in \cite{Leigh et al} that two-loop graphs involving W boson loop can induce electric dipole moment of $d_e$ of the order of the experimental bound ($10^{-27} cm$) in the multi-Higgs models provided there is an exchange of Higgs  in the Higgs propagator and the CP violation in the neutral Higgs sector is fairly maximal. The approach was given by S. Weinberg in \cite{weinberg_Higgs, weinberg_Higgs_2}. In these papers, he pointed out that dimension-six purely gluonic operator gives a large value for the EDM of the neutron, which is just below the
present experimental bound if one considers CP violation through the exchange of
Higgs particles, whose interactions involve one or more complex phases. The approach was extended by Barr and Zee who have found that Higgs exchange can also give an electric dipole moment to the electron of the order of experimental limits by considering an EDM operator involving a top quark also.  In this spirit, we present an analysis of EDM of electron/neutron involving W boson loop in the context of $\mu$ split-SUSY model which, as already discussed involves a light Higgs and a heavy Higgs doublet.

 In the notations of Weinberg, the CP-violating phase can appear from the neutral Higgs-boson exchange through imaginary terms in the amplitude and Higgs propagators are represented as: $
A(q^2)= \sqrt{2}G_f \sum_{n} \frac{Z_n}{q^2+ m^{2}_{H_n}}$,
where $Z_n$ is non-zero phase appearing due to exchange between Higgs doublet in the propagator.
We address this argument of generation of non-zero phase in the ${\cal N}=1$ gauged supergravity action.
We first provide the analysis of  required SM-like coupling involved in Figure~10 in the context of ${\cal N}=1$  gauged supergravity action.
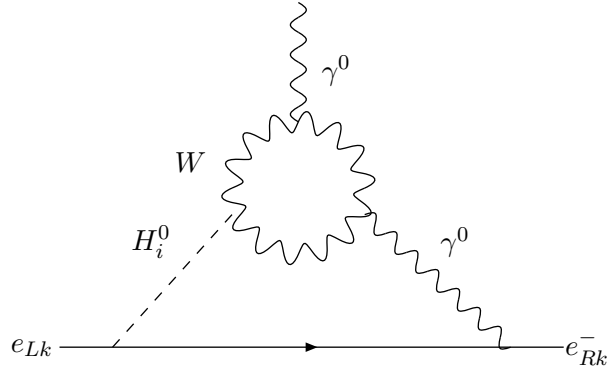
\begin{figure}
\begin{center}
\begin{picture}(100,200) (170,0)
   \ArrowLine(110,60)(300,60)
   \DashLine(130,60)(175,110){4}
   \PhotonArc(200,120)(25,0,180){4}{7}
   \PhotonArc(200,120)(25,180,0){4}{7}
   \Photon(225,110)(280,60){3}{7}
   \Photon(200,145)(200,190){3}{4}
    \Text(145,100)[]{$H^{0}_i$}
   \Text(215,165)[]{$\gamma^0$}
    \Text(160,130)[]{$W$}
   \Text(100,60)[]{{$e_{L k}$}}
   \Text(310,60)[]{{$e^{-}_{R k}$}}
   \Text(260,100)[]{{$\gamma^0$}}
  \end{picture}
\end{center}
\vskip -0.8in
\caption{Two-loop diagram involving W boson in the internal loop.}
\end{figure}

The contribution of $W^{+}$-photon-$W^{-}$ vertex is evaluated by CP-even interaction term given as \cite{tripleboson}$:
{\cal L}= - Re(f)A^{\mu}W^{-}_{\mu\nu}W^{+\nu}+ Re(f)W^{+\mu}W^{-\nu}F_{\mu\nu}$, where $W^{-}_{\mu\nu}=\partial_{\mu}W^{-}_{\nu}-\partial_{\nu}W^{-}_{\mu}$ and $F_{\mu\nu}=\partial_{\mu}A_{\nu}-\partial_{\nu}A_{\mu}$ and $ Re(f)$ is gauge kinetic function, which in our set-up is given by 'big' divisor volume modulus $Re(T_B)\sim{\cal V}^{\frac{1}{18}}\equiv {\cal O}(1)$ for Calabi-Yau ${\cal V}=10^5$.
Therefore, the volume suppression corresponding to this interaction vertex is:
$C_{W^{+}W^{-}\gamma}\equiv {\cal V}^{\frac{1}{18}}\equiv {\cal O}(1).$

The effective $W^{+}$-Higgs-$W^{-}$ vertex can be evaluated in the effective supergravity action as follows.
 Consider the gauge kinetic term: $Re(T) F^2$ and then choose the term $C_{1{\bar3}} a_1 a_{\bar3}$ in $Re(T_B)$ with the understanding that one first gives  VEV to the predominantly $SU(2)_L$-doublet valued $a_1$, then one picks out the ${\cal Z}$-dependent contribution in $a_3$ and also use the value of the intersection component $C_{1{\bar 3}}$. One will therefore consider:
$C_{1{\bar 3}} \langle a_1\rangle{\cal V}^{-\frac{7}{5}}{\cal Z}
\frac{p_1\cdot p_2}{\sqrt{K_{{\cal A}_1{\bar{\cal A}}_1}}\sqrt{K_{{\cal Z}{\bar{\cal Z}}}}\left(\sqrt{Re(T)}\right)^2}$; $K_{{\cal Z}{\bar{\cal Z}}}|_{M_s}\sim10^{-5}, K_{{\cal A}_1{\bar{\cal A}}_1}|_{M_s}\sim10^4$ which at the EW scale we will assume to be $\frac{10^{-5}}{({\cal O}(1))^2}$ and $\frac{10^4}{({\cal O}(1))^2}$. For non-relativistic gauge bosons, $p_1\cdot p_2\sim M_{W/Z}^2, Re(T)|_{EW} \sim {\cal O}(1)M_P\sim v {\cal V}^3\ {\rm GeV}, C_{1{\bar 3}}\sim {\cal V}^{\frac{29}{18}}, \langle a_1\rangle|_{EW}\sim{\cal O}(1){\cal V}^{-\frac{2}{9}}M_P$ (related to the requirement of obtaining ${\cal O}(10^2)$ GeV $W/Z$-boson mass at the EW scale- see \cite{gravitino_DM}). We thus obtain the following:
${\cal V}^{\frac{29}{18}}\times({\cal O}(1))^2 \times {\cal O}(1) \times {\cal V}^{-\frac{2}{9}} \times {\cal V}^{-\frac{7}{5}} M_{W/Z}^2 \times \frac{\sqrt{10}}{({\cal O}(1) \times v \times {\cal V}^3 )}  \sim
({\cal O}(1))^2 \times \sqrt{10}{\cal V}^{-3} \frac{M_{W/Z\ {\rm in GeV}}^2}{v ({\rm GeV})}$.
Now, in the superspace notation, the kinetic terms for the gauge field are generically written as:
$\int d^2 \theta f_{ab}(\Phi) W^a W^b$ where $W^a$ is the gauge-invariant super-field strength and $W=W^a T^a$ for a non-abelian group - as $f_{ab}$ is an apriori arbitrary holomorphic function of $\Phi$. Consider hence $\Phi=T, f\sim e^T$ and look at $\int d^2\theta (T)^{2m+1}_{\theta,{\bar\theta}=0} W^2$ which will consist of $({\cal O}(1)^2 \times C_{1{\bar 3}} \langle a_1 {\bar a}_3\rangle )^{2m}  \times \sqrt{10} \times {\cal V}^{-3}
\frac{M_{W/Z\ {\rm in\ GeV}}^2}{GeV}$, which, e.g., for $m=2$ yields
$({\cal O}(1))^2 \times \sqrt{10} \times  {\cal V}^{\frac{8}{3}-3} \times \frac{M_{W/Z\ {\rm in\ GeV}}^2}{v ({\rm GeV})}$ or for ${\cal V}\sim10^4$, one obtains ${\cal O}(1)\frac{M^2_{W/Z ({\rm in\ GeV})}}{v ({\rm in\ GeV})}$.
Utilizing this, at EW scale,
$ C_{W^{+}H^{0}_{i}W^{-}}\equiv  \frac{M^{2}_{W}}{v}e^{i \phi_{W}}.$
 The value of complex Yukawa coupling to be used to evaluate EDM corresponding to Figure 10 have already been obtained in \cite{gravitino_DM} and given as: ${\hat Y}_{H^{0}_i e_{L}e^{c}_{R}} \sim {\cal V}^{-\frac{47}{45}}e^{i\phi_{y_e}}$ and of ${\hat Y}_{H^{0}_i u_{L}u^{c}_{R}} \sim {\cal V}^{-\frac{19}{18}}e^{i\phi_{y_u}}$. The matrix amplitude as well as analytical expression for W boson related loop diagrams has been worked out in \cite{Leigh et al}. We utilize the same in a modified form to get the numerical estimate of EDM corresponding to a loop diagram given in Figure~10.
\beqn
&& {\hskip -0.7in}\frac{d}{e}|_W = \frac{\alpha}{(4 \pi)^3 M^{2}_W}C_{W^{+}W^{-}\gamma}\sum_{i} Im ({\hat Y}_{H^{0}_i e_{L}e^{c}_{R}}  C_{W^{+}H^{0}_{i}W^{-}} )[5g(z^{W}_{i})+3f(z^{W}_{i})\ +\frac{3}{4}(g(z^{W}_{i})+h(z^{W}_{i}))],
\eeqn
where f(z) and g(z) are already defined in subsection {\bf 4.1} and
\beqn
h(z)=\frac{z}{2}\int^{1}_{0}dx \frac{1}{x(1-x)-z}\left(\frac{z}{x(1-x)-z}\ln\left(\frac{x(1-x)-z}{x}\right)-1\right);
\eeqn
where $z^{W}_{i}=\frac{m^{2}_{H^{0}_i}}{m^{2}_W}$. Considering $(\phi_{W}-\phi_{y_e})= (0,\frac{\pi}{2}]$;  using the values $m_{H^{0}_{1}}=125 GeV$ and $m_{H^{0}_{2}}={\cal V}^{\frac{59}{72}}m_{\frac{3}{2}}$, we get:
$f\Bigl(\frac{m^{2}_{H^{0}_i}}{m^{2}_W}\Bigr)= g\left(\frac{m^{2}_{H^{0}_i}}{m^{2}_W}\right)= h\Bigl(\frac{m^{2}_{H^{0}_i}}{m^{2}_W}\Bigr)={\cal O}(1)$,
and the EDM result for electron will be given as:
\beqn
\frac{d_e}{e}|_W \sim \frac{\alpha}{(4 \pi)^3} \frac{1}{v} \times {\cal V}^{-\frac{47}{45}}\equiv10^{13} GeV^{-1}\equiv 10^{-27} cm.
\eeqn
Similarly, by considering $(\phi_{W}-\phi_{y_u})= (0,\frac{\pi}{2}]$;  EDM of neutron will be given as:
\beqn
\frac{d_n}{e}|_W \sim \frac{\alpha}{(4 \pi)^3} \frac{1}{v} \times {\cal V}^{-\frac{19}{18}}\equiv10^{13} GeV^{-1}\equiv 10^{-27} cm.
\eeqn
\subsection{Two-Loop Level Rainbow Type Contribution to Electric Dipole Moment}
The two-loop level analysis of the supersymmetric effects to the fermion electric dipole moment has been extended by considering rainbow diagrams in addition to famous Barr-Zee diagrams with the expectation that they might give a significant contribution to fermionic EDM. The importance of these diagrams is discussed in detail in \cite{Pilaftsis}. In this subsection, we estimate the contribution of two-loop rainbow type of diagrams involving R-parity conserving supersymmetric interaction vertices and R-parity violating vertices. The CP violating phases  appear from the diagonalized eigenstates in the inner loop as well as from complex effective Yukawa couplings in the higgsino sector. The Feynman diagrams have been classified based on different types of inner one-loop insertions. One corresponds to one-loop effective higgsino-gaugino-gauge boson vertex and the other corresponds to one-loop effective higgsino-gaugino transition. The matrix amplitudes as well analytic expressions to estimate the EDM for above rainbow diagrams are calculated in \cite{Yamanaka_rainbow} to the first order in the external momentum carried by the gauge boson. We utilize their expressions to get the order of magnitude of EDM of electron as well as neutron in our case.
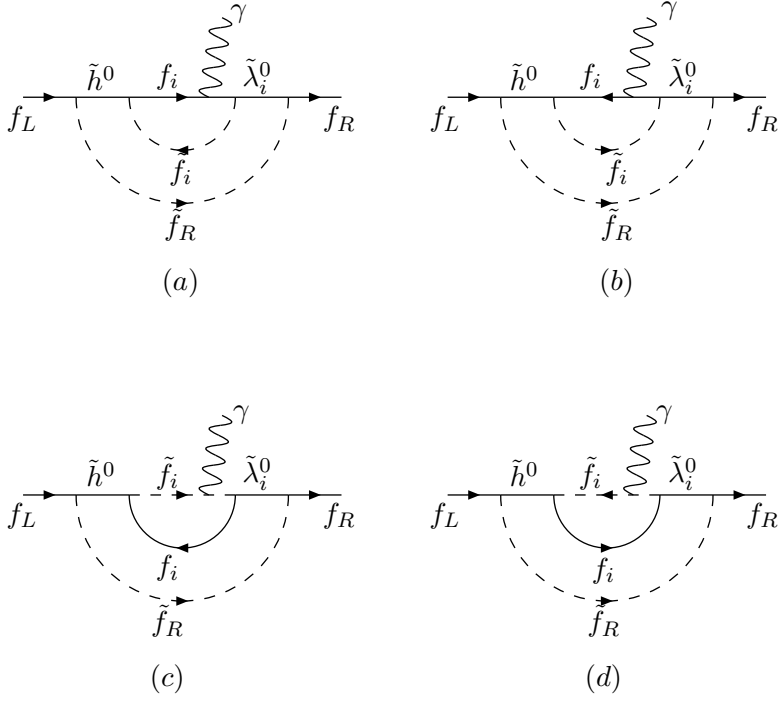
\begin{figure}
\begin{center}
\begin{picture}(345,250) (80,-200)
   \ArrowLine(95,50)(115,50)
   \Line(115,50)(135,50)
   \ArrowLine(135,50)(175,50)
   \Line(175,50)(195,50)
   \ArrowLine(195,50)(215,50)
   \DashArrowArc(155,50)(40,180,0){4}
     \DashArrowArcn(155,50)(20,0,180){4}
   \Photon(165,50)(170,80){4}{4}
   \Text(125,58)[]{${\tilde h}^{0}$}
      \Text(150,58)[]{$f_i$}
   \Text(184,58)[]{${\tilde \lambda}^{0}_{i}$}
   \Text(95,42)[]{{$f_{L}$}}
   \Text(177,82)[]{{$\gamma$}}
   \Text(215,42)[]{{$f_{R}$}}
   \Text(155,23)[]{{${\tilde f_i}$}}
   \Text(155,2)[]{{${\tilde f_R}$}}
   \Text(155,-20)[]{{$(a)$}}
   \ArrowLine(255,50)(275,50)
   \Line(275,50)(295,50)
   \ArrowLine(335,50)(295,50)
   \Line(335,50)(355,50)
   \ArrowLine(355,50)(375,50)
   \DashArrowArc(315,50)(40,180,0){4}
     \DashArrowArc(315,50)(20,180,0){4}
   \Photon(325,50)(330,80){4}{4}
   \Text(285,58)[]{${\tilde h}^{0}$}
    \Text(310,58)[]{$f_i$}
   \Text(345,58)[]{${\tilde \lambda}^{0}_{i}$}
   \Text(255,42)[]{{$f_{L}$}}
   \Text(340,83)[]{{$\gamma$}}
   \Text(375,42)[]{{$f_{R}$}}
     \Text(320,23)[]{{${\tilde f_i}$}}
      \Text(320,2)[]{{${\tilde f_R}$}}
   \Text(320,-20)[]{{$(b)$}}
   \ArrowLine(95,-100)(115,-100)
   \Line(115,-100)(135,-100)
   \DashArrowLine(135,-100)(175,-100){4}
   \Line(175,-100)(195,-100)
   \ArrowLine(195,-100)(215,-100)
   \DashArrowArc(155,-100)(40,180,0){4}
     \ArrowArcn(155,-100)(20,0,180)
   \Photon(165,-100)(170,-70){4}{4}
   \Text(125,-92)[]{${\tilde h}^{0}$}
      \Text(150,-92)[]{${\tilde f_i}$}
   \Text(184,-92)[]{${\tilde \lambda}^{0}_{i}$}
   \Text(95,-108)[]{{$f_{L}$}}
   \Text(178,-70)[]{{$\gamma$}}
   \Text(215,-108)[]{{$f_{R}$}}
   \Text(150,-128)[]{{${ f_i}$}}
   \Text(150,-148)[]{{${\tilde f_R}$}}
   \Text(150,-170)[]{{$(c)$}}
   \ArrowLine(255,-100)(275,-100)
   \Line(275,-100)(295,-100)
   \DashArrowLine(335,-100)(295,-100){4}
   \Line(335,-100)(355,-100)
   \ArrowLine(355,-100)(375,-100)
   \DashArrowArc(315,-100)(40,180,0){4}
     \ArrowArc(315,-100)(20,180,0)
   \Photon(325,-100)(330,-70){4}{4}
   \Text(285,-92)[]{${\tilde h}^{0}$}
      \Text(310,-92)[]{${\tilde f_i}$}
   \Text(345,-92)[]{${\tilde \lambda}^{0}_{i}$}
   \Text(255,-108)[]{{$f_{L}$}}
   \Text(338,-70)[]{{$\gamma$}}
   \Text(375,-108)[]{{$f_{R}$}}
     \Text(315,-129)[]{{${ f_i}$}}
      \Text(315,-148)[]{{${\tilde f_R}$}}
   \Text(315,-170)[]{{$(d)$}}
   \end{picture}
  \end{center}
  \vskip -0.3in
\caption{Two-loop level rainbow type diagrams involving higgsino-gaugino-gauge boson vertex.}
 \end{figure}

{{\bf R-parity conserving rainbow type contribution:}} For the loop diagrams given in Figure 11 and Figure 12, the result of EDM will given by the following formulae respectively:

\begin{eqnarray}
d_f^1
&\approx &
\sum
\frac{n_c (Q_f+ Q'_f) C_{{\tilde h^{0}}f_L {\tilde f}^{*}_R}  C_{{\tilde h^{0}}f_i f^{*}_i}}{64\pi^3} \sum_{n=1,2} |m_{\lambda^{0}_n}| \sin (\delta_f -\theta_n ) \times \frac{e ({g'}^{(n)}_{\tilde f_L} -{g'}^{(n)}_{\tilde f_R})  }{4\pi} \sin \theta_f \cos \theta_f
\nonumber\\
&&
\times \sum_{\tilde f= \tilde f_L,\tilde f_R} \hspace{-1em} s \, {g'}^{(n)}_{\tilde f}
\Bigl[
F' (|m_{\lambda_n}|^2 , |\mu|^2 , m_{\tilde f_{L/R}}^2 , m_{\tilde f_{2}}^2) -F' (|m_{\lambda^{0}_n}|^2 , |\mu|^2 , m_{\tilde f_{L/R}}^2, m_{\tilde f_{1}}^2)
\Bigr]
,
\label{eq:edmhgg}
\\
d_f^2
&=&
\sum_f
\frac{n_c Q'_{f_R} C_{{\tilde h^{0}}f_L{\tilde f}^{*}_R}  C_{{\tilde h^{0}}f_i f^{*}_i }}{64\pi^3}  \sum_{n=1,2} |m_{\lambda^{0}_n}| \sin (\delta_f -\theta_n)
 \times \frac{ e ({g'}^{(n)}_{\tilde f_L} +{g'}^{(n)}_{\tilde f_R})}{4\pi} \sin \theta_f \cos \theta_f
\nonumber\\
&&
\times \sum_{\tilde f= \tilde f_L,\tilde f_R} \hspace{-1em}  {g'}^{(n)}_{\tilde f} m_{\tilde F}^2
\Bigl[
F''(|m_{\lambda_n}|^2 , |\mu|^2 ,m_{\tilde f_{L/R}}^2, , m_{\tilde f_{1}}^2 )
-F''(|m_{\lambda_n}|^2 , |\mu|^2 , m_{\tilde f_{L/R}}^2, , m_{\tilde f_{2}}^2 )
\Bigr]
 ,\ \
\label{eq:edmhg}
\end{eqnarray}
where $n_c=3$ for the inner quark-squark loop and $n_c=1$ for the inner lepton-slepton loop. The fields $\tilde f_1$ and $\tilde f_2$ correspond to mass eigenstates of the sfermion $\tilde f$. The value of constant $s$ is $+1$ for left-handed sfermion $\tilde f_L$ and $-1$ for right-handed sfermion $\tilde f_R$.
The effective electric charges are given by $Q'_{f}= C_{ f_i f^{*}_i \gamma}Q_{f}$ and $Q'_{f_R}= C_{ f_R f^{*}_R\gamma}$.  The interaction vertices $C_{{\tilde h^{0}}f_L f^{*}_R} $ and $C_{{\tilde h^{0}}f f^{*} }$ correspond to effective Yukawa couplings. ${g'}^{(n)}_{\tilde f_L}$ and ${g'}^{(n)}_{\tilde f_R}$ denote effective gauge couplings corresponding to supersymmetric sfermions and the functions $F'$ and $F''$ are defined in \cite{Yamanaka_rainbow}.
 \begin{figure}
\begin{center}
\begin{picture}(345,120) (80,-50)
   \ArrowLine(95,50)(115,50)
   \Line(115,50)(135,50)
   \ArrowLine(175,50)(135,50)
   \Line(175,50)(195,50)
   \ArrowLine(195,50)(215,50)
   \DashArrowArc(155,50)(40,180,0){4}
     \DashArrowArc(155,50)(20,180,0){4}
   \Photon(165,11)(172,-18){4}{4}
   \Text(125,58)[]{${\tilde h}^{0}$}
    \Text(150,58)[]{$f_i$}
   \Text(184,58)[]{${\tilde \lambda}^{0}_{i}$}
   \Text(95,42)[]{{$f_{L}$}}
   \Text(178,-18)[]{{$\gamma$}}
   \Text(215,42)[]{{$f_{R}$}}
   \Text(155,21)[]{{${\tilde f_i}$}}
   \Text(155,2)[]{{${\tilde f_R}$}}
   \Text(155,-20)[]{{$(a)$}}
   \ArrowLine(255,50)(275,50)
   \Line(275,50)(295,50)
   \ArrowLine(295,50)(335,50)
   \Line(335,50)(355,50)
   \ArrowLine(355,50)(375,50)
   \DashArrowArc(315,50)(40,180,0){4}
     \DashArrowArcn(315,50)(20,0,180){4}
   \Photon(327,11)(340,-18){4}{4}
   \Text(285,58)[]{${\tilde h}^{0}$}
    \Text(310,58)[]{$f_i$}
   \Text(345,58)[]{${\tilde \lambda}^{0}_{i}$}
   \Text(255,42)[]{{$f_{L}$}}
   \Text(347,-20)[]{{$\gamma$}}
   \Text(375,42)[]{{$f_{R}$}}
     \Text(320,21)[]{{${\tilde f_i}$}}
      \Text(320,2)[]{{${\tilde f_R}$}}
   \Text(320,-20)[]{{$(b)$}}
      \end{picture}
  \end{center}
  \vskip -0.2in
\caption{Two-loop level rainbow type diagrams involving higgsino-gaugino transition.}
 \end{figure}
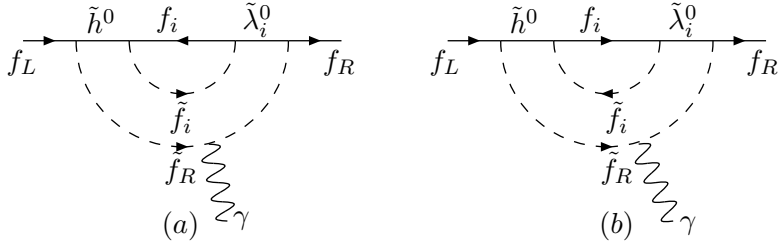

The effective Yukawa's as well as gauge interaction vertices are already calculated in {{\bf section 4}}. The magnitude of the values of the same are:
\begin{eqnarray}
&& |C^{{\tilde h^{0}}e_L {\tilde e}^{*}_R}| \equiv  {\cal V}^{-\frac{9}{5}}, |C^{{\tilde h^{0}}e_i {{\tilde e}_i}^{*} }| \equiv {\cal V}^{-\frac{10}{9}}, |C^{\tilde e_i \tilde e_i \gamma}|\equiv {\tilde f}{\cal V}^{\frac{53}{45}},  |C^{{\tilde h^{0}}u_L {\tilde u}^{*}_R}| \equiv  {\cal V}^{-\frac{5}{3}}, |C^{{\tilde h^{0}}u_i {{\tilde u}_i}^{*} }|\equiv {\cal V}^{-\frac{10}{9}}, \nonumber\\
&& |C^{\tilde u_i \tilde e_i \gamma}|\equiv {\tilde f}{\cal V}^{\frac{53}{45}},   {g'}^{(n)}_{\tilde e_R} \equiv |C^{e_R {\tilde e}^{*}_R {\lambda}^{0}_i}| \equiv  {\tilde f} {\cal V}^{-\frac{3}{5}} , {g'}^{(n)}_{\tilde e}\equiv  |C^{e_i {\tilde e}^{*}_i {\lambda}^{0}_i}| \equiv {\tilde f}{\cal V}^{-\frac{3}{5}},
 \nonumber\\
&&  {g'}^{(n)}_{\tilde e_R} \equiv |C^{u_R {\tilde u}^{*}_R {\lambda}^{0}_i}| \equiv  {\tilde f} {\cal V}^{-\frac{3}{5}} , {g'}^{(n)}_{\tilde u}\equiv  |C^{u_i {\tilde u}^{*}_i {\lambda}^{0}_i}| \equiv {\tilde f}{\cal V}^{-\frac{3}{5}}; ~~i=1,2.
\end{eqnarray}
Using the value of $ m_{\lambda^{0}_1}= m_{\lambda^{0}_2}= {\cal V}^{\frac{2}{3}}m_{\frac{3}{2}}$, $m^{2}_{\tilde e_{1}}=m^{2}_{\tilde u_{1}}=  {\cal V}m^{2}_{\frac{3}{2}} + m^{2}_{{\tilde e}_{12}}$, $m^{2}_{\tilde e_{2}}=m^{2}_{\tilde u_{2}}=  {\cal V}m^{2}_{\frac{3}{2}} - m^{2}_{{\tilde e}_{12}}$,  $m_{\tilde h^{0}}\equiv  {\cal V}^{\frac{59}{72}}m_{\frac{3}{2}}$, we get
\begin{eqnarray}
&& F' (| m_{\lambda_n}|^2 , |\mu|^2 , m_{\tilde f_{L/R}}^2, m_{\tilde f_{2}}^2) - F' (|m_{\lambda_n}|^2 , |\mu|^2 , m_{\tilde f_{L/R}}^2 , m_{\tilde f_{1}}^2)\equiv  10^{-24}, \nonumber\\
 && F' (| m_{\lambda_n}|^2 , |\mu|^2 , m_{\tilde f_{L/R}}^2, m_{\tilde f_{2}}^2) - F' (|m_{\lambda_n}|^2 , |\mu|^2 , m_{\tilde f_{L/R}}^2 , m_{\tilde f_{1}}^2)\equiv  10^{-43}.
 \end{eqnarray}
 Incorporating the above results in the analytical expression as given in (\ref{eq:edmhgg}) and (\ref{eq:edmhg})  (with the assumption that value of phase factor associated with all effective R-parity conserving Yukawa couplings are of ${\cal O}(1)$),
  \begin{eqnarray}
d_e^1/e
&\equiv &
{\tilde f}^{3} {\cal V}^{-\frac{18}{5}} \times  {\cal V}^{\frac{2}{3}}m_{\frac{3}{2}} \times (10^{-24}) GeV^{-2} \equiv {10}^{-57} cm.
\nonumber\\
d_e^2/e
&\equiv &
{\tilde f}^{3} {\cal V}^{-\frac{18}{5}}\times  {\cal V}^{\frac{5}{3}}m^{3}_{\frac{3}{2}} \times (10^{-43} )GeV^{-4} \equiv {10}^{-55} cm.
\label{eq:edmhg1}
\end{eqnarray}
and similarly,
\begin{eqnarray}
d_u^1/e
&\equiv &
{\tilde f}^{3} {\cal V}^{-\frac{10}{3}} \times  {\cal V}^{\frac{2}{3}}m_{\frac{3}{2}} \times (10^{-24}) GeV^{-2} \equiv {10}^{-56} cm.
\nonumber\\
d_u^2/e
&\equiv&
{\tilde f}^{3} {\cal V}^{-\frac{10}{3}}\times  {\cal V}^{\frac{5}{3}}m^{3}_{\frac{3}{2}} \times (10^{-43}) GeV^{-4} \equiv {10}^{-54} cm.
\label{eq:edmhg2}
\end{eqnarray}

So, the final EDM of electron as well as quark or neutron in case of R-parity conserving supersymmetric Feynman diagrams are given as:
 \begin{equation}
{d_e}/{e} = d^{1}_e/{e} + {d^{2}_e}/{e} \equiv {10}^{-55} cm, {d_u}/{e} = d^{1}_u/{e} + {d^{2}_u}/{e} \equiv {10}^{-54} cm.
\end{equation}
{{\bf R-parity violating rainbow type contribution:}}
The similar kind of Feynman diagrams can be drawn by replacing the neutral higgsino component with Dirac massless neutrino in Figure 11 and Figure 12.
The formulae of the EDM of the fermion f for two types of Feynman diagrams as defined in \cite{Yamanaka_Rparityvi} are given below:
\begin{eqnarray}
d^1_{f}
&=&
\sum_{n=1,2}
{\rm Im} ( C_{{\tilde h^{0}}f_L f^{*}_R}  C_{{\tilde h^{0}}f_i f^{*}_i } e^{i(\theta_n -\delta_{f_j})})
\frac{(Q_f+ Q'_f)n_c }{64 \pi^3} |m_{\lambda^{0}_n }| \frac{e (g^{(n)}_{\tilde f_L} -g^{(n)}_{\tilde f_R}) }{4\pi} \sin \theta_{f_j} \cos \theta_{f_j}
\nonumber\\
&&\hspace{5em} \times
\sum_{\tilde f= \tilde f_L , \tilde f_R} s\, g^{(n)}_{\tilde f}
\left[ f'(|m_{\lambda^{0}_n}|^2, 0 , m_{\tilde f_{L/R}}^2, m_{\tilde f_{1j}}^2 \, )
-f'(|m_{\lambda_n}|^2, 0 , m_{\tilde f_{L/R}}^2, m_{\tilde f_{2j}}^2 \, ) \right]
\, ,
\label{eq:edmnugg}
\\
d^2_{f}
&=&
-\sum_{n=1,2}
{\rm Im} ( C_{{\tilde h^{0}}f_L f^{*}_R}  C_{{\tilde h^{0}}f_i f^{*}_i }e^{i(\theta_n -\delta_{f_j})} )
\frac{Q'_{f_R}  n_c }{64 \pi^3} |m_{\lambda^{0}_n }| \frac{e (g^{(n)}_{\tilde f_L} +g^{(n)}_{\tilde f_R}) }{4\pi} \sin \theta_{f_j} \cos \theta_{f_j}
\nonumber\\
&&\hspace{2em} \times
\sum_{\tilde f= \tilde f_L , \tilde f_R} g^{(n)}_{\tilde f} m_{\tilde f_{k}}^2
\left[ f''(|m_{\lambda^{0}_n}|^2, 0 ,m_{\tilde f_{L/R}}^2, m_{\tilde f_{1}}^2 \, )
-f''(|m_{\lambda_n}|^2, 0 , m_{\tilde f_{L/R}}^2, m_{\tilde f_{2}}^2 \, ) \right]
\label{eq:edmnug}
\end{eqnarray}
The interaction vertices $C_{\nu^{0} f_L {\tilde f}^{*}_R} $ and $C_{\nu^0 f_i f^{*}_i }$ correspond to effective R-parity violating couplings. ${g'}^{(n)}_{\tilde f_L}$ and ${g'}^{(n)}_{\tilde f_R}$ denote effective gauge couplings corresponding to supersymmetric sfermions. The functions $F'$ and $F''$ are defined in  \cite{Yamanaka_Rparityvi}.

 Using the value of $ m_{\lambda^{0}_1}= m_{\lambda^{0}_2}= {\cal V}^{\frac{2}{3}}m_{\frac{3}{2}}$, $m^{2}_{\tilde e_{1}}=m^{2}_{\tilde u_{1}}=  {\cal V}m^{2}_{\frac{3}{2}} + m^{2}_{{\tilde e}_{12}}$, $m^{2}_{\tilde e_{2}}=m^{2}_{\tilde u_{2}}=  {\cal V}m^{2}_{\frac{3}{2}} - m^{2}_{{\tilde e}_{12}}$, we get
\begin{eqnarray}
&& F' (| m_{\lambda^{0}_n}|^2 ,0 , m_{\tilde f_{L/R}}^2, m_{\tilde f_{2}}^2) - F' (|m_{\lambda^{0}_n}|^2 , |\mu|^2 , m_{\tilde f_{L/R}}^2 , m_{\tilde f_{1}}^2)\equiv  10^{-22}, \nonumber\\
 && F' (| m_{\lambda^{0}_n}|^2 ,0 , m_{\tilde f_{L/R}}^2, m_{\tilde f_{2}}^2) - F' (|m_{\lambda^{0}_n}|^2 , |\mu|^2 , m_{\tilde f_{L/R}}^2 , m_{\tilde f_{1}}^2)\equiv  10^{-42}.
 \end{eqnarray}
The contribution of R-parity violating vertices are already calculated in \cite{gravitino_DM} in the context of ${\cal N}=1$ gauged supergravity action. The values of the same are as follows:
\begin{eqnarray}
&& |C^{\nu^{0} e_L {\tilde e}^{*}_R}| \equiv |C^{\nu^{0} e_i {{\tilde e}_i}^{*} }| \equiv {\cal V}^{-\frac{5}{3}}, |C^{\nu^{0} u_L {\tilde u}^{*}_R}| \equiv |C^{\nu^{0} u_i {{\tilde u}_i}^{*} }| \equiv {\cal V}^{-\frac{5}{3}}  ; i=1,2 .
 \end{eqnarray}
 Incorporating values of above-mentioned R-parity violating interaction vertices and the values of effective gauge couplings in the analytic expressions given in equations (\ref{eq:edmnugg}) and (\ref{eq:edmnug})(with the assumption that value of phase factor associated with all effective R-parity violating Yukawa couplings are of ${\cal O}(1)$),
\begin{eqnarray}
d_e^1/e= d_u^1/e
&\equiv &
{\tilde f}^{3} {\cal V}^{-\frac{10}{3}} \times  {\cal V}^{\frac{2}{3}}m_{\frac{3}{2}} \times (10^{-22}) GeV^{-2} \equiv {10}^{-53} cm.
\nonumber\\
d_e^1/e=d_u^2/e
&\equiv&
{\tilde f}^{3} {\cal V}^{-\frac{10}{3}}\times  {\cal V}^{\frac{5}{3}}m^{3}_{\frac{3}{2}} \times (10^{-42}) GeV^{-4} \equiv {10}^{-52} cm.
\label{eq:edmhg2}
\end{eqnarray}

So, the final EDM of electron as well as quark or neutron in case of R-parity violating Feynman diagrams are given as:
 \begin{equation}
{d_n}/{e} = {d_e}/{e} \equiv {10}^{-52} cm.
\end{equation}

The results of all two-loop diagrams contributing to EDM of electron/neutron are summarized in a table given below:

 \begin{table}[h]
\label{table:decay_lifetime}
\caption{Results of EDM of electron/neutron for all possible two-loop diagrams}  
\centering  
{\hskip -0.1in}
\begin{tabular}{l c c rrrrrrr}  
\hline\hline                       
 Two-loop particle exchange& Origin of complex phase & $d_{e}$(e cm)  & $d_n$(e cm)
\\ [1.0ex]
\hline
{$ h^{0}_i \gamma f$} &  Complex effective Yukawa couplings &  $10^{-36}$& $10^{-36}$    \\[1ex]
{$h^{0}_i \gamma \chi^{\pm}_{i}$} & ''& $10^{-47}$ & $ 10^{-47} $ & \\[1ex]
{${\tilde f} f \gamma$}&''  & $10^{-70}$& $ 10^{-70} $   \\[1ex]
{${\tilde f} ^{0}_i h^{0}_i \gamma$} &''& $10^{-29}$ & $ 10^{-29}  $   \\[1ex]
{$\gamma W^{\pm} h^{0}_i$}&Higgs exchange & $10^{-27}$ & $ 10^{-27} $   \\[1ex]
{${\tilde h^{0}} {\tilde f}{\lambda^{0}_i}$}(Rainbow type)& Diagonalized sfermion mass  & $10^{-55} $ & $10^{-54}$ \\[1ex]
&   eigenstates and effective Yukawas & &  \\[1ex]
{$\nu^0{\tilde f}{\lambda^{0}_i}$} (Rainbow type)& '' & $10^{-52} $ & $10^{-52}$ \\[1ex]
  \hline                          
\end{tabular}
\label{tab:PPer1}
\end{table}
\section{Summary and Conclusions}

To summarize, we have performed a quantitative order-of-magnitude analysis of EDM of electron and neutron in a phenomenological model which provides a local realization of large volume $D3/D7$ $\mu$-split supersymmetry that could possibly, locally be obtained, in the framework of four Wilson line moduli living on the world volume of fluxed stacks of space-time filling $D7$-branes wrapped around the `big divisor' and two position moduli of a mobile space-time filling $D3$-brane restricted to (nearly) a special Lagrangian three-cycle of a Swiss-Cheese Calabi-Yau. The proposed phenomenological model is governed by super-heavy gaugino and higgsino mass parameter in addition to heavy sfermion masses  except one light Higgs (obtained by considering linear combination of eigenstates of Higgs doublets at EW scale). Because of the presence of heavy gaugino/higgsino mass parameter, one can not ignore one-loop diagrams mediated by gaugino/higgsino's and sfermions as compared to 2-loop diagrams as traditional split SUSY models do. Keeping this in mind, we have taken into account the complete set of one-loop graphs and the dominant Higgs-mediated Barr-Zee diagrams. The non-zero CP-violating phase corresponding to dimension-five non-renormalizable EDM operator can be made to appear at one-loop and two-loop level from  off-diagonal component of scalar mass matrix and complex effective Yukawa couplings respectively in the context of ${\cal N}=1$ gauged supergravity action. We have considered the order of phases to exist in (0,$\frac{\pi}{2}$]. We have also shown that for a given choice of vevs of Wilson line as well as position moduli, the phases corresponding to effective Yukawa couplings do not change in the renormalization group flow from string scale down to Electro-Weak scale.  The relevant interaction vertices have been calculated in the context of ${\cal N}=1$ effective gauged supergravity action. Having  described the aforementioned model we estimate all possible one-loop as well as two-loop diagrams. In the one-loop graphs involving sfermions, the neutralino-mediated loop diagrams give the dominant contributions to the electron(neutron) EDM values  as compared to gaugino-mediated one-loop diagrams and the diagrams involving R-parity violating vertices because in ${\cal N}=1$ gauged supergravity, gaugino interaction vertices are dependent on suppressed dilute non-abelian fluxes and contribution of R-parity violating vertices are generally suppressed. However, all of the three-loop diagrams give a very suppressed contribution to electron and neutron EDM. Next, we consider one-loop diagrams involving Higgs and other supersymmetric/SM particles. By considering Standard Model like fermions with Higgs in a loop, we get the electron EDM estimate ($d_{e}/e\equiv 10^{-34}$ cm) and neutron EDM estimate ($d_{n}/e\equiv10^{-33}$ cm) considerably higher than the value predicted by Standard Model. Interestingly, by considering one-loop diagrams involving chargino and Higgs, the  electron EDM value turns out to be ($d_e/e\equiv 10^{-32}cm$) i.e one gets a healthy EDM of electron even in the presence of super-heavy chargino in the loop.   All of the above one-loop diagrams involve MSSM-like superfields and CP-violating phases appear from visible sector fields only. For a full-fledged analysis, we have taken into account goldstino supermultiplet also as the physical degrees of freedom in one-loop diagrams. As sgoldstino corresponds to the bosonic component of the superfield corresponding to which there is a supersymmetry breaking and same occurs maximally in our large volume $D3/D7$ model via complex `big' divisor volume modulus ($\tau_B$), we have identified the sgoldstino field with complex $\tau_B$ field.  Since, the fermionic component goldstino gets absorbed into the gravitino and becomes a longitudinal component of the massive gravitino, we basically consider one-loop diagrams involving gravitino and sgoldstino in the loop. In such kind of loop diagrams, CP-violating phases appear from hidden sector fields. However, by evaluating the matrix amplitudes of these loop diagrams, we get a very suppressed contribution of electron and neutron EDM. The results of all one-loop diagrams are summarized in Table \ref{tab:PPer}. In case of two-loop diagrams, we have evaluated contribution of Barr-Zee diagrams involving sfermions/fermions in an internal loop and mediated via $\gamma h$ exchange, and R-parity violating diagram involving fermions but mediated via ${\tilde f} h$ exchange. Here, the two-loop Barr-Zee diagrams involving heavy sfermions and a light Higgs  give a most dominant contribution of EDM (${d_{(e/n)}}/{e}\equiv 10^{-29} cm$) as compared to 2-loop diagrams involving only SM-like particles. With substantial fine tuning in Calabi-Yau volume, one can hope to produce EDM results same as experimental limits. Next, inspired by the approach given in \cite{ Leigh et al, Barr_Zee, weinberg_Higgs, weinberg_Higgs_2} to obtain large EDM value(almost same as an experimental bound) from Barr-Zee diagrams involving top quarks and W bosons loop in multi-Higgs models, we have provided an estimate of the same using two Higgs doublets given in the context of $\mu$-split-SUSY. By showing the possibility of obtaining the numerical estimate of all SM-like vertices relevant for these diagrams  to be same as their standard values in the context of ${\cal N}=1$ gauged supergravity model, we also produce EDM ($d_{(e,n)}/e\equiv  10^{-27}$) cm in case of a Barr-Zee diagram involving $W$ Boson.  As evaluated explicitly,  we show that two-loop rainbow diagrams give a very suppressed contribution as compared to Barr-Zee diagrams. The results of all two-loop diagrams are summarized in Table \ref{tab:PPer1}.Thus, we conclude that in our large volume $D3/D7$ $\mu$-split SUSY model, despite the presence of very heavy supersymmetric scalars/fermions in the loops, we are able to produce a contribution to electric dipole moment of  both electron as well as neutron close to experimental bound at two-loop level and a sizable contribution even at one-loop level.

All of the above results have been obtained in the context of the model which can be constructed locally near a particular nearly special Lagrangian three-cycle  of a Swiss-Cheese Calabi-Yau three-fold. It would be interesting to determine the global embedding of our model. Further, in the $D3/D7$ set-up described above, we have shown the possibility of identification of  Wilson line moduli only with first or second generation quarks and leptons. By extending the set-up to include Wilson line moduli identifiable with second and third generation quarks, one hopes  to obtain via the one-loop and two-loop Barr-Zee diagrams involving fermions, the value of electron/neutron EDM very close to experimental bound for a given choice of the internal complex three-fold volume.

\section*{Acknowledgements}

MD is supported by a CSIR Senior Research Fellowship. AM would like to thank Syrcause University, Johns Hopkins University and the Abdus Salam ICTP (under the regular associateship program) for their kind hospitality and support where part of this work was done, and would also thank
Nima Arkani-Hamed for useful comments and suggestions when he had presented some preliminary results of this paper in a seminar at Johns Hopkins. We acknowledge participation of S.Serrao in the earlier stages of the project.

\appendix
\section{Chargino Mass Matrix}

The chargino mass matrix is formed by mixing (charged)wino's and higgsino after electroweak symmetry breaking. In ${\cal N}=1$ gauged supergravity, the interaction vertex corresponding to Higgs-gaugino-higgsino term is given by ${\cal L}= g_{YM}g_{T^B{\cal Z}^i}X^B\tilde{H}^i\lambda^{i} + \partial_{{\cal Z}_i}T_B D^{B}\tilde{H}^i\lambda^{i} $ where $\lambda^{i}$ corresponds to gaugino (such as the bino/wino). Expanding the same in the fluctuations linear in $Z_i$, we have
\begin{equation}
\label{eq:gYM_ReT_II}
g_{YM}g_{T^B{\cal Z}^i}X^{B}= {\tilde f} {\cal V}^{-2}\frac{{\cal Z}_i}{M_P}, (\partial_{{\cal Z}_i}T_B) D^{B}\equiv    {\tilde f} {\cal V}^{-\frac{4}{3}}\frac{{\cal Z}_i}{M_P}
\end{equation}
After giving a VEV to ${\cal Z}_i$, the interaction vertex corresponding to mixing between gaugino and higgsino will be given as:
\begin{equation}
\label{eq:gYM_ReT_II}
C^{\tilde \lambda^{-}-{{\tilde H}^{-}_{d}}}/C^{\tilde \lambda^{+}-{{\tilde H}^{+}_{u}}}=\frac{ {\tilde f} {\cal V}^{-\frac{4}{3}}}{\sqrt{K_{{\cal Z}_i {\cal Z}_i}K_{{\cal Z}_i {\cal Z}_i}}}\equiv    {\tilde f} {\cal V}^{-\frac{1}{3}}\frac{v}{M_P},{\rm where}~v=246GeV.
\end{equation}
For  higgsino-doublets
${\tilde H_u}=\left({\tilde H}^{0}_u , {\tilde H}^{+}_u \right), {\tilde H}_d=\left({\tilde H}^{-}_d, {\tilde H}^{0}_d \right)$,
the chargino mass matrix is given as:
\begin{equation}
 M_{{\tilde \chi}^{-}}=\left(\begin{array}{ccccccc}
  M^{2}_{{\tilde H}^{-}_{d}} & C^{\tilde \lambda^{-}-{{\tilde H}^{-}_{d}}}\\
C^{\tilde \lambda^{-}-{{\tilde H}^{-}_{d}}} &  M^{2}_{{\tilde \lambda}^{-}}
 \end{array} \right),  M_{{\tilde \chi}^{+}}=\left(\begin{array}{ccccccc}
  M^{2}_{{\tilde H}^{+}_{u}} & C^{\tilde \lambda^{+}-{{\tilde H}^{+}_{u}}}\\
C^{\tilde \lambda^{+}-{{\tilde H}^{+}_{u}}} &  M^{2}_{{\tilde \lambda}^{+}}
 \end{array} \right)
\end{equation}

 Incorporating value of  $M_{{\tilde \lambda}^{+}}=M_{{\tilde \lambda}^{-}}= {\cal V}^{\frac{2}{3}} m_{\frac{3}{2}}$, $M_{{{\tilde H}^{-}}_{d}}=  M_{{{\tilde H}^{+}}_{u}}= {\cal V}^{\frac{59}{72}}m_{\frac{3}{2}}$ and $m_{\frac{3}{2}}= {\cal V}^{-2}M_P$  at electroweak scale, we have
\begin{equation}
 M_{{\tilde \chi}^{\pm}}=\left(\begin{array}{ccccccc}
  {\cal V}^{-\frac{4}{3}} & \frac{v}{M_P} {\tilde f}{\cal V}^{-\frac{1}{3}}\\
 \frac{v}{M_P}{\tilde f}{\cal V}^{-\frac{1}{3}}  & {\cal V}^{-\frac{85}{72}}
 \end{array} \right),
\end{equation}

 giving eigenvalues
\begin{eqnarray}
\label{eq:charginoevalues}
& & \Biggl\{\frac{{M^{2}_P} {\cal V}^{4/3}+{M^{2}_P} {\cal V}^{85/72}-\sqrt{{M^{4}_P} {\cal V}^{8/3}-2 {M^{4}_P} {\cal V}^{181/72}+{M^{4}_P} {\cal V}^{85/36}+4 {\tilde f}^2{M^{2}_P} v^2 {\cal V}^{109/36}}}{2
   {M^{2}_P} {\cal V}^{157/72}},\nonumber\\
& & \frac{{M^{2}_P} {\cal V}^{4/3}+{M^{2} _P}{\cal V}^{85/72}+\sqrt{{M^{4}_P} {\cal V}^{8/3}-2 {M^{4}_P} {\cal V}^{181/72}+{M^{4}_P}{\cal V}^{85/36}+4 {\tilde f}^2 {M^{2}_P} {v}^2 {\cal V}^{109/36}}}{2
   {M^{2}_P} {\cal V}^{157/72}}\Biggr\}M_P, \nonumber\\
\end{eqnarray}
and normalized eigenvectors:
\begin{eqnarray}
\label{eq:charginoI}
&& {\tilde \chi}^{+}_{1}= -{\tilde H}^{+}_{u}+\left(\frac{v}{M_P}{\tilde f}{\cal V}^{\frac{5}{6}}\right){\tilde \lambda}^{+}_{i},{\tilde \chi}^{-}_{1}=  -{\tilde H}^{-}_{d}+\left(\frac{v}{M_P}{\tilde f}{\cal V}^{\frac{5}{6}}\right){\tilde \lambda}^{-}_{i},~ {\rm and}~m_{{\tilde \chi}^{\pm}_{1}}\equiv  {\cal V}^{\frac{59}{72}}m_{\frac{3}{2}}\nonumber\\
&&  {\tilde \chi}^{+}_{2}= {\tilde \lambda}^{+}_{i} +\left(\frac{v}{M_P}{\tilde f}{\cal V}^{\frac{5}{6}}\right){\tilde H}^{+}_{u},{\tilde \chi}^{-}_{2}= {\tilde \lambda}^{-}_{i} +\left(\frac{v}{M_P}{\tilde f}{\cal V}^{\frac{5}{6}}\right) {\tilde H}^{-}_{d},~ {\rm and}~m_{{\tilde \chi}^{\pm}_{2}}\equiv  {\cal V}^{\frac{2}{3}}m_{\frac{3}{2}}.
\end{eqnarray}

\end{document}